\global\def\draftcontrol{0}

   \def\versionno{ Exponential Networks }

\catcode`\@=11

\expandafter\ifx\csname draftcontrol\endcsname\relax\global\def\draftcontrol{0} 
\fi 

{\count255=\time\divide\count255 by 60 
\xdef\hourmin{\number\count255} 
\multiply\count255 by-60\advance\count255 by\time 
\xdef\hourmin{\hourmin:\ifnum\count255<10 0\fi\the\count255}} 
\def\draftdate{\number\month/\number\day/\number\year\ \ \ \hourmin } 

\newcommand\makepapertitle{\par
  \begingroup 
    \renewcommand\thefootnote{\@fnsymbol\c@footnote}%
    \def\@makefnmark{\rlap{\@textsuperscript{\normalfont\@thefnmark}}}%
    \long\def\@makefntext##1{\parindent 1em\noindent 
            \hb@xt@1.8em{%
                \hss\@textsuperscript{\normalfont\@thefnmark}}##1}%
     \newpage 
     \global\@topnum\z@   
     \@makepapertitle 
     \thispagestyle{empty}\@thanks 
  \endgroup 
  \setcounter{footnote}{0}%
  \global\let\thanks\relax 
  \global\let\makepapertitle\relax 
  \global\let\@makepapertitle\relax 
  \global\let\@thanks\@empty 
  \global\let\@author\@empty 
  \global\let\@date\@empty 
  \global\let\@title\@empty 
  \global\let\title\relax 
  \global\let\author\relax 
  \global\let\date\relax 
  \global\let\and\relax 
  \def\version{\let\version\@version\@gobble} 
} 
\def\@makepapertitle{%
  \newpage 
   \ifnum\draftcontrol=1 {} 
   \version\versionno 
   \vskip 5.5em%
   \else 
   \hfill\hbox to 4.0cm {\parbox{5.5cm}{\@pubnum}\hss}%
   \vskip 5.5em%
   \fi 
   \begin{center}%
   \let \footnote \thanks 
      {\hskip -0\textwidth \hbox to 1\textwidth%
        {\centerline{\Large\bf{\noindent\@title}}}}%
     \vskip 1.7em%
     {\normalsize
       \lineskip .5em%
       \begin{tabular}[t]{c}%
         \@author 
       \end{tabular}\par}%
     \vskip 1.2em%
     {\@bstract}%
     \end{center}%
     \vfill
     \@date\\[-.2cm]\noindent
     \rule{12em}{.05em}\par\noindent
     \@email%
   \par 
} 

\gdef\@pubnum{} 
\def\pubnum#1{%
  \gdef\@pubnum{#1}} 

\gdef\@bstract{} 
\def\Abstract#1{%
  \gdef\@bstract{%
   \parbox{\textwidth-0pc}{%
   \centerline{\bf Abstract}\penalty1000 
   \noindent
   \renewcommand\baselinestretch{1.0} 
   {#1}}} 
} 

\gdef\@email{}
\def\email#1{%
   \gdef\@email{%
   \small {#1}}
}

\def\ps@paper{\let\@mkboth\@gobbletwo%
     \ifnum\draftcontrol=1 
        \def\@oddfoot{\hbox to \textwidth{\tiny \versionno \hfil\tiny\draftdate}%
        \hskip -\textwidth \hbox to \textwidth{\hfil\rm\thepage\hfil}}%
     \else\def\@oddfoot{\hbox to \textwidth{\hfil\rm\thepage\hfil}} 
     \fi 
     \let\@evenfoot\@oddfoot 
} 

\newenvironment{acknowledgments}{%
\vskip 3.25ex 
\addcontentsline{toc}{section}{Acknowledgments}
\noindent {\bf Acknowledgments} 
} 

\def\@version#1{\ifnum\draftcontrol=1 
\typeout{}\typeout{#1}\typeout{} 
\vskip3mm\centerline{\hbox{\fbox{\normalsize{\tt DRAFT -- #1 -- } 
                   {\draftdate}}}}\vskip3mm 
\fi} 
\let\version\@version 

\documentclass[12pt,letterpaper]{article} 
\pdfoutput=1

\usepackage[colorlinks=false]{hyperref}
\usepackage{amsmath,bm,amsfonts,amssymb,array,calc,amsthm,rotating}
\usepackage{epsfig}
\usepackage{graphicx}
\usepackage{color}
\usepackage{amscd}
\usepackage{tikz-cd}[2013/12/13]
\usepackage{caption,subcaption}
\usepackage{verbatim}
\usetikzlibrary{calc}

\ifnum\draftcontrol=1
\usepackage[color,notref]{showkeys}

\definecolor{refkey}{rgb}{0,0.3,0}
\definecolor{labelkey}{rgb}{0,0,.7}
\def\SK@@ref#1>#2\SK@{%
  \leavevmode\vbox to\z@{{%
    \vss
    \SK@refcolor
    \rlap{\vrule\raise .75em%
       \hbox{\underbar{\normalfont\tiny#2}}}}}}
\fi

\definecolor{Mblue}{rgb}{0.368417, 0.506779, 0.709798}
\definecolor{Morange}{rgb}{0.880722, 0.611041, 0.142051}
\definecolor{Mgreen}{rgb}{0.560181, 0.691569, 0.194885}
\definecolor{D4}{rgb}{0,0,1}

\renewcommand\baselinestretch{1.25} 
\renewcommand\section{\@startsection {section}{1}{\z@}%
                                   {-3.5ex \@plus -1ex \@minus -.2ex}%
                                   {2.3ex \@plus.2ex}%
                                   {\normalfont\large\bfseries}} 
\newcommand\ssection{\@startsection {section}{1}{\z@}%
                                   {-0.5ex \@plus -1ex \@minus -.2ex}%
                                   {2.3ex \@plus.2ex}%
                                   {\normalfont\large\bfseries}} 
\renewcommand\subsection{\@startsection{subsection}{2}{\z@}%
                                   {-3.25ex\@plus -1ex \@minus -.2ex}%
                                   {1.5ex \@plus .2ex}%
                                   {\normalfont\normalsize\bfseries}} 
\newcommand\ssubsection{\@startsection{subsection}{2}{\z@}%
                                   {-0.5ex\@plus -1ex \@minus -.2ex}%
                                   {1.5ex \@plus .2ex}%
                                   {\normalfont\normalsize\bfseries}} 
\renewcommand\subsubsection{\@startsection{subsubsection}{3}{\z@}%
                                   {-3.25ex\@plus -1ex \@minus -.2ex}%
                                   {1.5ex \@plus .2ex}%
                                   {\normalfont\normalsize\it}} 
\renewcommand\paragraph{\@startsection{paragraph}{4}{\z@}%
                                   {-3.25ex\@plus -1ex \@minus -.2ex}%
                                   {1.5ex \@plus .2ex}%
                                   {\normalfont\normalsize\it}} 
\renewcommand\subparagraph{\@startsection{subparagraph}{5}{\z@}%
                                   {-1.25ex\@plus -1ex \@minus -.2ex}%
                                   {0ex \@plus .2ex}%
                                   {\normalfont\normalsize\it}}

\numberwithin{equation}{section} 

\long\def\@makecaption#1#2{%
  \vskip\abovecaptionskip
  \sbox\@tempboxa{{\bf #1:} #2}%
  \ifdim \wd\@tempboxa >\hsize
    {\small\bf #1:} {\small #2}\par
  \else
    \global \@minipagefalse
    \hb@xt@\hsize{\hfil\box\@tempboxa\hfil}%
  \fi
  \vskip\belowcaptionskip}

\renewcommand*\l@section[2]{%
  \ifnum \c@tocdepth >\z@
    \addpenalty\@secpenalty
    \addvspace{.1em \@plus\p@}%
    \setlength\@tempdima{1.5em}%
    \begingroup
      \parindent \z@ \rightskip \@pnumwidth
      \parfillskip -\@pnumwidth
      \leavevmode \bfseries
      \advance\leftskip\@tempdima
      \hskip -\leftskip
      #1\nobreak\hfil \nobreak\hb@xt@\@pnumwidth{\hss #2}\par
    \endgroup
  \fi}
\renewcommand*\l@subsection{\addvspace{.0em \@plus\p@}\@dottedtocline{2}{1.5em}{2.3em}}
\renewcommand*\l@subsubsection{\addvspace{-.2em \@plus\p@}\@dottedtocline{3}{3.8em}{3.2em}}

\def\hepth#1{\href{http://xxx.arxiv.org/abs/hep-th/#1}{{arXiv:hep-th/#1}}}
\def\mathph#1{\href{http://xxx.arxiv.org/abs/math-ph/#1}{{arXiv:math-ph/#1}}}

\def\alggeom#1{\href{http://xxx.arxiv.org/abs/alg-geom/#1}{{arXiv:alg-geom/#1}}}
\def\arxiv#1#2{\href{http://xxx.arxiv.org/abs/#1}{{arXiv:#1 [#2]}}}
\let\arXiv\arxiv

\definecolor{refcol}{rgb}{0.2,0.2,0.8}
\definecolor{eqcol}{rgb}{.6,0,0}
\definecolor{purple}{cmyk}{0,1,0,0}

\gdef\@citecolor{refcol}
\gdef\@linkcolor{eqcol}
\def\colorlinkspurple{\gdef\@urlcolor{purple}}
\def\colorlinksblue{\gdef\@urlcolor{blue}}
\def\colorlinksred{\gdef\@urlcolor{red}}

\def\ie{{\it i.e.}}

\def\revise#1       {\raisebox{-0em}{\rule{3pt}{1em}}%
                     \marginpar{\raisebox{.5em}{\vrule width3pt\ 
                     \vrule width0pt height 0pt depth0.5em 
                     \hbox to 0cm{\hspace{0cm}{%
                     \parbox[t]{4em}{\raggedright\footnotesize{#1}}}\hss}}}}

\def\cala         {{\cal A}}

\def\cali         {{\cal I}}

\def\caln         {{\cal N}}

\def\cals         {{\cal S}}

\def\calw         {{\cal W}}

\def\complex      {{\mathbb C}} 
 
\def\projective   {{\mathbb P}} 
 
\def\reals        {{\mathbb R}} 
\def\zet          {{\mathbb Z}} 
\def\CP{\complex\projective}

\def\ee           {{\it e}} 
\def\ii           {{\it i}}

\def\Re           {{\rm Re\hskip0.1em}} 
\def\Im           {{\rm Im\hskip0.1em}}

\newcommand\topa[2]{\genfrac{}{}{0pt}{2}{\scriptstyle #1}{\scriptstyle #2}}

\def\sqr#1#2{{\vcenter{\vbox{\hrule height.#2pt   
 \hbox{\vrule width.#2pt height#1pt \kern#1pt 
 \vrule width.#2pt}\hrule height.#2pt}}}}

\let\C\complex
\let\R\reals
\let\Z\zet
\let\cO\calo
\let\cW\calw
\newcommand{\fB}{\mathfrak{B}}
\newcommand{\pp}[2]{\frac{\partial #1}{\partial #2}}
\newcommand{\dd}[2]{\frac{d #1}{d #2}}

\DeclareMathOperator{\Tr}{Tr}

\DeclareMathOperator{\Hom}{Hom}

\DeclareMathOperator{\Rep}{Rep}
\DeclareMathOperator{\Mod}{Mod}

\def\Dtwobar{\overline{{\textrm D}2}}
\def\D0bar{\overline{{\textrm D}0}}

\catcode`\@=12

\begin{document}


\title{Exponential Networks and Representations of Quivers}

\date{November 2016}

\author{
Richard Eager$^{\dag}$, Sam Alexandre\ Selmani$^{\dag\ddag}$ and
Johannes Walcher$^{\dag}$ \\[0.2cm]
\it $^{\dag}$ Mathematical Institute, \\
\it Heidelberg University, Heidelberg, Germany \\[.1cm]
\it $^{\ddag}$ Department of Physics, \\ 
\it McGill University,
\it Montr\'eal, Qu\'ebec, Canada}

\email{eager@mathi.uni-heidelberg.de, 
sam.selmani@physics.mcgill.ca,  walcher@uni-heidelberg.de}

\Abstract{We study the geometric description of BPS states in supersymmetric 
theories with eight supercharges in terms of geodesic networks on suitable
spectral curves. We lift and extend several constructions of 
Gaiotto-Moore-Neitzke from gauge theory to local Calabi-Yau threefolds 
and related models. The differential is multi-valued on the covering
curve and features a new type of logarithmic singularity in order to account 
for D0-branes and non-compact 
D4-branes, respectively.
We describe local rules for the three-way junctions of BPS 
trajectories relative to a particular 
framing of the curve. We 
reproduce BPS quivers of local geometries and illustrate the 
wall-crossing of finite-mass bound states in several new examples. 
We describe first steps toward understanding the spectrum of
framed BPS states in terms of such ``exponential networks.''
}

\makepapertitle


\version\versionno

\vskip 1em

\newpage

\tableofcontents

\newpage


\hfill\parbox[b]{10cm}{{\it
It is a capital mistake to theorize before you have all the evidence. 
It biases the judgment.} (Sherlock Holmes)}

\section{Introduction}

The spectrum of BPS states plays a prominent role in the study of quantum 
mechanical theories with extended supersymmetry and in the interest of
such theories for mathematics. Of particular significance are theories 
with eight real supercharges, such as four-dimensional gauge theories with 
$\caln=2$ supersymmetry, or compactifications of M-theory or type II 
string theory on Calabi-Yau threefolds.

In such models, the intrinsic representation data of the supersymmetry 
algebra (BPS charges and masses, their monodromy and singularities, the 
chiral metric) fit together in such a tightly constrained way over the moduli space 
of vacua that its geometric structure can be recovered from a clever 
combination of local flatness and global consistency conditions. Typically,
this data can be solved for by studying the classical variation 
of an auxiliary spectral geometry. For string/M-theory, this is the 
Calabi-Yau manifold itself (or rather, its mirror), and for gauge theory, 
Seiberg-Witten geometry. 
These connections are extremely well understood, admit generalizations 
to gravitational and higher-derivative corrections of the effective theory, 
and include relations to classical and quantum integrable systems and a 
variety of interesting mathematics.

On the other hand, determining the representation content, \ie, describing
the actual BPS subspace of the Hilbert space, is much more subtle, and it is 
not in general clear to what extent this data is determined by the properties of 
the vacuum  manifold alone. This has to do with the fact that while the graded 
dimensions of the space of BPS states (the BPS degeneracies) are locally constant 
over the moduli space, they can jump discontinuously at the crossing of certain 
real co-dimension-one walls. There is by now a lot of circumstantial evidence 
that wall-crossing is not incompatible with the idea that the BPS spectrum
is in fact determined by the effective low-energy dynamics. First of all,
the location of the walls of course follows from the properties of the charge 
lattice (the central charge), which is determined by special geometry. 
Secondly, the change of the BPS spectrum across the walls can be studied
from the dynamics of bound states in the effective theory \cite{mode} and 
is subject to the fully general formula of Kontsevich-Soibelman
\cite{ks1}. The first physics derivation of this Kontsevich-Soibelman 
wall-crossing formula \cite{gmn0} exploits
precisely the tension between the discontinuous changes in the BPS degeneracies
and the smoothness of the hyperk\"ahler metric to which they contribute.
In special cases, these constraints allow for a full calculation of the
BPS spectrum \cite{gmn00,gmn12}.
Moreover, at least for strings on Calabi-Yau, the
OSV conjecture \cite{osv} offers an even more general relation between the 
BPS degeneracy and the topological (string) partition function whose
asymptotic expansion captures the perturbative corrections to the low-energy
theory.

With or without assuming that these intricate consistency conditions can 
ultimately be completely solved, it is fruitful to also investigate 
the BPS spectrum more directly from the point of view of the spectral 
geometry. In string compactifications, for instance, BPS states arise 
by wrapping D-branes on 
supersymmetric cycles in the Calabi-Yau, and their degeneracies are encoded
in the cohomology of the associated moduli spaces. It is then not only satisfying to 
identify precisely the problem to which the wall-crossing formulas provide
an answer, but the explicit solution to some subclass also provides valuable 
complementary information to check the various conjectures.

\medskip

The present paper grew out of attempts to generalize the geometric
description of BPS states in  four-dimensional $\caln=2$ supersymmetric gauge 
theories that was given by Klemm-Lerche-Mayr-Vafa-Warner 
\cite{klmvw}\footnote{See citations of \cite{klmvw} for other work
done in the 1990's.} and that has received renewed interest in recent
years following the work of Gaiotto-Moore-Neitzke on spectral networks 
\cite{gmn12}. This approach, which we will review below, can be seen to arise 
in a suitable limit from the geometric realization of the gauge theories, 
either by 2-branes ending on 5-branes in M-theory, or by dimensional 
reduction of 3-branes wrapping supersymmetric cycles in type IIB string 
theory.

The main idea and motivation for the generalization we are seeking
can be explained from that second point view: The type IIB
local geometries are the mirror manifolds of the toric Calabi-Yau 
manifolds that geometrically engineer the gauge theory in type IIA.
In this context, it is known that the 3-fold geometry reduces to an
effective one-dimensional description even before taking the gauge
theory limit to the Seiberg-Witten curve, and that this holds also
for local toric geometries that do not admit a gauge theory 
interpretation. Among the evidence for this statement we mention
the coincidence of the period calculation \cite{batyrev,hova}, the 
evaluation of D-brane probes \cite{agva}, the so-called remodeling
conjecture \cite{remodel}, as well as modularity in its various forms.
Our work will provide additional corroboration.

In the A-model, BPS states arise from B-branes. Their counting is, 
in many instances, rather well understood mathematically in terms of the 
cohomology of moduli of coherent sheaves, and many of the conjectures 
that we alluded to above have been checked and verified.
It is known in principle that the problem whose solution reproduces the 
BPS state counting in the B-model is related to the moduli of special 
Lagrangian submanifolds (stable A-branes).  Making this explicit is, 
however, complicated by the need to complexify the moduli space to resolve 
certain uncontrolled singularities, and by the obstructions by holomorphic 
disks whose effects on the problem are still not completely understood.

We will not be able to fully fill this important gap in this paper.
However, we will give, in some simple examples, a {\it proof of principle}
that BPS degeneracies in local Calabi-Yau manifolds can be understood
from the B-model perspective in terms of a calibrated geometry that
is the reduction of special Lagrangian geometry to the mirror curve,
suitably corrected by holomorphic disks.

To this end, we will study the analogue of spectral networks in the 
simplest possible examples of local Calabi-Yau manifolds. We will attempt
to reproduce as much as possible of the BPS spectrum that is known
from the A-model for these geometries. A useful tool that 
is shared by the A- and B-model is the description of D-brane bound states 
in terms of the representation theory of so-called BPS quivers 
\cite{dfr,cv,accerv}. This theory also plays an important role in our story.

An interesting payoff of our work is a kind of ``reversed'' perspective on mirror 
symmetry \`a la Strominger-Yau-Zaslow for local Calabi-Yau manifolds. Recall
that in the SYZ picture, the mirror manifold is interpreted as the moduli 
space of a particular special Lagrangian 3-torus. This picture is 
in principle completely symmetric between A- and B-model for any given 
mirror pair. In the local case, however, one usually starts from A-branes
on the toric side, and reconstructs the mirror as a Landau-Ginzburg model 
from the obstruction theory of the toric fibers. In our examples, we 
will start from a particular calibrated submanifold in the
B-model (as we will see, a suitable spectral network), whose moduli space
is related (after complexification) to the original toric manifold.

The paper is organized as follows. We begin in section \ref{review} with
a brief review of the work that we will generalize, and a summary
of the new features that derive from the exponential nature of the 
differential on the local mirror curves. We give further mathematical 
details in section \ref{mathi}, and an overview of the current state of
our theory in section \ref{executive}. In our first example, section
\ref{conifold}, we reproduce the finite BPS spectrum of the conifold by
exploiting a new junction rule for BPS trajectories. The main feature
appears in section \ref{localP2}, in which we produce exponential networks
for a large class of BPS states on local $\projective^2$. We concentrate
on states with a reasonable quiver representation, and study their 
wall-crossing under variation of the stability condition. In section \ref{C3},
we return to what should be regarded as the simplest example of a Calabi-Yau, 
$\complex^3$, and describe our best attempts at framed BPS states in this
model. Along the way, we study the moduli space of a distinguished
state that is mirror to a single pure D0-brane (a calibrated
version of what is known in symplectic geometry as the ``Seidel
Lagrangian'' \cite{seidel1,auroux}), and show that this moduli space retracts to the toric
diagram of the A-model Calabi-Yau.

\section{BPS Trajectories, Quivers, and Mirror Symmetry}
\label{review}

The main goal of this paper is to provide a new, B-model, perspective on BPS 
states of local Calabi-Yau manifolds by combining and generalizing the following 
lines of research: 
\\
(i) The description of BPS states in four-dimensional $\caln=2$ supersymmetric
gauge theories (of ``class $\cals$'') in terms of {\it spectral networks} on
Riemann surfaces given by Gaiotto-Moore-Neitzke \cite{gmn12}, and follow-up
work.
\\
(ii) Local mirror symmetry as consolidated by Hori-Vafa \cite{hova}, which
identifies the mirror of local toric Calabi-Yau manifolds with conic
bundles over $(\complex^\times)^2$ degenerating over a Riemann surface 
called the {\it mirror curve}.
\\
(iii) The wealth of knowledge about BPS states in these models that has
accumulated over the past fifteen years. We will rely in particular on the relation
to the representation theory of suitable {\it BPS quivers}, which in the case of
(``complete'') $\caln=2$ gauge theories has been related to the 
spectral curve perspective by Alim et al.\ \cite{accerv}.
We begin with brief reviews of each of these topics.

\subsection{Spectral networks}
\label{GMNreview}

The solution of $\caln=2$ supersymmetric gauge theories in four dimensions in
terms of a suitable ``spectral'' (Seiberg-Witten) curve includes a fruitful
representation of their spectrum of massive BPS states. The basic idea is to 
embed the gauge theory in a higher-dimensional setup where the spectral geometry 
becomes part of the space-time, and BPS particles in four dimensions are realized 
geometrically as extended objects calibrated by the Seiberg-Witten differential. 
This approach was pioneered in \cite{klmvw} and championed by \cite{gmn12}.
For more on the early development of the subject see the review \cite{lerche} 
and references therein. 

\paragraph{Theories of class $\cals$}

The large class of such theories studied in \cite{gmn12} are defined as 
the result of dimensional reduction (with a partial topological twist) of the 
six-dimensional $(2,0)$ theories associated to an ADE Lie algebra $\mathfrak{g}$ 
on a punctured Riemann surface $C$ with certain defect data at the punctures.
In the embedding in M-theory, the theories arise on the world-volume of 
$k={\rm rank}(\mathfrak{g})$ M5-branes wrapped on $\reals^{3,1}\times C\subset 
\reals^{3,1}\times T^*C\times \reals^3$. At a generic point on the Coulomb
branch, the IR description involves a single M5-brane wrapped on $\reals^{3,1}
\times\Sigma$, where $\Sigma\to C$ is the spectral cover
\begin{equation}
\label{spectralcover}
\left\{\lambda: \det(\phi-\lambda I)=0\right\}\subset T^*C,
\end{equation}
and $\phi$ is a $\mathfrak{g}$-valued 1-form on $C$ that parametrizes the
moduli space of vacua. For convenience, we will take $\mathfrak{g}=\mathfrak{su}(k)$
in what follows.

The six-dimensional theory contains string-like excitations that arise as boundaries
of M2-branes ending on the stack of M5-branes. When extended along paths of $C$, these 
strings give rise to particles upon dimensional reduction to four dimensions. For states 
of finite mass, the M2-brane should have finite spatial volume. This means that with respect 
to a local trivialization of the spectral cover, the paths are labelled locally by a pair 
of integers $i,j\in \{1,\ldots,k\}$. Such a string is locally BPS if it saturates 
the condition
\begin{equation}
M=\int |\lambda_{(ij)}|\geq\left\vert\int\lambda_{(ij)}\right\vert=|Z|,
\end{equation}
where $\lambda_{(ij)}\equiv\lambda_i-\lambda_j$ and $\lambda_i$ is the restriction 
of the Liouville 1-form on $T^*C$ to the $i$th sheet. This condition is satisfied
if and only if $\lambda_{(ij)}=e^{i\vartheta}|\lambda_{(ij)}|$ for some phase $\vartheta$ and
some orientation of the path. (Equivalently, the condition is that $\Im(e^{-i\vartheta}
\lambda_{(ij)})=0$, and we use $\Re(e^{-i\vartheta}\lambda_{(ij)})$ as volume form.)
Following \cite{gmn12}, we call such a locally minimizing path an $(ij)$ trajectory 
of phase $\vartheta$.

\begin{figure}[htbp]
\centering
\includegraphics[scale=1]{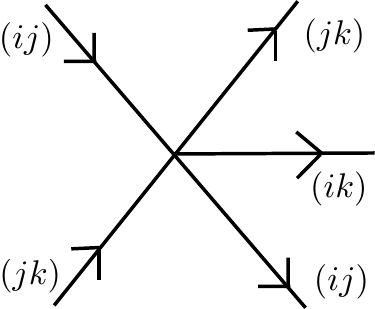}
\caption{Birth of an $(ik)$-trajectory at the intersection of an 
$(ij)$- and a $(jk)$-trajectory.} 
\label{fig:collideg}
\end{figure}

The spectral network is simply the ``life story'' of such BPS trajectories drawn
on the Riemann surface $C$. To describe it, we assume for simplicity that all branch 
points of the covering \eqref{spectralcover} are simple. Then, from an $ij$ branch 
point $z_0\in C$ of the spectral cover emanate three trajectories for any phase 
$\vartheta$. This can be seen by writing $\lambda_{(ij)} \sim\sqrt{z-z_0}$ and noting 
that
\begin{equation}
\int_{z_0}^{z}\lambda_{ij}\sim \int_{z_0}^{z}\sqrt{z-z_0}dz \sim (z-z_0)^{3/2}
\sim e^{i\vartheta}t
\end{equation}
has three independent solutions $z(t)=z_0+\bigl(e^{i\vartheta t}\bigr)^{1/3}$. 
Also note that depending on the placement of the 
branch cut, two of these trajectories are of type $(ij)$, and one is of type $(ji)$.
As the trajectories emanating from the various branch points evolve around $C$, it 
is possible that they meet. The pronouncement is that when an $(ij)$- and $(jk)$-trajectory
meet, an $(ik)$-trajectory is born, as illustrated in Figure \ref{fig:collideg}.
The collection of all trajectories emerging from the branch points and born in 
collisions is called the spectral network of phase $\vartheta$.

\begin{figure}[t]
\centering
\includegraphics[scale=1]{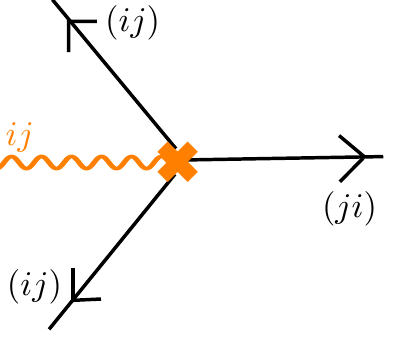}
\caption{Local structure of BPS trajectories near an $ij$-branch point.} 
\label{fig:localg}
\end{figure}

\begin{figure}[htbp]
\begin{center}
\begin{subfigure}[b]{0.45\textwidth}
\includegraphics[width=\textwidth]{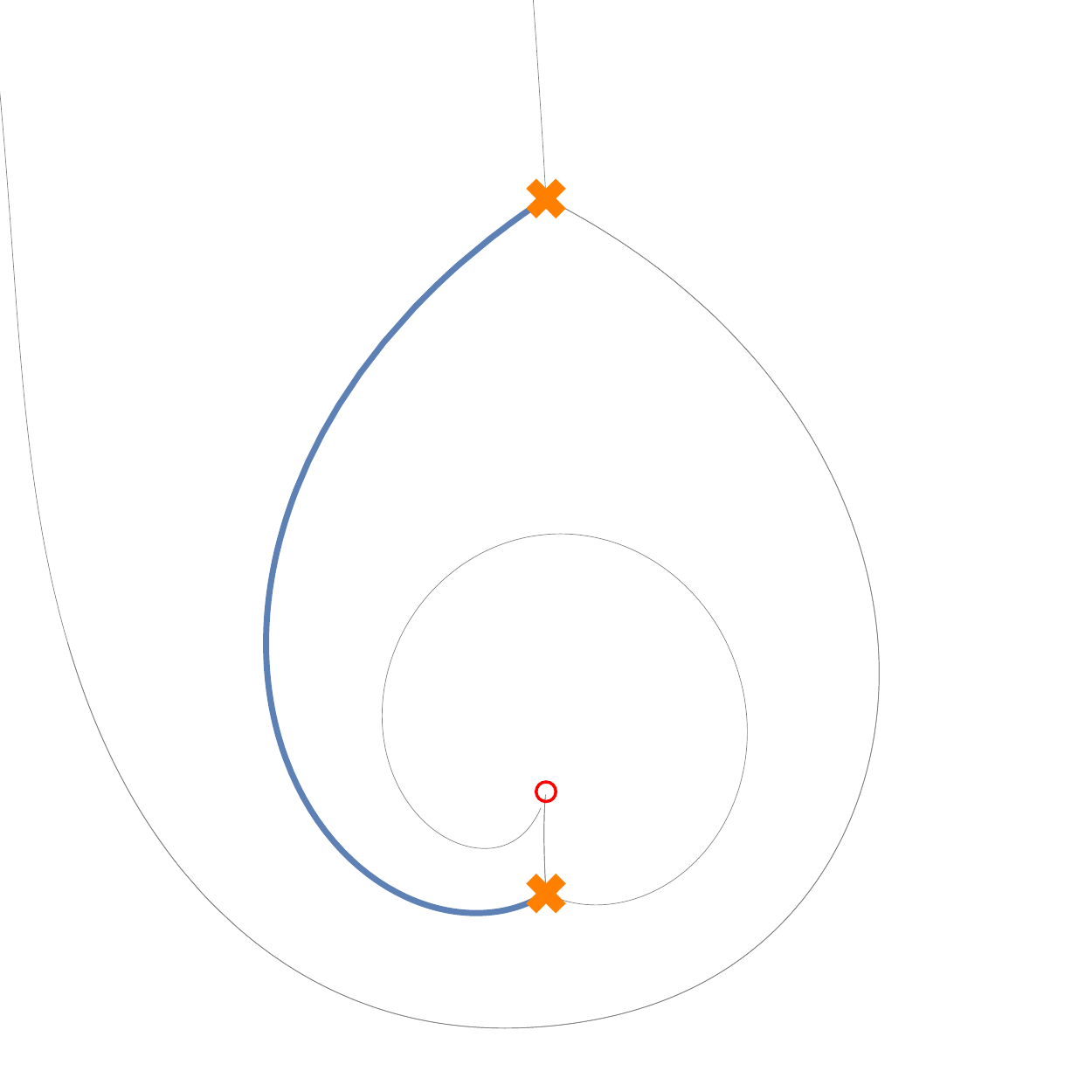}
\caption{Monopole of charge $(1,0)$.}
\label{fig:monopole}
\end{subfigure}
\begin{subfigure}[b]{0.45\textwidth}
\includegraphics[width=\textwidth]{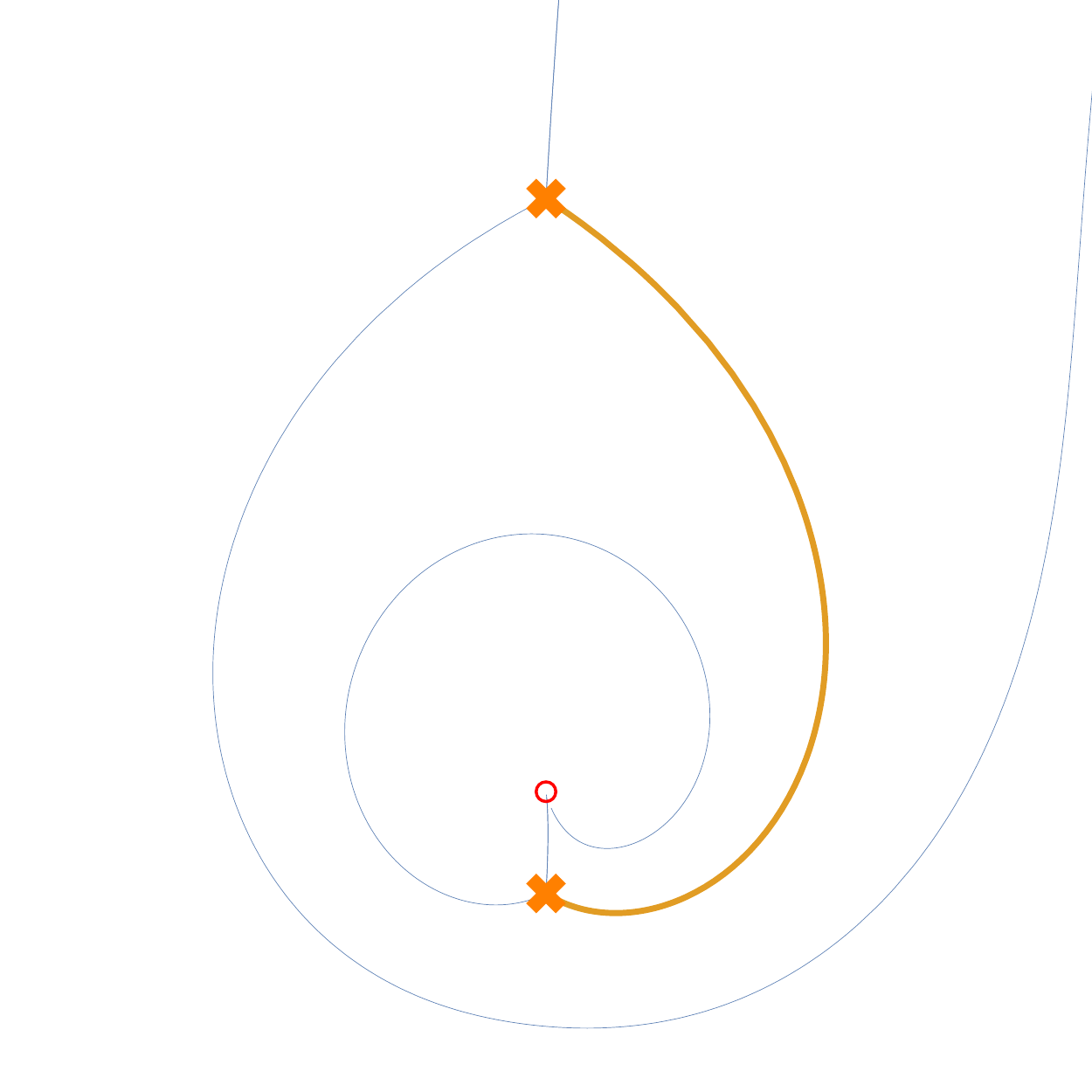}
\caption{Dyon of charge $(0,1)$.}
\label{fig:dyon}
\end{subfigure}
\begin{subfigure}[b]{0.45\textwidth}
\includegraphics[width=\textwidth]{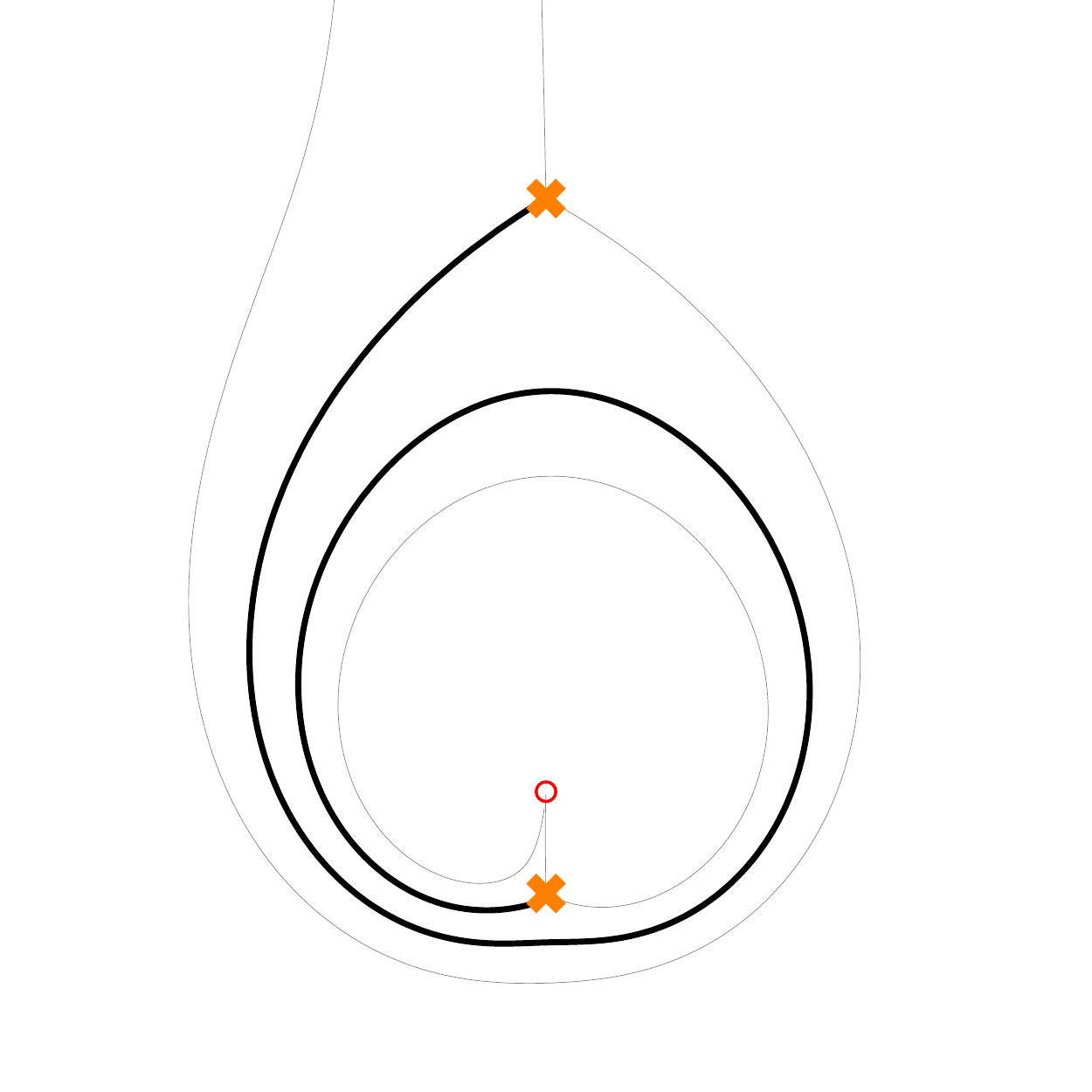}
\caption{A bound state with charge $(2,1)$.}
\label{fig:kroneckerbound21}
\end{subfigure}
\begin{subfigure}[b]{0.45\textwidth}
\includegraphics[width=\textwidth]{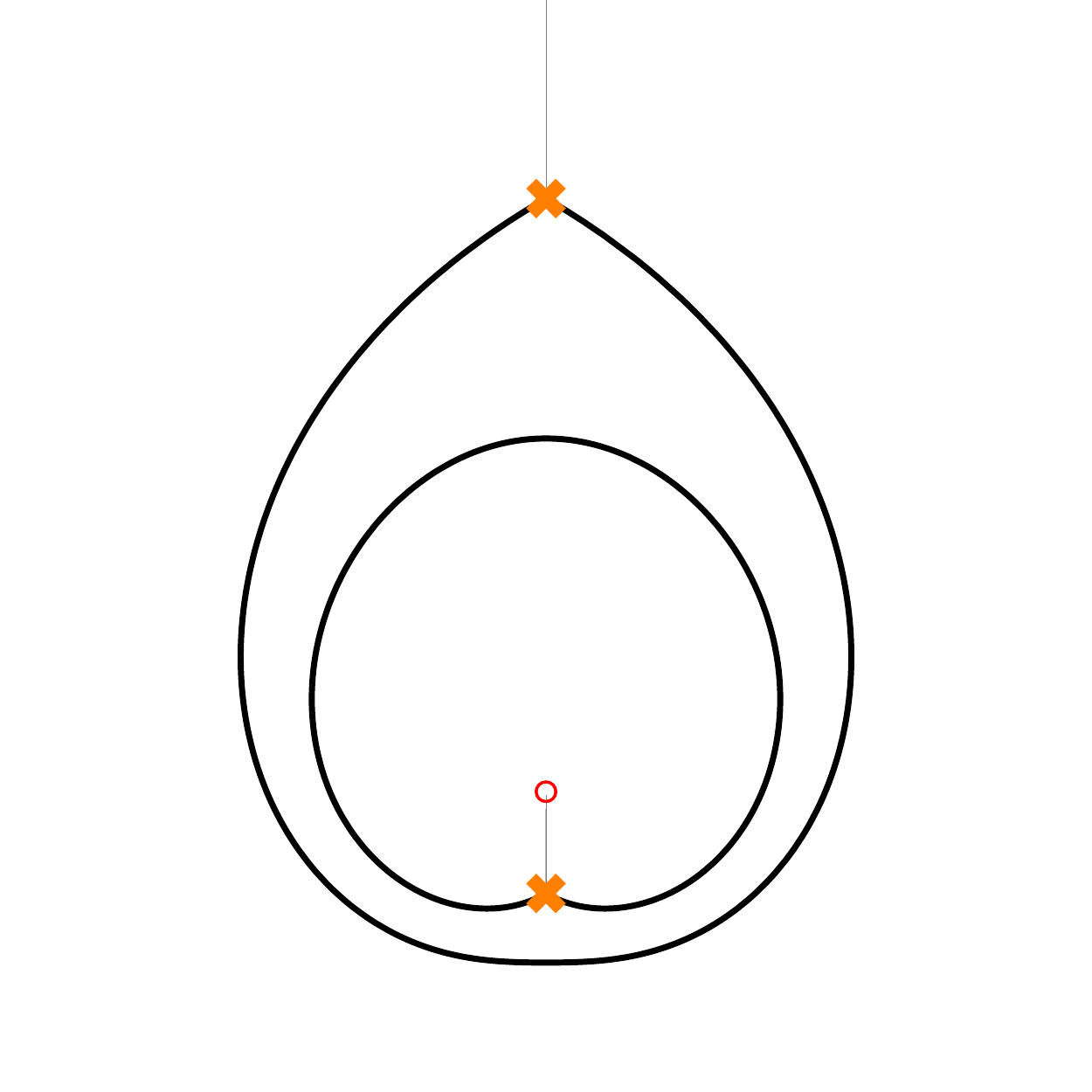}
\caption{Closed loops with charge $(1,1)$.}
\label{fig:vectm}
\end{subfigure}
\caption{Finite webs in the pure $SU(2)$ gauge theory at $u=2i$.}
\label{fig:finitewebs}
\end{center}
\vskip 1cm
\end{figure}

Generically, the trajectories will eventually be attracted to one of the
punctures of $C$ and crash. At special values of $\vartheta$, however, it may happen 
that some of the BPS trajectories terminate at a branch point or collide head-on. 
This gives rise to a closed subset of the plane with the geometry of a trivalent
graph of finite total length that we call a ``finite web'' following 
\cite{gmn12} (see Figure \ref{fig:finitewebs} for relevant examples). A finite web 
corresponds to a state of a finite mass BPS particle in four dimensions. Under the
identification of the lattice of electric-magnetic charges of the 4d theory with
$H_1(\Sigma,\zet)$, the charge of a finite web is the homology class of its 
canonical lift to $\Sigma$ determined by the labelling on the strands. 
The Dirac-Schwinger-Zwanziger pairing between electric and magnetic charges is 
identified with the intersection pairing on homology. In the M-theory picture, 
the BPS particles arise from M2-branes ending on the M5 branes, connecting the two 
lifts pointwise along the finite web. The BPS nature of the junction $(ij)+(jk) \to 
(ik)$ can be verified in this setup, see e.g.\ \cite{keshav}. 

To determine BPS degeneracies from the counting of finite webs with fixed charge,
one has to take into account that finite webs might exist in continuous families, 
of which the spectral network only produces some ``critical members'' (as is the 
case for example in Figure 
\ref{fig:vectm}), whereas the generic member does not pass through the branch points,
but is still locally calibrated and satisfies the junction rules. These deformations 
of the finite webs realize geometrically the zero modes of BPS particles in 4d. In 
such a situation, the BPS degeneracies should be determined by quantizing those zero modes.
It appears, however, that the information about the degeneracies can in fact be read
off purely from the critical pictures that arise from the spectral network without
deformations. The prescription of \cite{gmn12} for calculating these degeneracies
results from considering not only BPS particles but also line and surface defects, 
and thoroughly analyzing the consistency of the wall-crossing behavior of particles 
bound to them. Indeed, the curve $C$ is identified as the parameter space of UV
couplings of a canonical surface defect and finite webs with an open end at the point 
$z$ are identified as a particle bound to 
the surface defect $S_z$. Line defects, which can be thought of as infinitely heavy 
particles arising from M2 branes stretching infinitely in one cotangent direction, 
are represented by the (uncalibrated) homotopy class of a path on $C$.

In this paper we supply evidence that a similar story holds for BPS degeneracies
of D-branes in toric Calabi-Yau 3-folds. In these models, it is natural to propose 
that the finite webs arising from the network should be viewed as the fixed points 
of the given torus action on the associated D-brane moduli spaces. We have not yet fully
fleshed out this correspondence, but we expect that a generalization of 
the analysis of \cite{gmn12} including line and surface defects will lead directly 
to a complete and systematic algorithm for determining BPS degeneracies in these 
models as well \cite{wip}.

\paragraph{A useful heuristics: D-branes on the torus}

One of the premises of our analysis is the relation between the combinatorics
of spectral networks and their geometric deformations on the one hand, and the
dynamics of the associated BPS states on the other. For an example of this,
consider a D2-brane wrapping a torus, with a D0-brane sitting somewhere on it. 
Condensation of the open string tachyon in this system corresponds to dissolving 
the D0-brane into flux. This can be seen more graphically in the T-dual picture,
in which the D2 and D0-branes become a pair of perpendicular D$1$-branes. The
tachyonic D$1$-D$1$'-string is localized at their intersection and its condensation
corresponds to resolving the intersection, resulting in a D$1$-brane at an angle
(if we must, call it a simple ``network'' of D$1$ branes). Upon T-duality in the 
same direction as the first, the result is mapped back to a D2 brane with magnetic 
flux on it. 

\begin{figure}[tbp]
\centering
\includegraphics[scale=1]{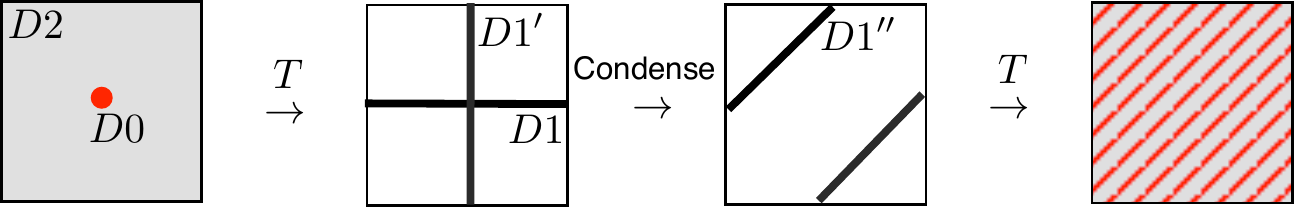}
\caption{T-duality picture of bound state formation.}
\label{fig:tachyons}
\end{figure}

This would essentially be the story of ``spectral networks and mirror symmetry for 
Calabi-Yau 1-folds''. We will also find useful the heuristic interpretation of 
resolving intersections as condensing tachyonic fields, especially in relation 
to the quiver description.

\paragraph{The pure \texorpdfstring{${\it SU}(2)$}{SU(2)} theory}

\begin{table}[bp]
\begin{center}
\begin{tabular}{l|l}
Weak Coupling & Strong Coupling \\
 \hline
 \hline
Dyons $\pm \mathbf{m} + n \mathbf{e}$ & Dyon $\mathbf{m} + \mathbf{e}$ \\
\hline
W-boson $\pm 2 \mathbf{e}$ & Monopole $\mathbf{m}$
\end{tabular}
\caption{BPS spectrum of pure $\mathcal{N} = 2$ theory.}
\end{center}
\label{tab:PureSpectrum}
\end{table}%
For another illustrative example, consider the original ``pure glue'' $SU(2)$ 
Seiberg-Witten theory \cite{sw1}. The charge lattice is just $\Gamma\cong 
\Z_{\it magnetic} \times \Z_{\it electric}$, with basis ${\bf m}$ and ${\bf e}$. 
The moduli space of vacua is the complex $u$-plane, with singular locus 
$\{\infty,1,-1\}$ dominated by the lightest particles of charge $2 \mathbf{e}$, 
$\mathbf{m}$ and $\mathbf{m} + \mathbf{e},$ respectively. In the weak coupling 
regime (large $|u|$), there is an infinite number of stable BPS particles with 
charges $\pm \mathbf{m} + n \mathbf{e},\thinspace n \in \zet$, as well as the $W$ 
bosons with charges $\pm 2 \mathbf{e}$. Famously, there is a (topologically circular) 
line of marginal stability passing through $u=1$ and $u=-1$, on the other side of which 
the stable spectrum  consists of only the monopole of charge $\pm\mathbf{m}$ and dyon of 
charge $\pm(\mathbf{m} + \mathbf{e})$. The central charge of a BPS particle with charge 
$g \mathbf{m} + q \mathbf{e}$ is given by $Z_u(g \mathbf{m} + q \mathbf{e})=g 
a_D(u) +q a(u)$ where $a(u)$, $a_D(u)$ are certain hypergeometric functions, arising
as periods of the elliptic curve \ref{swcurve}. The monopole and dyon are naturally 
thought of as the basic states of which the others are bound states.  To simplify 
the exposition in the rest of the paper we represent the charge of the monopole as 
$(1,0)$ and the charge of the dyon as $(0,1).$  In terms of the previous 
electric-magnetic charge basis, $(n,m)=  -n \mathbf{m} +m(\mathbf{m} + \mathbf{e})$.
 
The geometric realization is as follows. The spectral curve $\Sigma\to C$ is a genus 1 
double cover with two punctures and two simple ramification points:
\begin{equation}
\label{swcurve}
\begin{split}
\Sigma=\Bigl\{y^2=x+\frac 1x&-2u\Bigr\} \longrightarrow C = \C^\times\ni x
\\
\lambda_{\it SW} &= y \frac{dx}{x}
\end{split}
\end{equation}
With two branches, there is only one type of strand on $C$, so junctions do not occur. 
There are two ``elementary'' finite webs shown in Figures \ref{fig:monopole}-\ref{fig:dyon}.
These are identified (up to monodromy) via their central charge with the monopole 
and the dyon, which exist at any $u$. At large $|u|$, there is an infinite family of 
``spirals'' formed by concatenating $k$ copies of one of the elementary webs with $k+1$ 
copies of the other, separating them from the branch points and straightening them
out in the process. These have electric-magnetic charge $(k,k+1)=\mathbf{m}+(k+1)
\mathbf{e}$ and $(k+1,k)=-\mathbf{m}+(k+1)\mathbf{e}$, respectively.

There are also bound states of one copy of each of the two basic states that 
are realized by closed loops.
The closed loops actually exist in a family interpolating between the two loops attached at 
either branch point. The Hilbert space of 1-particle states associated to this family 
of loops is in principle the cohomology of its moduli space. The algorithm of 
Gaiotto-Moore-Neitzke 
\cite{gmn12} gives the invariant trace over this Hilbert space as $-2$, reflecting 
the contribution of a BPS vector multiplet.

Note that the existence of the bound states depends geometrically on the ability
to locally shorten the web by detaching the strands from the branch point, because
the angle between the two elementary webs is less that $2\pi/3$. In the strong
coupling regime, this is no longer possible, and all bound states cease to exist.

\subsection{Mirror curves for toric Calabi-Yau manifolds}
\label{mirrorsymmetry}

It is well known that the embedding of $\caln=2$ gauge theories of
class $\cals$ into M-theory by wrapping M5-branes on punctured Riemann 
surfaces is related, by a sequence of dualities, to {\it geometric 
engineering} of gauge theories in type II string theory. In
this approach, the gauge dynamics arises from D-branes wrapping
vanishing cycles in local singularities of Calabi-Yau manifolds. One
typically starts in type IIA  (where computations are done in the 
``A-model'') with a toric Calabi-Yau threefold, $X$, which is equivalent 
by mirror symmetry to type IIB (the ``B-model'') on a Calabi-Yau of 
the form
\begin{equation}
\label{form}
Y = \{ H(x,y) = uv \}\subset (\complex^\times)^2 \times \complex^2
\end{equation}
Here, $H(x,y)$ is a certain Laurent polynomial determined by the toric
data. For a suitable design, the curve \eqref{spectralcover} arises in 
a scaling limit from the mirror 
curve $H(x,y)=0$ which is the locus where the conic fibration $Y\to
(\complex^\times)^2$ degenerates. It is in fact in this context that 
the description of BPS states in terms of ``geodesics on the Seiberg-Witten 
curve'' was originally derived in \cite{klmvw}. We review the setup here
in order to emphasize the points in which the full result departs from
the gauge theory limit. We start in the A-model with the gauged linear 
sigma model (GLSM) description of the toric threefold.

\paragraph{Local mirror symmetry}

We let $r$ be the ``rank'' of the Abelian gauge group $U(1)^r$, and 
$Q_i^\alpha$ the charges of the $r+3$ chiral fields. We assume that these
charges satisfy the Calabi-Yau condition
\begin{equation}
\sum_{i=1}^{r+3} Q_i^\alpha=0
\end{equation}
and denote the Fayet-Iliopoulos couplings by $r^\alpha$ for each
$\alpha=1,\ldots, r$. The toric manifold then arises as the solution of 
the D-term constraints
\begin{equation}
\label{dterm}
\sum Q^\alpha_i|z_i|^2 = r^\alpha
\end{equation}
on the lowest components $z_i$ of the chiral fields, taken modulo $U(1)^r$ 
gauge equivalence. The space can be described more mathematically as the symplectic
reduction or GIT quotient $X=\complex^{r+3}//U(1)^r$ with stability specified
by the $r^\alpha$, which become (the real part of) the K\"ahler parameters of $X$. 
In the process, the $z_i$ become homogeneous coordinates on $X$.

In \cite{hova}, Hori and Vafa derived the theory mirror to $X$ by applying 
T-duality to the argument of all the chiral fields in the GLSM. They showed 
that in terms of the variables 
\begin{equation}
y_i = \exp(-|z_i|^2+ \ii \cdots) \,,
\end{equation}
the mirror of the GLSM is the Landau-Ginzburg theory with superpotential
\begin{equation}
\label{LGpotential}
W = \sum_{i=1}^{r+3} y_i
\end{equation}
on the solution of a complexification of \eqref{dterm},
\begin{equation}
\label{subject}
\prod_{i=1}^{r+3} y_i^{Q_i^\alpha} = q^\alpha
\end{equation}
where $q^\alpha = \exp(-t^\alpha)$ are the exponentiated and complexified
K\"ahler parameters. 

By now, mirror symmetry between $X$ and this Landau-Ginzburg model has 
been checked in great detail, and the equivalence of the topological models 
has the status of a mathematical theorem. Ultimately, the duality of course 
also holds at the level of the full physical theory, including the BPS 
spectrum in space-time. The resulting mathematical statements are however 
more difficult to check directly, mostly because stability of D-branes 
in Landau-Ginzburg models still is only partially understood \cite{stability}. 

In this paper, we will use the relationship between the Landau-Ginzburg model 
\eqref{LGpotential} and the Calabi-Yau hypersurface \eqref{form} in the form
in which local mirror symmetry was initially discovered. While this reduction 
is slightly less than fully rigorous at this point (although its validity at 
the topological level is beyond doubt), it puts us in a position to perform 
some explicit checks of the BPS spectrum. The easiest way to see 
the relation is to consider the evaluation of the periods: The statement that 
the fundamental variables are $\log y_i$ means that the holomorphic volume 
form is the residue of
\begin{equation}
\label{asfar}
\prod_{i=1}^{r+3} \frac{dy_i}{y_i} \exp(-W) \,,
\end{equation}
on the solutions of \eqref{subject}. Solving these equations in terms of three 
of the variables, $y_1$, $y_2$, $y_3$, and factoring out one of them by homogeneity, 
we define $H$ by the equation
\begin{equation}
\label{whichone}
W(y_1,y_2,y_3,q^\alpha) = y_1 H(x,y,q^\alpha)
\end{equation}
For a contour along which the integral converges, we can insert a Gaussian to 
rewrite the periods of \eqref{asfar} as
\begin{equation}
\label{griffiths}
\begin{split}
\frac{1}{(2\pi\ii)^{r+3}}\int \prod \frac{dy_i}{y_i} \exp(-W) &=
\frac{1}{(2\pi\ii)^4}
\int \frac{dx}{x}\frac{dy}{y} dy_1 du dv \exp(-y_1 H + y_1 u v) \\
&= 
\frac{1}{(2\pi\ii)^4}\int \frac{dx}x\frac{dy}{y} \frac{du \,dv}{H-uv} \\
&=
\frac{1}{(2\pi\ii)^3} \int_{H=uv} \frac{dx}x\frac{dy}{y}\frac{du}u
\end{split}
\end{equation}
where the last step is the Griffiths-Poincar\'e residue that gives us
the standard form of the holomorphic three-form on the hypersurface
\eqref{form}.

Granting \eqref{form}, the study of supersymmetric A-branes in $Y$ can
be further reduced to the ``mirror curve'' 
\begin{equation}
\label{onthecurve}
\Sigma = \{ H(x,y)=0\} \subset (\complex^\times)^2
\end{equation}
endowed with the differential
\begin{equation}
\label{way}
\lambda=\log y \frac{dx}{x} = \log y \, d\log x
\end{equation}
by ``integrating over the fibers'' of the map $Y\to(\complex^\times)^2$ sending 
$(x,y,u,v)$ to $(x,y)$ \cite{klmvw}.  Over each point in the $(x,y)$-plane
the equation $H(x,y)=uv$, viewed as an equation on $(u,v)\in\complex^2$, describes
an affine conic that is reducible precisely when $H(x,y)=0$. The generic conic
has the topology of a cylinder $S^1\times\reals$, and a ``minimal'' $S^1$
given by the intersection of $uv=H$ with $v=\bar u H/|H|$, in other words
$|u|^2= |H|$. This $S^1$ shrinks to a point precisely on the curve
$\Sigma$, so that tracing the $S^1$ along any path in $(x,y)$-space that 
intersects $\Sigma$ precisely at the beginning and end of the path gives 
rise to a two-sphere $S^2$. Assuming that the path begins and ends at
the same value of $x$, but at possibly different values of $y_1$ and $y_2$,
we can evaluate the integral
\begin{equation}
\label{eq:wind}
\frac{1}{2\pi\ii} \int_{S^2} \frac{du}{u} \frac{dy}{y} = \log y_1-\log y_2 
\end{equation}
which gives the differential \eqref{way} up to factors of $2\pi\ii$. We'll
work in the normalization \eqref{way} in the following. 

\paragraph{Initial observations}

Note that in \eqref{eq:wind} we have to allow $\log y_1$ and $\log y_2$ to stand 
for different branches of the logarithm 
if the path winds around the origin in the $y$-plane. More formally,
the mirror curve description of toric Calabi-Yau threefolds is an instance of 
a branched covering $\Sigma\to C$ embedded in $(T^*C)^\times = T^*C\setminus C$, 
where $C\hookrightarrow T^*C$ is the zero-section, with a holomorphic symplectic 
form that in a local coordinate $(z,y=\partial_z)$ on $T^*C$ takes the form 
$\omega=dz\wedge \frac{dy}{y}$. In contrast to the ordinary spectral cover
\eqref{spectralcover}, this form is not exact, although the {\it exponential} 
of the local ``Liouville form'' $\lambda= \log y\, dz$ is still single-valued.
We emphasize that it would be a mistake to replace $\Sigma$ by an infinite
covering on which $\lambda$ is well-defined -- only the winding number in the
fiber direction is detected by \eqref{eq:wind}, and not the absolute choice
of branch itself.

As far as we know, the geometry associated with such ``exponential differentials'' 
has not been studied in the literature so far, although the problems arising from 
the winding in the fiber direction were mentioned back in \cite{mayrwarner}.
We note however, that these multi-valued differentials
play a central role in what is known as the ``remodeling'' description of the 
topological string on local Calabi-Yau manifolds \cite{bkmp}. In this approach, the
formalism of topological recursions developed by Eynard-Orantin \cite{eyor}
is lifted to curves in $(\C^\times)^2$, with appropriate modifications of the
local residue calculus at the branch points. Given the striking similarities with
the gauge theory setup, it is very natural to expect that ``exponential'' versions
of spectral networks will capture the BPS spectrum in the same fashion.

Another way to understand the close analogy between mirror curves \eqref{form} and 
the spectral curves for gauge theories \eqref{spectralcover} is through the 
interpretation of these curves as ``IR moduli spaces'' of defects of the
respective $4d$ $\caln=2$ theories. This interpretation gives an alternative
derivation of the differential \eqref{way} by reduction along the non-compact cycles
instead of the compact cycles in the fibers of \eqref{form}.
Following the original approach of Aganagic and Vafa \cite{agva}, consider
a probe brane given by one of the two components, say $v=0$ of the singular fiber 
over some given point $(x_*,y_*)$ of the mirror curve. Even though such a 
brane is holomorphic for any point on the curve, a non-trivial superpotential
arises on account of the non-compactness of the cycle on which the brane is
wrapped. We fix one of $\complex^\times$ coordinates, say $y$, at infinity on the brane 
world-volume, which is identified with the $u$-plane. The other coordinate, say $x$, 
is treated as a holomorphic modulus.  Then, the superpotential is calculated as a 
chain integral over the three-chain $\Gamma$ of the form
\begin{equation}
x = x(t)\,,\qquad y=y(u,\bar u,t)\,,\qquad 
H\bigl(x(t),y(u,\bar u,t)\bigr) = u \cdot v(u,\bar u,t)
\end{equation}
with $u\in\complex$, $t\in[0,1]$ and functions $x$, $y$ subject to the conditions
\begin{equation}
\begin{split}
\bigl(x(0),y(u,\bar u,0)\bigr) &= (x_*,y_*)\,,\text{for all $u$}  \\
y(u,\bar u,t) & \to y_*\,,\text{for $|u|\to\infty$, all $t$}\\
H(x(t),y(0,0,t)) &= 0 \,,\text{for $u=0$, all $t$}
\end{split}
\end{equation}
With these conditions, and assuming a radially symmetric profile for $y(u,\bar u,t)
=y(r,t)$ for simplicity, one easily finds \cite{agva}
\begin{equation}
\begin{split}
\int_\Gamma \frac{dx}{x}\frac{dy}{y}\frac{du}{u} &= 
\int 
\frac{dt}{x}
\frac{\partial x}{\partial t} 
\frac{d\bar u}{y}
\frac{\partial y}{\partial \bar u} 
\frac{du}{u}  \\
&=
(2\pi\ii)\,\int_0^1 dt\int_0^\infty dr \frac{1}{y}\frac{\partial y}{\partial r} 
\frac{1}{x}\frac{\partial x}{\partial t}\\
&=
(2\pi\ii)\,\int \bigl(\log y-\log y_*\bigr) \frac{dx}{x}
\end{split}
\end{equation}
where the last integral is a contour integral on $\Sigma$, as claimed.

\paragraph{Framing}

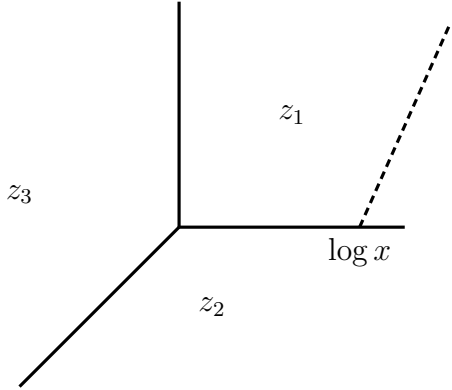
\begin{figure}[t]
\begin{center}
\begin{tikzpicture}[scale=3] 
\draw[very thick] (0,0) -- (1,0);
\draw[very thick] (0,0) -- (0,1);
\draw[very thick] (-.707,-.707) -- (0,0);
\draw[very thick, densely dashed] (0.8,0) -- (1.2,0.9);
\node at (-.707,-.707/2+.5) {$z_3$};
\node at (-.707/2+.5,-.707/2) {$z_2$};
\node at (.5,.5) {$z_1$};
\node[below] at (0.8,0) {$\log x$};
\end{tikzpicture}
\caption{A toric brane in the neighborhood of a vertex, anywhere
 in the toric diagram.}
\label{vertex}
\end{center}
\end{figure}

An important aspect of this derivation is the dependence of the differential
on the mirror curve on the brane that is used as a probe, a degree of freedom
known as ``framing'' \cite{akv}. In the A-model, the toric brane (of topology 
$S^1\times\reals^2$) that is mirror to the $v=0$ fibers is specified by a point 
on the toric diagram (the projection of a one-dimensional linear subspace of 
the base of the toric fibration). The 
vertex of the toric diagram closest to that point is surrounded by three faces, 
corresponding to toric divisors say $z_1$, $z_2$, $z_3$ as in Fig.\ \ref{vertex}. 
Then, modulo the D-term equations \eqref{dterm}, the brane is specified by
\begin{equation}
\label{deform}
|z_2|^2-|z_1|^2 = 0 \,, 
\end{equation}
and the modulus is $\sim |z_3|^2-|z_1|^2$. The semi-classical regime is when the
brane is far from the vertex (which requires in particular, if the brane sits on 
an internal leg, that leg to be long). In the quantum regime, captured by the
mirror, \eqref{deform} ceases to vanish, and the modulus is subject to the
framing ambiguity
\begin{equation}
-{\rm Re}(\log x) = |z_3|^2-|z_1|^2 + f (|z_2|^2-|z_1|^2)
\end{equation}
(Note that $f$ disappears under \eqref{deform}!) 
In other words, the good variables to use in \eqref{whichone} are $x$ and $y$  which are defined by
\begin{equation}
\label{goodcoords}
x = \frac{y_3}{y_1} \bigl(\frac{y_2}{y_1}\bigr)^f \,,\qquad y = \frac{y_2}{y_1},
\end{equation} 
and in these variables the differential on $\Sigma$ is 
given by \eqref{way}.
Alternatively, we may also use $y=y_1/y_2$, with necessary changes.

The idea of the spectral network approach to BPS states is that open webs
capture the degeneracy of solitons bound to defects represented by their
endpoints, whereas closed finite webs correspond to the ``vanilla'', or
purely 4d, BPS particles. As a result, the degeneracy of the finite webs that 
we construct from our exponential networks should be independent of the
framing, even though the respective differentials might differ by exact
terms. This framing independence provides an important check on our 
formalism.

\subsection{BPS Quivers}
\label{BPSQ}

The third starting point of our investigation is the description of BPS 
spectra in terms of quivers, which also has a long history going back to
\cite{domo} in the physics literature and builds on earlier mathematical 
work by Nakajima, Kronheimer, and ultimately Gabriel
 and Kac \cite{nakajima1,nakakron,kac1,gabriel}. The basic
physical idea is to study BPS states and their interactions from the point 
of view of the effective theory on their world-volume (supersymmetric 
quantum mechanics in the case of BPS particles). This point of view is
particularly convenient to understand the formation of bound states in 
terms of ``tachyon condensation'' in the effective theory and the decay
into constituents in terms of Higgs-Coulomb transitions induced by
the variation of couplings in the background space-time theory.

The quiver description arises when, under certain conditions, it is possible to 
build up the entire spectrum of BPS states by bound state formation out of a 
finite number of ``basic states''. These basic states, which as a minimum 
requirement must generate the BPS charge lattice of the theory under consideration, 
form the nodes of the quiver diagram. The gauge group of the world-volume
theory on some integral combination of the basis states is a product over
the nodes of the given rank. The chiral fields in bifundamental representations 
that allow the formation of bound states form the arrows of the quiver. 
The gauge-invariant superpotential is a formal sum of traces over closed loops 
in the quiver, and the D-terms depend on Fayet-Iliopoulos parameters associated
with the $U(1)$ factors at each node. The supersymmetric vacua correspond
to stable representations of the quiver algebra \cite{dfrStability}.

\paragraph{Mathematical recapitulation}

To explain this identification and establish some notation, we state that our 
{\it quiver} $Q$ is specified by a tuple $(Q_0, Q_1, h: Q_1 \rightarrow Q_0, t: 
Q_1 \rightarrow Q_0)$, where the finite sets $Q_0$ and $Q_1$ collect the nodes
and arrows, respectively, and the maps $h$ and $t$ specify the head and 
tail of an arrow. Given this, a {\it representation} $M$ of $Q$ is specified by 
\cite{MR2197389}
\\
$\bullet$ A  finite-dimensional $\C-$vector space $M_v$ for each node $v$ in 
$Q_0$ .
\\
$\bullet$ A $\C$-linear map $\varphi_{\alpha} : M_v \rightarrow M_w$
for each arrow $\alpha: v \rightarrow w$ in $Q_1.$ 
\\
For any representation $M$ of $Q$, the assignment of the dimension
$n_v$ of $M_v$ to each vertex $v\in Q_0$ is called the {\it dimension vector}
of $M$. Given an ordering $v_1, \dots, v_{k}$ of the vertices of $Q$, the
dimension vector has components $\left( \dim(M_{v_1}), \dots, \dim(M_{v_k}) 
\right)$, but we will often denote it simply by $n(M)$.

Given two representations $(M_v, \varphi_{\alpha})$ and $(M'_v, \varphi'_{\beta})$ 
of a quiver $Q,$ a {\it morphism} of quiver representations $f: M \rightarrow M'$ 
is a family $f = (f_v)_{v \in Q_0}$ of $\C-$linear maps
\begin{equation}
\begin{CD}
M_v 		@>\varphi_{\alpha} >> 	M_w \\
@VV f_vV			@VVf_wV \\
M'_v		@> \varphi'_{\beta} >>	M'_w
\end{CD}
\end{equation}
such that $\varphi'_{\beta} \circ f_v = f_w \circ  \varphi_{\alpha}.$ The category 
$\Rep(Q)$ of representations of a quiver $Q$ is equivalent to the category of modules 
over the path algebra $\C Q-\Mod.$  In particular $\Rep(Q)$ is a category with 
kernels and cokernels.  A representation $L$ over a quiver $Q$ is a  
{\it subrepresentation} of a representation $M$ if there is an injective morphism 
$i : L \rightarrow M.$  More concretely, this means that $L_v \subseteq M_v$ and 
the homomorphisms $f: L_v \rightarrow L_w$ are induced from the restriction of 
homomorphisms $f': M_v \rightarrow M_w.$

\begin{figure}[t]
\begin{center}
\begin{tikzpicture}[scale=1] 
\tikzset{ 
my loop/.style={->,to path={ 
.. controls +(-45:2) and +(45:2) .. (\tikztotarget) \tikztonodes}}
} 
\tikzset{ 
left loop/.style={->,to path={ 
.. controls +(225:2) and +(135:2) .. (\tikztotarget) \tikztonodes}}
} 

\path (0,0) node[draw,shape=circle,color=Mblue] (v0) {$v_1$}; 
\path (4,0) node[draw,shape=circle,color=Morange] (v1) {$v_2$}; 
\path[->, thick, bend left = 10] (v0) edge node[above, yshift = -0.3ex] {$a_1$} (v1);
\path[->, thick, bend left = -10] (v0) edge node[below,yshift=-0.3ex] {$a_2$} (v1);
\end{tikzpicture}
\caption{The Kronecker-2 quiver.}
\label{fig:purequiver}
\end{center}
\end{figure}
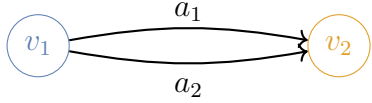

As a simple example, consider the generic representations
\begin{equation}
\label{kron2}
\begin{tikzcd}
  \textcolor{Mblue} \C \arrow[r, shift left, "a_1" above] \arrow[r, shift right, "a_2" below]
& \textcolor{Morange} \C  
\end{tikzcd}
\end{equation}
of the Kronecker-2 quiver 
from Fig.\ \ref{fig:purequiver} with dimension vector $(1,1)$.
For $a_1 \neq 0$ or $a_2 \neq 0$ there are no sub-representations with dimension 
vector $(1,0)$ since it is impossible for the square 
\begin{equation}
\begin{tikzcd}
  \textcolor{Mblue} \C \arrow[r, shift left, "a_1" above] \arrow[r, shift right, "a_2" below]
& \textcolor{Morange} \C  \\
  \textcolor{Mblue} \C \arrow[r, shift left, "0" above] \arrow[r, shift right, "0" below] 
\arrow[u, "\psi"]
    & \textcolor{Morange} 0 \arrow[u, "0"] 
\end{tikzcd}
\end{equation}
to be commutative.
However there are sub-representations with dimension vector $(0,1)$ which we 
will consider shortly. 

A representation is called {\it indecomposable} if it cannot be expressed as the 
direct sum of two non-zero representations. For example, for any $\lambda\neq 0$,
the representation
\begin{equation}
\label{indecss}
\begin{tikzcd}[ampersand replacement=\&]
\textcolor{Mblue} \C^2 \ar[r, shift left, "{\begin{pmatrix} 1 & 0 \\ 0 & 1 \end{pmatrix}}" above]
         \ar[r, shift right, "{\begin{pmatrix} 1 & \lambda \\ 0 & 1 \end{pmatrix}}" below]
    \& \textcolor{Morange} \C^2
\end{tikzcd}
\end{equation}
of the Kronecker-2 quiver is indecomposable \cite{schiffler}. It is
however not irreducible in the familiar sense of representation theory,
since it admits \eqref{kron2} with $a_1=a_2=1$ as a non-trivial sub-representation 
in an obvious way.

The complex algebraic group $G_\C= \prod_{v} {\it GL}(n_v,\C)$ acts on the space of 
quiver representations of fixed dimension vector. It appears physically as the
complexification of the gauge group $G$ of the effective world-volume theory. In this 
theory, (classical) supersymmetric vacua correspond to solutions of the D-term 
constraints\footnote{In 
the presence of a superpotential, the relevant representations are those of the quiver 
path algebra with relations. This plays a role in our examples later, but for now, we 
assume that the F-term constraints have been solved.} modulo the action of $G$.
Equivalently, one may consider the space of those $G_\C$ orbits that contain 
a solution of the D-terms. In other words, on each $G_\C$ orbit the solution 
of the D-terms (if one exists) is unique up to the action of $G$. Mathematically,
this is the correspondence between the symplectic and algebraic quotient
constructions of moduli spaces, which was established for quiver representations 
by King \cite{king}. In this context, the ``good'' $G_\C$ orbits are those 
that are ``stable'', in the sense of Mumford's numerical criterion, with respect 
to a character of $G_\C$ that is related physically to the FI-parameters entering 
the D-term constraints.

More concretely, a King stability condition for quiver representations is specified
by a map $\theta:Q_0\to\R$. An indecomposable representation $M$ of $Q$ is 
$\theta$-{\it semistable} if
\begin{equation}
\label{funny}
\theta(M)= \sum_{v} \theta_v n_v(M) = 0 
\end{equation}
and for every sub-representation $M'$ of $M$,
\begin{equation}
\label{serious}
\theta(M')=\sum_{v} \theta_v n_v(M') \ge 0.
\end{equation}
The representation $M$ is $\theta$-{\it stable} if additionally the only sub-representations 
$M'$ with $\theta(M') = 0$ are $M$ and $0$.
As an example, we again consider the representations of the Kronecker-2 quiver 
with dimension vector $(1,1)$. As we have seen, the generic such representation
is indecomposable. To see that it is stable if and only if $\theta_1=-\theta_2>0$, it suffices to
consider 
\begin{equation}
\begin{tikzcd}
 \textcolor{Mblue} \C \arrow[r, shift left, "a" above] \arrow[r, shift right, "b" below]
& \textcolor{Morange} \C  \\
  \textcolor{Mblue} 0 \arrow[r, shift left, "0" above] \arrow[r, shift right, "0" below] 
\arrow[u, "0"]
    & \textcolor{Morange} \C \arrow[u, "\psi"] 
\end{tikzcd}
\end{equation}
which is the only potentially destabilizing subrepresentation.
On the other hand, the indecomposable representation \eqref{indecss} is 
$\theta$-semi-stable, but not $\theta$-stable, since for any choice of $\theta_1=-\theta_2$  
the subrepresentation \eqref{kron2} with $a_1=a_2=1$ has $\theta=\sum_v \theta_v n_v=0$ 
as well.

\paragraph{D-brane Bound States}
The precise relation between (semi-)stable quiver representations and D-brane bound
states was obtained in \cite{dfrStability}. The punchline is that solutions of 
the D-flatness conditions of the world-volume theory correspond to direct sums of 
representations, each of which is $\theta$-stable in the above sense with respect 
to the same $\theta$. For a one-particle bound state, only the center of mass $U(1)$ 
should remain unbroken. This means that the space of endomorphisms of the associated 
representation should be one-dimensional (\ie, the representation should be ``Schur''). 
Semi-stable representations correspond to marginally bound states.

An important entry in this dictionary is the identification between the FI-parameters 
$\zeta_v$ for $GL(n_v)$ on the vertex $v\in Q_0$ and the stability parameters 
$\theta_v$. The seeming subtlety arises from the (trivial) fact that to determine 
the supersymmetric ground states, it is in general neither sufficient nor necessary
that the D-term potential vanishes. On the one hand, setting the gaugino variation
for the unbroken gauge group to zero requires a combination of the FI parameters to
vanish. On the other hand, the minimum energy configuration may preserve a 
non-linearly realized supersymmetry with a constant shift of the FI parameters.

For illustration, consider the toy model \cite{km} of D-brane bound state 
formation 
from two (stable) constituents interacting via a single massless chiral multiplet 
$\phi$ of charge $(-1,1)$ under the $U(1)_1\times U(1)_2$ gauge group. The D-term 
potential for the anti-diagonal $U(1)_-\subset U(1)_1\times U(1)_2$ is
\begin{equation}
V_D = \left( |\phi|^2 - \zeta \right)^2
\end{equation}
where $\zeta=\zeta_1-\zeta_2$ is the difference of the FI parameters in the original
basis. Then, for $\zeta < 0$, the minimum of the potential at $\phi=0$ has positive
energy and the ``space-time'' $\mathcal{N} = 1$ supersymmetry is broken as in 
the Fayet model. No bound state forms in this case. In the marginal case, $\zeta = 0$, 
both $U(1)_{1}$ and $U(1)_{2}$ are unbroken. Finally, for $\zeta>0$, the vacuum
$\phi=0$ is an unstable maximum with a tachyonic mode. The bound state corresponds 
to the stable minimum at $|\phi|^2=\zeta$ (modulo $U(1)_-$ gauge transformations) 
of zero energy. Naively, supersymmetry is still broken since the gaugino variation
\begin{equation}
\delta_\alpha\lambda_\beta^{(-)} = \{Q_\alpha,\lambda^{(-)}_\beta\} 
\sim \epsilon_{\alpha\beta} D^{(-)} + \bigl(F_{\alpha\beta}^{(-)}\bigr)^+ 
= \zeta\neq 0 
\end{equation}
is non-vanishing in the vacuum. However, under a suitable linear combination of the 
original supersymmetry with a constant one of the form
\begin{equation}
\delta_\alpha \lambda_\beta = \epsilon_{\alpha\beta}
\end{equation}
the gaugino variation does in fact vanish. This modification has the
effect of a constant shift of the D-terms.

For a bound state made up of a finite number of constituents with a single unbroken
$U(1)$, the criterion is that all D-terms should be equal \cite{dfrStability}.
In the quiver notation, the D-term potential is
\begin{equation}
V  = \sum_{v\in Q_0} \left(D_v - \zeta_v \right)^2\,,
\end{equation}
where
\begin{equation}
D_v = \sum_{\topa{\alpha\in Q_1}{h(\alpha)=v}} |\varphi_\alpha|^2- 
\sum_{\topa{\alpha\in Q_1}{t(\alpha)=v}}
|\varphi_\alpha|^2.
\end{equation}
Following \cite{dfrStability}, we add zero to this expression in the form 
$ 0 = - \theta_v + \theta_v$ and then choose $D_v$ such that $D_v = \theta_v$.
\begin{equation}
\begin{split}
V & = \sum_{v} \left(D_v - \theta_v + \theta_v - \zeta_v \right)^2 \\
   & = \sum_{v} n_v \left( \theta_v - \zeta_v \right)^2. \\
\end{split}
\end{equation}
The $\theta_v$ minimizing this expression subject to the constraint
\begin{equation}
\sum_v \theta_v = 0
\end{equation}
arising from the trace of D-flatness are
\begin{equation}
\label{shifted}
\theta_v = \zeta_v - \frac{\sum_{w} n_w \zeta_w}{\sum_{w} n_w}.
\end{equation}
In the end, all the D-terms have been shifted by the constant 
\begin{equation}
\frac{\sum_{v} n_v \zeta_v}{\sum_{v} n_v}.
\end{equation}
The shifted $\theta_v$ are identified as the King-stability parameters -- 
clearly, $\sum \theta_v n_v=0$ (see \eqref{funny}), and it only remains to check
the condition \eqref{serious} on all subrepresentations.

\paragraph{Spectrum of pure $SU(2)$ theory from stable representations}

To illustrate this procedure, let us rederive in the quiver formalism \cite{fm}
the BPS spectrum of pure glue $\caln=2$ Seiberg-Witten theory that we obtained 
at the end of subsection \ref{GMNreview} using spectral networks. The relevant 
BPS quiver follows from using the monopole and dyon as the basic states. These 
states, which are stable over the entire moduli space, are represented by the 
elementary webs of Figs.\ \ref{fig:monopole} and \ref{fig:dyon} and correspond 
to the nodes of the BPS quiver. Upon superimposing the two pictures, the webs 
intersect once at each branch point. The corresponding intersections occur in 
the same relative orientation on $\Sigma$, so we draw two arrows in the same 
direction between the nodes. Since there are no closed loops in the quiver the superpotential must
be zero. The resulting quiver is the Kronecker-2 quiver from Fig.\ 
\ref{fig:purequiver}.  For a more systematic derivation of BPS quivers for
a large class of $\caln=2$ gauge theories, see \cite{hiv,accerv}.

The starting point for the full representation theory of the Kronecker-2
quiver are the theorems of Gabriel and Ka\v c, which guarantee
that the indecomposable representations of the 2-Kronecker quiver are
precisely the generic representations of dimension vector $(n_1,n_2)$
with
\begin{equation}
n_1^2+n_2^2-2n_1n_2 \le 1.
\end{equation}
Of those, $(n_1,n_2)=(k,k\pm 1)$ correspond to the positive real roots of the 
quiver viewed as Dynkin diagram, and $n_1=n_2=1$ to the imaginary root.

Depending on $u$, the phases of the central charges determining the stability 
parameters (without the shift \eqref{shifted})
are $\theta_1=\arg\bigl(Z_u(-\mathbf{m})\bigr)=\arg(-a_D(u))$ and 
$\theta_2=\arg\bigl(Z_u(\mathbf{m}+\mathbf{e})\bigr) = \arg(a_D(u)+a(u))$.

Then, in the strong coupling regime, $\theta_1<\theta_2$, and the only 
stable representations are the simple modules of dimension vectors $(1,0)$ and 
$(0,1)$.
The simple representation of dimension vector $(1,0)$ is given by
\begin{equation}
\begin{tikzcd}
  \textcolor{Mblue} \C \arrow[r, shift left, "0" above] \arrow[r, shift right, "0" below]
& \textcolor{Morange} 0 
\end{tikzcd}
\end{equation}
and corresponds to the monopole in pure $SU(2)$ Seiberg-Witten theory. The dimension 
vector of the representation is the same as the charge basis introduced in 
section \ref{GMNreview}. The simple representation of dimension vector $(0,1)$ 
is given by
\begin{equation}
\begin{tikzcd}
  \textcolor{Mblue} 0 \arrow[r, shift left, "0" above] \arrow[r, shift right, "0" below]
& \textcolor{Morange} \C  
\end{tikzcd}
\end{equation}
and corresponds to the dyon.

In the weak coupling regime, $\theta_1 > \theta_2$, the dimension vectors with 
stable representations are
\\
$\bullet$ $(k, k + 1)= \mathbf{m} + (k+1)\mathbf{e}$
\\
$\bullet$ $(k+1,k) = - \mathbf{m} + k \mathbf{e}$
\\
$\bullet$ $(1,1)= 2 \mathbf{e}$.
\\
Together with their negatives (which give the anti-particles), these are
all the roots listed above. The representations for the first two dimension 
vectors are unique up to gauge transformations, so their moduli space is a point. 
A representative representation with dimension vector $(1,2)$ is
\begin{equation}
\begin{tikzcd}
 \textcolor{Mblue}  \C^{1} \arrow[r, shift left, "A_1" above] \arrow[r, shift right, "A_2" below]
& \textcolor{Morange} \C^{2}  
\end{tikzcd}
\end{equation}
with
\begin{equation}
A_1=\begin{pmatrix} 1 \\ 0 \end{pmatrix}
\qquad
A_2= \begin{pmatrix} 0 \\ 1 \end{pmatrix}.
\end{equation}
The generic indecomposable representation with dimension vector $(k+1,k)$ can be 
brought to the form
\begin{equation}
\label{zeroone}
A_1 = 
\begin{pmatrix}
 1  &  0               	   &   \cdots  &  0 &  0 \\
  0 & 1                  	   &     &  0 &  0 \\
  \vdots   &     	            &  \ddots      & \vdots &    \vdots  \\
 0    &     0                        &    \cdots  & 1 &  0 \\
\end{pmatrix}
\qquad
A_2 = 
\begin{pmatrix}
0 &  1  &  0               	   &   \cdots  &  0  \\
0&   0 & 1                  	   &     &  0  \\
\vdots &  \vdots   &     	            &  \ddots      & \vdots   \\
0 &  0    &     0                        &    \cdots  & 1 \\
\end{pmatrix}
\end{equation}
by a suitable complexified gauge transformation. The stable representations of 
dimension $(1,1)$ have moduli space $\mathbb{P}^1$, parametrized by the ratio 
$[a_1:a_2]$ in \eqref{kron2}.

\begin{figure}[htbp]
\begin{center}
\includegraphics[width=0.8\textwidth]{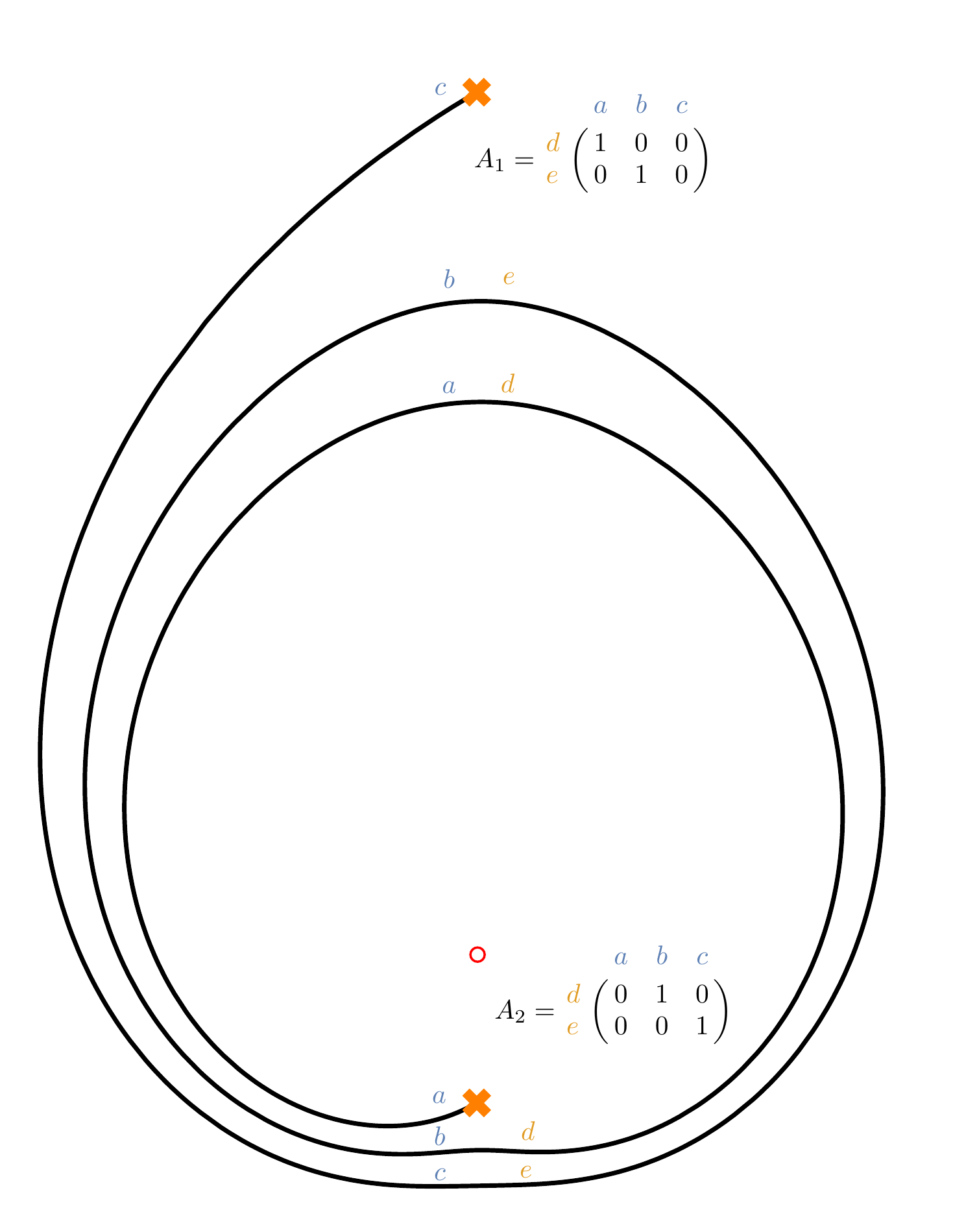}
\caption{$(3,2)$ bound state and its corresponding quiver representation.}
\label{fig:boundq}
\end{center}
\end{figure}

Interestingly, the correspondence between the intersection pattern of the 
elementary webs and the associated BPS quiver extends to the representation theory
between the position of non-zero entries in \eqref{zeroone} and the connection
pattern of the spectral networks such as Fig.\ \ref{fig:kroneckerbound21}:
A $1$ in the $i$-th column and $j$-th row in $A_1$ corresponds to the fact
that (counting from the inside out) the $i$-th strand on the left is
connected at the top to the $j$-th strand on the right. Similarly, 
a $1$ in the $i$-th column and $j$-th row of $A_2$ is related to a
connection at the bottom of the figure. With slightly different labelling,
this is illustrated in Fig.\ \ref{fig:boundq}.

\paragraph{String Modules}

This correspondence between the connectivity of the network and the non-zero 
entries in the representation matrices will provide important clues later
on, so we elaborate a bit further on the special nature of the representations
\eqref{zeroone}, known as ``string modules'' \cite{br}. 

In order to define string modules we start with a few relevant definitions.  A 
{\it walk} in a quiver is an unoriented path, or 
more formally, a sequence of vertices in the quiver connected by arrows in either 
direction,
\begin{equation}
\begin{tikzcd}
  v_1 \arrow[leftrightarrow, "\alpha_{1}"]{r} & v_2  \arrow[leftrightarrow, "\alpha_{2}"]{r} 
&\cdots  \arrow[leftrightarrow, "\alpha_{n-1}"]{r}  & v_n. 
\end{tikzcd}
\end{equation}
A {\it string} in the path algebra of a quiver $\C Q,$ is a walk which avoids traversing 
sequences of arrows of the form
\begin{equation}
\begin{tikzcd}
v_1  & v_2  \arrow["\beta",swap]{l} \arrow["\beta"]{r} & v_3
\end{tikzcd}
\; \text{or} \;
\begin{tikzcd}
 \arrow["\beta_{1}"]{r}  &\cdots  \arrow[ "\beta_s"]{r}  & ,
\end{tikzcd}
\end{equation}
or their duals, where $\beta_1, \dots, \beta_s \in \partial \calw$ is a zero-relation 
coming from the F-term relations.  The first forbidden sequence is that the inverse of 
an arrow can not be immediately succeeded by the arrow, or conversely that an arrow can 
not be immediately succeeded by its inverse.  A {\it string module} is obtained from 
a string by replacing each vertex with a copy of $\C$ and representing each arrow by 
the identity morphism.  Each vertex has vector space $\C^{n_v}$ where $n_v$ is the number 
of times the vertex $v$ appears in the string and the morphisms are determined by their 
actions on arrows. Conversely, decomposing $\C^{n_v}$  in $n_v$ copies of $\C$ with
a separate node for each copy amounts to thinking about this particular class of 
representations in terms of an ``abelianized'' quiver.

In the Kronecker-2 example, the representation in \eqref{zeroone} with $k=2$ can
be identified with the string module
\begin{equation}
\begin{tikzcd}
    &  \textcolor{Morange}  \C      &               & \textcolor{Morange} \C   & \\ 
\textcolor{Mblue}  \C  \arrow[ru, "1"] &       	 &        \textcolor{Mblue}  \C  
\arrow[lu, "1",swap] \arrow[ru,"1"]  &      & \textcolor{Mblue}  \C \arrow[lu, "1",swap].
\end{tikzcd}
\end{equation}
corresponding to the string
\begin{tikzcd}
  v_1 \arrow["A_{1}"]{r} & v_2  &  v_1 \arrow["A_2", swap]{l} \arrow[ "A_1"]{r} & v_2   
& v_1 \arrow["A_2", swap]{l}
\end{tikzcd}
In general, there are $k$ arrows pointing to the right from $A_1$ and 
$k$ arrows pointing to the left from $A_2.$

\section{More on Quivers and D-branes}
\label{mathi}

In the introduction we recalled that BPS states arise in string compactifications 
by wrapping D-branes on supersymmetric cycles in the Calabi-Yau, and their degeneracies 
are encoded in the cohomology of the associated moduli spaces.  We here give a bit of
further background on the types of supersymmetric cycles, their effective world-volume 
theory, and the cohomology of their moduli spaces. We then elaborate on
the special class of quiver representations that we will find realized in
terms of our exponential networks.

\subsection{Supersymmetric cycles redux}

Supersymmetric cycles are, by definition, cycles such that the world-volume theory of 
a brane wrapping the cycle is supersymmetric. Two conditions to be a supersymmetric 
cycle in Calabi-Yau 3-folds were found from a space-time perspective in \cite{bbs} 
and from supersymmetric string world-sheet boundary conditions preserving A or 
B-type supersymmetry in \cite{ooy}.  The first possibility is that the cycle is an 
even-dimensional holomorphic submanifold, carrying a stable holomorphic vector bundle.
The second is that the cycle is a middle-dimensional (in this case three-dimensional) 
cycle, such that $\Re \bigl(e^{-i\vartheta} \Omega  \vert_{L}\bigr)$ is its volume 
form, where  $\Omega$ is the holomorphic volume form on the Calabi-Yau.

The interactions of BPS states obtained from string compactifications are described 
by an effective quiver quantum mechanics.  The form of the effective theory of the 
massless modes can be determined  using the topological A and B-models.  A-branes 
in the B-model wrap special Lagrangian cycles and their F-term interactions are 
mathematically described by the Fukaya category. On the other hand, D-term equations
are related to mathematical considerations of stability and are controlled by the
B-model. Similarly, B-branes in the A-model wrap holomorphic cycles. F-terms are
captured by the derived category of coherent sheaves, and D-terms by the A-model.

In type IIB string theory on $\R^{1,3} \times Y$ with $Y$ a local Calabi-Yau 
and D3 branes on $\R^{1,3} \times L_i$ that wrap a special Lagrangian 
$L_i \in Y$ and the time component of $\R^{1,3},$ we can choose a basis of 
branes. These branes are the BPS particles in a 4d $\mathcal{N} = 2$ theory.  
Their interactions are described by quiver quantum mechanics with four supercharges. 
The quiver quantum mechanics has gauge group $\prod U(n_i)$ where $n_i$ is 
the number of D3 branes wrapping the special Lagrangians $L_i.$  

Similarly, in type IIA string theory on $\R^{1,3}\times X$, where $X$ is another
local Calabi-Yau (say, the mirror of $Y$), D0-branes at points, D2-branes on holomorphic 
curves, and D4-branes on compact 4-cycles give rise to finite mass BPS particles in 
spacetime. Natural bases of B-branes are those of fixed dimension at large 
volume, or fractional branes at an orbifold point. D2/D4-branes on non-compact 
holomorphic cycles, as well as a D6-brane wrapping all of $X$, correspond to 
infinitely massive objects that can provide framing to the lighter states.

\subsection{A-branes in the B-model}
\label{quiversAB}

\begin{figure}
\begin{center}
\includegraphics[scale=0.55]{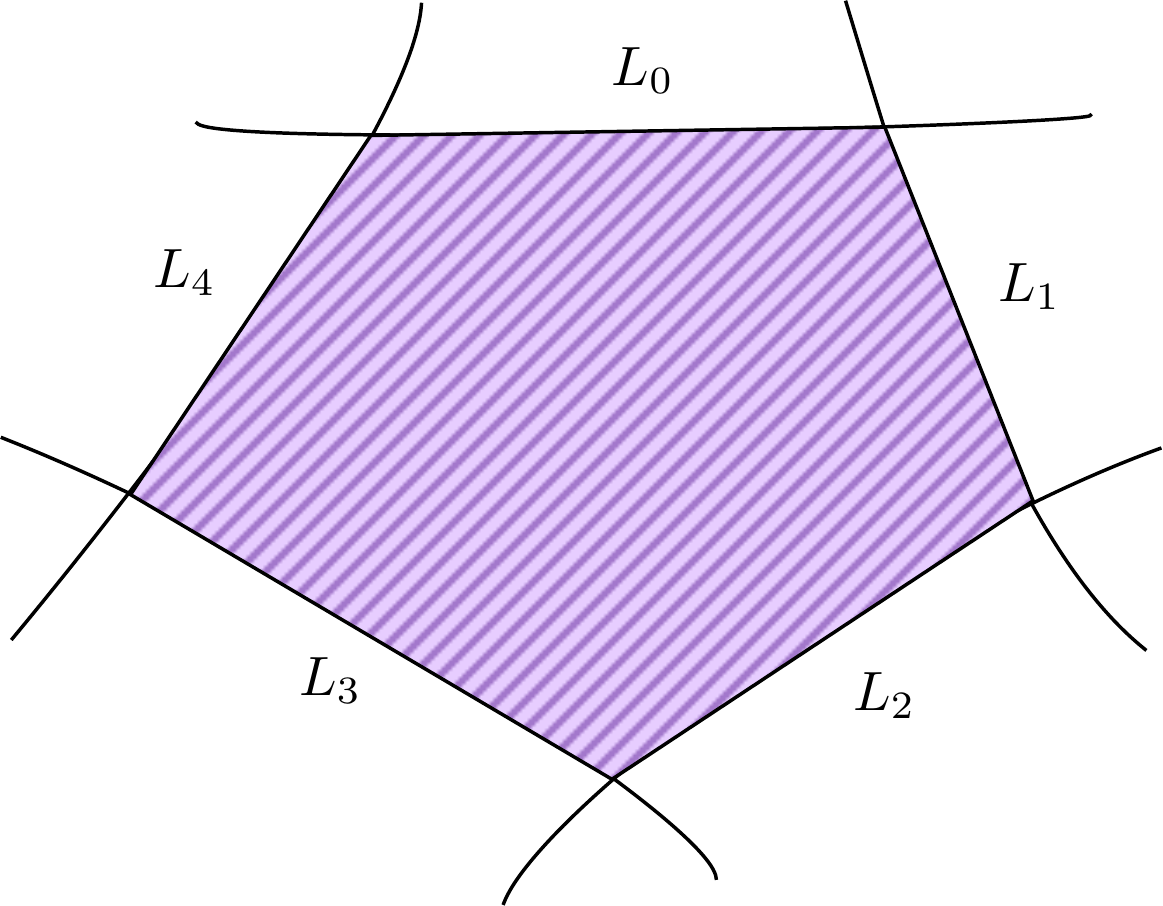}
\caption{Holomorphic disk bounded by five Lagrangians.}
\label{fig:holodisk}
\end{center}
\end{figure}

In the Fukaya category of a Calabi-Yau manifold $Y$, the objects are special 
Lagrangians $L_i$.  The space of morphisms between two transversely intersecting 
special Lagrangians $L_i$ and $L_j$ is 
\begin{equation}
\Hom(L_i, L_j) = \bigoplus_{p \in L_i \cap L_j } \C \cdot \langle p \rangle
\end{equation}
where the sum is over all intersection points $p$
\footnote{
For special Lagrangians with flat bundles, the $\Hom$-spaces are
\begin{equation}
\Hom((L_i,E_i), (L_j,E_j)) = \bigoplus_{p \in L_i \cap L_j } \Hom(E_i \vert_p, E_j 
\vert_p) \cdot \langle p \rangle.
\end{equation}
}.
The space $\Hom(L_i, L_j)$ is the space of massless open strings stretching between 
branes wrapped on the cycles $L_i$ and $L_j.$  The space of morphisms 
is not associative.  Instead the morphisms are only associative ``up to 
homotopy''.  This weaker structure is called $A_{\infty}$ since there can 
be arbitrarily high order of failure of strict associativity.  The 
$A_{\infty}$-structure is specified by the composition maps
\begin{equation}
m^k : \Hom(L_{k-1}, L_k) \otimes \dots \otimes \Hom(L_0, L_1) \rightarrow 
\Hom(L_0, L_k)[2 - k].
\end{equation}
which we now define.
\cite{FOOO} define a moduli space $\mathcal{M}$ of holomorphic maps $u: D \rightarrow Y$
from the disk $D$ with $k + 1$ marked points $p_{0,1}, \dots, p_{k-1,k}, p_{k,0}$
such that the image of the boundary intervals $[p_{k-1,k}, p_{k, k+1}]$ under the map
are contained in the corresponding Lagrangians $L_k$ (see Fig.\ \ref{fig:holodisk}).  
The $A_{\infty}$-maps are given in terms of the signed areas of the holomorphic disks,
$$m^k( \langle p_{0,1} \rangle, \dots \langle p_{k,0} \rangle) = 
\sum_{u \in \mathcal{M}} \pm q^{\int u^{*} \omega} \langle p_{0,k} \rangle.$$

The $A_{\infty}$-maps encode the Yukawa interactions between massless open strings.
When $Y$ is Calabi-Yau of dimension $d$, there is a trace-map
\begin{equation}
\gamma: \cala \rightarrow \C
\end{equation}
of degree $-d$ on the algebra $\cala$ of massless open strings obtained from the 
direct sum of all of the $\Hom$-spaces. Using the trace-map, the $A_{\infty}$-maps 
can be encoded in a superpotential
\begin{equation}
\calw = \Tr \left( \sum_{k=2}^{\infty} \sum_{i_0, i_1, \dots, i_k} 
\frac{c_{i_0, i_1, \dots i_k}}{k+1} \phi_{i_0, i_1} \phi_{i_1, i_2} \dots \phi_{i_k, i_0}\right)
\end{equation}
where 
\begin{equation}
c_{i_0, i_1, \dots i_k} = \gamma \left( m^2 \left( m^k \left(\langle p_{i_0,i_1} 
\rangle, \dots,  \langle p_{i_{k-1},i_k} \rangle \right), \langle p_{i_k,i_0} 
\rangle  \right) \right)
\end{equation}
and $\phi_{i,j} \in \Hom(L_i, L_j)$ are the massless fields in the quiver quantum 
mechanics.  In the subsequent sections, we will simplify our discussion and say 
that the superpotential is obtained by summing contributions by holomorphic disks.

\subsection{Quiver representations}
\label{quiverReps}
 
The geometric origin of the BPS quivers that we consider in this paper is
reflected in special properties of their representation theory. Already in
our discussion of the Kronecker-2 quiver, we saw that the representations
with dimension vector $(k\pm 1,k)$ are ``string modules'' in the terminology 
of \cite{br}, while the (in general, semi-stable) representations with dimension 
vector $(n,n)$ are so-called ``band modules''. As pointed out in 
\cite{TinkerToys}, this relation extends to all theories of class $\cals$ with 
$\mathfrak{g}=\mathfrak{su}(2)$: According to \cite{cv}, the BPS spectra of these
theories can be studied in terms of triangulated surfaces, and it is a
general result \cite{Gentle} that for quivers from triangulated surfaces \cite{fst}, 
all representations are either string or band modules. From the spacetime
perspective, these are hyper- and vector-multiplets, respectively \cite{TinkerToys}.

In more general situations, such as those involving mirror curves of the form 
\eqref{onthecurve}, string and band modules will not be enough. We here develop
a graphical representation of certain special classes of quiver representations
for the specific cases of the ADHM and Kronecker-3 quivers. For Kronecker-3
quiver, these representations cover the class of ``tree modules'' which were
discussed in \cite{ringel}, and can be seen to account for an exponential
growth of BPS degeneracies \cite{Weist1, Weist2, Weist3}.

These results will be used later in sections \ref{conifold}, \ref{localP2} 
and \ref{C3}, where we will (re)produce such quiver representations 
from exponential networks.

\paragraph{Representations of the ADHM quiver}
\label{par:adhm}
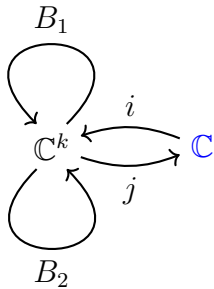
\begin{figure}[t]
\begin{center}
\begin{tikzpicture}[scale=1] 
\tikzset{ 
above loop/.style={->,to path={ 
.. controls +(45:2) and +(135:2) .. (\tikztotarget) \tikztonodes}}
} 
\tikzset{ 
below loop/.style={->,to path={ 
.. controls +(-135:2) and +(-45:2) .. (\tikztotarget) \tikztonodes}}
}        
\path (0,0) node (v0) {$\C^k$}; 
\path (2,0) node (v1) {${\color[rgb]{0.000000,0.000000,1.000000}\C}$}; 
\draw[thick, black] (.2,.3) edge[above loop] node[above] {$B_1$} (-.2,.3) ;
\draw[thick, black] (-.2,-.3) edge[below loop] node[below] {$B_2$} (.2,-.3) ;
 \path[->, thick, bend right = 20]  (v1)  edge node[above,yshift=0ex] {$i$} (v0) ;
 \path[->, thick, bend right = 20]  (v0)  edge node[below,yshift=0ex] {$j$} (v1) ;
\end{tikzpicture}
\caption{The ADHM quiver.} 
\label{fig:ADHM}
\end{center}
\end{figure}
The ADHM quiver is shown in Figure \ref{fig:ADHM} and has the relations $[B_1, B_2] + ij = 0.$  Representations of the ADHM quiver
correspond to points of the Hilbert scheme of points in $\C^2,$ see for example Theorem 
1.9 of \cite{Na}.  We briefly explain how to construct a quiver representation from a 
point in the Hilbert scheme.  A point in the Hilbert scheme can be represented by an 
ideal $\cali \subset \C[z_1, z_2]$ with $\C[z_1, z_2]/\cali$ which is finite of dimension 
$k.$  The ideal defines a $k$-dimensional vector space $V =  \C[z_1, z_2]/\cali.$
Multiplication by $z_{\alpha}$ modulo the ideal $I$ defines two endomorphisms 
$B_{\alpha}$ where $\alpha = 1, 2.$  Furthermore we set $i(1) = 1 \mod \cali$ and 
$j = 0.$

The torus fixed points of the Hilbert scheme of $k$ points in $\C^2$ correspond to 
partitions of $k$.  To each partition there is its associated Young diagram.  Given 
a Young diagram, we can equivalently construct a representation of the ADHM quiver 
by means of a ``covering'' quiver.  Place the Young diagram at 45 degrees.  For each 
box in the Young diagram place a vertex.  Then connect vertices that are up to the 
left or up to the right by arrows.  Finally add an additional vertex and arrow 
connecting that vertex to the vertex corresponding to the first box in the Young 
diagram.  Then we can associate a  representation by placing a copy of $\C$ at 
each vertex and the identity homomorphism for each arrow.

We illustrate the correspondence between torus fixed points, partitions, and quiver 
representations in the example of $k = 4$ points in $\C^2.$
There are five ideals with $\dim \C[z_1, z_2]/\cali = 4.$  The first three are
\\
$\bullet$ $\cali = \langle z_1^4, z_2 \rangle$, corresponding to the partition $(4)$,
with $\C[z_1, z_2]/\cali = \langle [1], [z_1], [z_1^2], [z_1^3] \rangle $,
and covering quiver and representation matrices given by
\begin{equation}
\begin{tikzcd}
\C  &      &       &     \\
     & \C \arrow[lu, "B_1",swap] &       &     \\
     &     &   \C \arrow[lu, "B_1",swap] &    \\
     &     &        & \C \arrow[lu, "B_1",swap] \\
     &	    &        &        {\color[rgb]{0.000000,0.000000,1.000000}\C}  \arrow[u, "i"]  
\end{tikzcd}
\end{equation}
\begin{equation}
B_1 = 
\begin{pmatrix}
0 & 0 & 0 & 0 \\
1 & 0 & 0 & 0 \\
0 & 1 & 0 & 0 \\
0 & 0 & 1 & 0 \\
\end{pmatrix},
\qquad
B_2 = 
\begin{pmatrix}
0 & 0 & 0 & 0 \\
0 & 0 & 0 & 0 \\
0 & 0 & 0 & 0 \\
0 & 0 & 0 & 0 \\
\end{pmatrix},
\qquad
i = 
\begin{pmatrix}
1 \\ 0 \\ 0 \\ 0
\end{pmatrix},
\qquad
j = 
\begin{pmatrix}
0 \\ 0 \\ 0 \\ 0
\end{pmatrix}.
\end{equation}
\\
$\bullet$ $\cali = \langle z_1^3, z_1 z_2, z_2 \rangle$ corresponding to $(3,1)$,
$\C[z_1, z_2]/\cali = \langle [1], [z_1], [z_1^2], [z_2] \rangle$, and
\begin{equation}
\begin{tikzcd}
\C  &      &       &     \\
     & \C \arrow[lu, "B_1",swap] &       &   \C  \\
     &     &   \C \arrow[lu, "B_1",swap] \arrow[ru,"B_2"]  &   \\
     &     &         {\color[rgb]{0.000000,0.000000,1.000000}\C} \arrow[u, "i"] &   \\
\end{tikzcd}
\end{equation}

\begin{equation}
B_1 = 
\begin{pmatrix}
0 & 0 & 0 & 0 \\
1 & 0 & 0 & 0 \\
0 & 1 & 0 & 0 \\
0 & 0 & 0 & 0 \\
\end{pmatrix},
\qquad
B_2 = 
\begin{pmatrix}
0 & 0 & 0 & 0 \\
0 & 0 & 0 & 0 \\
0 & 0 & 0 & 0 \\
1 & 0 & 0 & 0 \\
\end{pmatrix},
\qquad
i = 
\begin{pmatrix}
1 \\ 0 \\ 0 \\ 0
\end{pmatrix},
\qquad
j = 
\begin{pmatrix}
0 \\ 0 \\ 0 \\ 0
\end{pmatrix}.
\end{equation}
\\
$\bullet$ and $\cali = \langle z_1^2, z_2^2 \rangle$ with partition $(2,2)$,
$\C[z_1, z_2]/\cali = \langle [1], [z_1], [z_2], [z_1 z_2] \rangle$, and
\\
\begin{equation}
\begin{tikzcd}
          &         \C         &   \\
  \C    \arrow[ru,"B_2"]   &                    & \C    \arrow[lu,"B_1",swap]  \\ 
       	  &        \C  \arrow[lu, "B_1",swap] \arrow[ru,"B_2"]  & \\
	         	  &         {\color[rgb]{0.000000,0.000000,1.000000}\C}  \arrow[u, "i"]  &
\end{tikzcd}
\end{equation}
\begin{equation}
B_1 = 
\begin{pmatrix}
0 & 0 & 0 & 0 \\
1 & 0 & 0 & 0 \\
0 & 0 & 0 & 0 \\
0 & 0 & 1 & 0 \\
\end{pmatrix},
\qquad
B_2 = 
\begin{pmatrix}
0 & 0 & 0 & 0 \\
0 & 0 & 0 & 0 \\
1 & 0 & 0 & 0 \\
0 & 1 & 0 & 0 \\
\end{pmatrix},
\qquad
i = 
\begin{pmatrix}
1 \\ 0 \\ 0 \\ 0
\end{pmatrix},
\qquad
j = 
\begin{pmatrix}
0 \\ 0 \\ 0 \\ 0
\end{pmatrix}.
\end{equation}
The remaining two partitions $(1,1,1,1)$ and $(2,1,1)$ are the transposes of $(4)$ 
and $(3,1)$ respectively. Transposing partitions acts by vertical reflection on the 
covering quiver, interchanging the $B_1$ and $B_2$ arrows.  The final example of the 
partition $(2,2)$ contains two distinct paths from $\C$ to $\C.$  The representation 
must satisfy $B_1 B_2 = B_2 B_1$ and it does by construction.
\paragraph{Representations of the Kronecker-3 quiver}

\begin{figure}[t]
\begin{center}
\begin{tikzpicture}[scale=0.8] 
\path (0,0) node[draw,shape=circle,color=Morange] (v0) {$v_0$}; 
\path (0:4cm) node[draw,shape=circle,color=Mgreen] (v1) {$v_1$}; 
\path[->,thick, bend right = 20] (v0) edge node[below] {`$+$',`$-$',$B$} (v1);
\path[->,thick, bend right = 0] (v0) edge node[] {$$} (v1);
\path[->,thick, bend right = -20] (v0) edge node[above] {$$} (v1);
\end{tikzpicture}
\caption{Kronecker-3 quiver.} 
\label{fig:K3}
\end{center}
\end{figure}
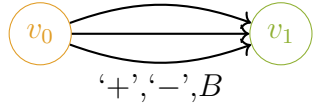
The Kronecker-3 quiver is shown in Fig.\ \ref{fig:K3}. It arises in a variety
of context as the arguably simplest example of a quiver of ``wild'' representation
type. As before, however, we can usefully abelianize the representations of the
Kronecker-3 quiver in terms of a ``covering quiver'' \cite{fr1,fr2}.

\begin{figure}[htbp]
\centering
\begin{tikzpicture}
\node[Morange] (pm) at (30:2cm) {$\C_2$};
\node[Morange] (pp) at  (150:2cm) {$\C_3$};
\node[Morange] (pb) at  (-90:2cm) {$\C_1$};
\node[Mgreen] (pmb) at ($(pm) +(90:2cm)$) {$\C_7$};
\node[Mgreen] (pmp) at ($(pm) +(-30:2cm)$) {$\C_5$};
\node[Mgreen] (ppb) at ($(pp) +(90:2cm)$) {$\C_8$};
\node[Mgreen] (ppm) at ($(pp) +(210:2cm)$) {$\C_3$};
\node[Mgreen] (pbp) at ($(pb) +(-30:2cm)$) {$\C_4$};
\node[Mgreen] (pbm) at ($(pb) +(-150:2cm)$) {$\C_1$};
\node[Mgreen] (z) at (0,0) {$\C^2_{2,6}$};
\draw[<-, very thick] (z) -- node[above] {$-$} node[below, xshift=1.5ex] 
{$\left( \begin{smallmatrix} 1 \\ 0\\ \end{smallmatrix} \right)$} (pm);
\draw[<-, very thick] (z) -- node[above] {$+$}  node[below, xshift=-1.5ex] 
{$\left( \begin{smallmatrix} 0 \\ 1 \\ \end{smallmatrix} \right)$} (pp);
\draw[<-, very thick] (z) -- node[right] {B}  node[left] {$\left( 
\begin{smallmatrix} 1 \\1\\ \end{smallmatrix} \right)$} (pb);
\draw[->, very thick] (pm) -- node[right] {$B$} (pmb);
\draw[->, very thick] (pm) -- node[above] {$+$} (pmp);
\draw[->, very thick] (pp) -- node[right] {$B$} (ppb);
\draw[->, very thick] (pp) -- node[above] {$-$} (ppm);
\draw[->, very thick] (pb) -- node[above] {$+$} (pbp);
\draw[->, very thick] (pb) -- node[above] {$-$} (pbm);
\end{tikzpicture}
\caption{The covering quiver for the (3,8) representation.}
\label{fig:tree38}
\end{figure}
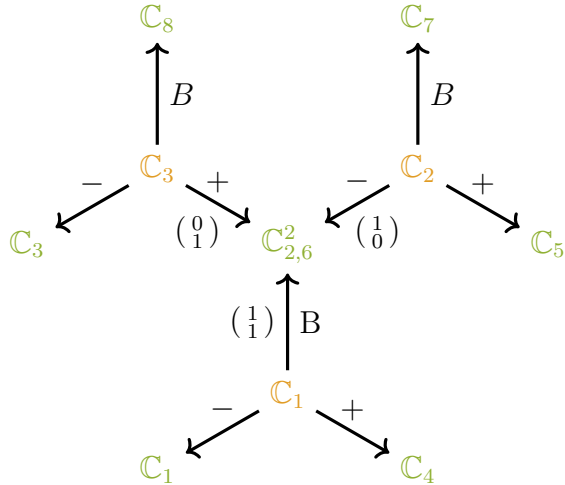

A much-studied family of representations are the ``Fibonacci representations''.  
Their dimension vector $(F_n, F_{n+2})$ is given by the $n$-th Fibonacci number 
$F_n$, which are defined recursively by $F_0 = 0, F_1 = 1$ and $F_n = F_{n-1} + 
F_{n-2}$ \cite{Leonardo}.  When $n$ is even, these dimension vectors are the 
Schur roots of the quiver and there exists a unique irreducible representation. 
For $n$ odd, the moduli space of representations is isomorphic to $\CP^2$. 
For $n=2$, \ie, dimension vector $(1,3)$, the representation is represented
graphically by
\begin{equation}
\label{fig:tree13}
\begin{tikzpicture}
\node[Mgreen] (pm) at (30:2cm) {$\C_1$};
\node[Mgreen] (pp) at  (150:2cm) {$\C_2$};
\node[Mgreen] (pb) at  (-90:2cm) {$\C_3$};
\node[Morange] (z) at (0,0) {$\C$};
\draw[->, very thick] (z) -- node[above] {$-$} (pm);
\draw[->, very thick] (z) -- node[above] {$+$}  (pp);
\draw[->, very thick] (z) -- node[right] {B} (pb);
\end{tikzpicture}
\end{equation}
with representation matrices given by
\begin{equation}
M_{-} =
\begin{pmatrix}
1 \\ 0 \\ 0 \\ 
\end{pmatrix},
\qquad
M_{+} =
\begin{pmatrix}
0 \\ 1 \\ 0 \\ 
\end{pmatrix},
\qquad
M_{B} =
\begin{pmatrix}
0 \\ 0 \\ 1 \\ 
\end{pmatrix}.
\qquad
\label{13rep}
\end{equation}
For $n=4$ the dimension vector is $(3,8)$ and the covering quiver is shown in Figure \ref{fig:tree38}.
From the covering quiver, we can read off the representation matrices
\begin{equation}
M_{-} = 
\begin{pmatrix}
1 & 0 & 0 \\
0 & 1 & 0 \\
0 & 0 & 1 \\
0 & 0 & 0 \\
0 & 0 & 0 \\
0 & 0 & 0 \\
0 & 0 & 0 \\
0 & 0 & 0 \\
\end{pmatrix},
\qquad
M_{+} = 
\begin{pmatrix}
0 & 0 & 0 \\
0 & 0 & 0 \\
0 & 0 & 0 \\
1 & 0 & 0 \\
0 & 1 & 0 \\
0 & 0 & 1 \\
0 & 0 & 0 \\
0 & 0 & 0 \\
\end{pmatrix},
\qquad
M_B = 
\begin{pmatrix}
0 & 0 & 0 \\
1 & 0 & 0 \\
0 & 0 & 0 \\
0 & 0 & 0 \\
0 & 0 & 0 \\
1 & 0 & 0 \\
0 & 1 & 0 \\
0 & 0 & 1 \\
\end{pmatrix}.
\end{equation}
The entries in the representation matrices are zero unless there is an arrow from $\C_i$ to $\C_j$.  In this case, there is a 1 in the $(i,j)$ position of
the corresponding representation matrix.  The central node labeled $\C^2_{2,6}$ is two-dimensional and we write the corresponding matrices.
The representations of the Kronecker-3 quiver shown in \eqref{fig:tree13}
and Figure \ref{fig:tree38} belong to the class of tree modules studied in \cite{ringel}.

\paragraph{Quantizing the moduli space of quiver representations}

In the examples we consider, most quivers have a moduli space of representations.  
The BPS particles are obtained by quantizing the moduli space.  In \cite{h&m} it 
is argued that $L^2$ cohomology is the appropriate cohomology for quantizing the 
moduli space.  One approach to computing the cohomology of quiver moduli spaces is to
count the number of points in the representation variety over finite fields 
and using the Weil conjectures \cite{Denef, reineckeHN} to extract the relevant 
cohomology groups.  A second approach is to use supersymmetric localization in 
quiver quantum mechanics \cite{beht1, beht2, hky, LeeYi}. In applications to 
the Kronecker quivers the localization calculations \cite{HKim, CordovaShao} 
reduce to a weighted sum over trees \cite{reineckeHN, Weist1, Weist2, Weist3}.
The relationship between tree modules and the appearance of trees in localization are related manifestations of abelianization.
The combinatorics of trees contributes to the exponential growth of BPS states.

\section{Exponential Networks}
\label{executive}

With the work reviewed in section \ref{review} in mind, we set out to investigate
networks of BPS trajectories on mirror curves $\pi:\Sigma\to C$ of the form
\eqref{onthecurve}. The main purpose for now is to describe the features that
are new compared to the gauge theory situation of \eqref{spectralcover},
but we also offer tentative geometric interpretations that will be corroborated
in the subsequent examples.

\subsection{New rules \ldots}

The first novelty is that the calibrating differential is only defined modulo 
$(2\pi\ii) d\log x$ on $\Sigma$. As a consequence, BPS trajectories on $C$ are
labelled locally by both a pair $(i,j)\in\pi^{-1}(x)$ of branches of the 
covering, and a winding number $n\in\Z$. Since this winding number is only
defined in relative terms (see \eqref{eq:wind}), we must indicate it as 
a subscript.\footnote{Formally, the labels live in an extension of the latticeoid
$\pi^{-1}(x)\times\pi^{-1}(x)$ by the integral winding number.}
Thus, a label $(i,j)_n$ for a trajectory on $C$ means that it is calibrated by 
\begin{equation}
\label{calibratedeq}
\lambda_{(ij)_n}:=\bigl(\log y_j-\log y_i + 2\pi\ii n\bigr) d\log x
\end{equation}
where the $y_i$ is the $i$-th local solution of $H(x,y)=0$. All examples we
consider in this paper involve a covering of degree $2$, and we will then
use $i,j\in\{+,-\}$. Being integral curves of the first order differential 
equation:
\begin{equation}
\label{geodesiceq}
e^{-i\vartheta} \lambda_{(ij)_n}  = dt,
\end{equation} 
$(ij)_n$ BPS trajectories are naturally oriented. Orientation reversal is implemented by interchanging 
$(ij)_n$ and $(ji)_{-n}$. 

\begin{figure}[tbp]
\centering
\includegraphics[width=0.6\textwidth]{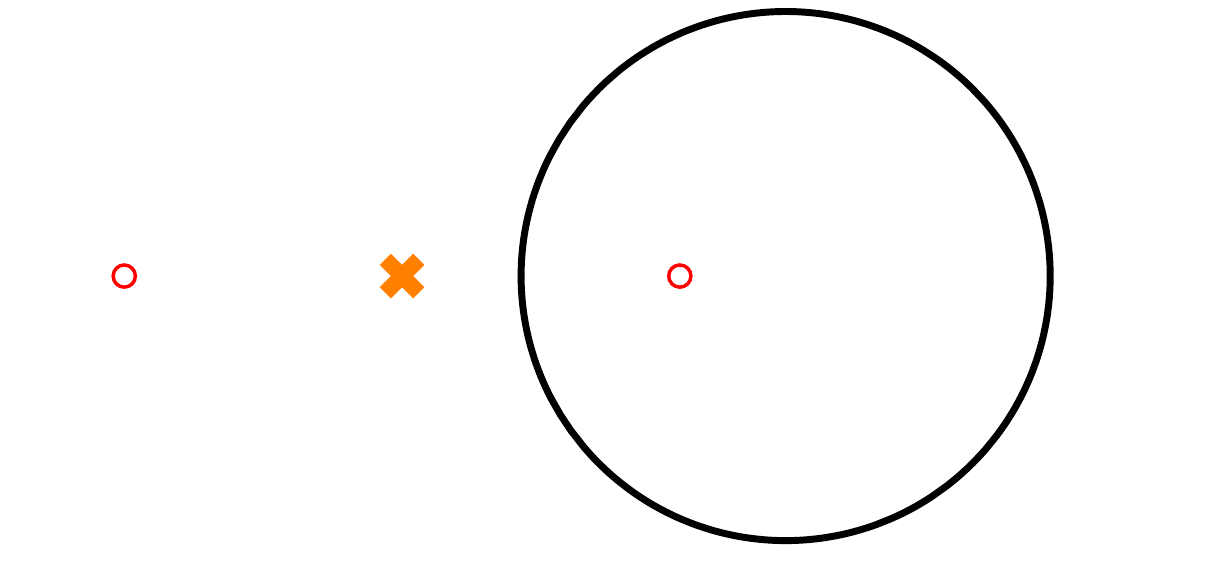}
\caption{D$0$-brane around a puncture.}
\label{fig:genericd02}
\end{figure}

Also note that $i=j$ is allowed if $n\neq 0$, and in fact this is an essential feature
for a successful physical/A-model interpretation of spectral networks of this
kind. Indeed, an easy warm up analysis of the BPS trajectories around a logarithmic branch point,
\ie, $x=0,\infty$, reveals a family of circular BPS trajectories of type $(++)_1$ 
or $(--)_1$ at $\vartheta=0$. These trajectories are shown in Figure
\ref{fig:genericd02}, and have constant length $4\pi^2$ in the normalization
\eqref{calibratedeq}. In the mirror picture, we interpret these trajectories 
as D0-branes localized near the non-compact leg of the toric diagram corresponding
to the punctures of $\Sigma$, see Fig.\ \ref{vertex}.

When $\vartheta$ is non-zero, the $(++)_1$ and $(--)_1$ trajectories near $x=0,\infty$
are logarithmic spirals of the form
\begin{equation}
\label{ncD2brane}
x(t) \sim \exp\bigl[\ee^{\ii\vartheta} t/(2\pi\ii)\bigr]
\end{equation}
which fall into the puncture as 
$t\to\pm\infty$ if $\vartheta > 0$ or emanate from the puncture if $\vartheta < 0.$  Under monodromy $\vartheta\to\vartheta+2\pi$, these trajectories 
come back to themselves up to the addition of a D0-brane. Based on this, we
interpret these trajectories as non-compact D2-branes extended along the
corresponding open leg of the toric diagram.

Perhaps the most interesting novelty compared to gauge theory is the presence of
``double logarithmic'' singularities in the differential, of the type
\begin{equation}
\lambda_{(ij)_n}  \sim \log x\; d\log x = \frac 12 d\bigl(\log x\bigr)^2
\end{equation}
when $y \sim x \to 0$ on one of the branches at the puncture. 
The corresponding trajectories look like
\begin{equation}
\label{ncD4brane}
x(t) \sim \exp\bigl[\ee^{\ii\vartheta/2} \sqrt{t}\bigr]
\end{equation}
which also spiral into/out of the puncture, but a slower rate than the D2-brane.
Cutting off the divergence at $|x_0|=\ee^{u}$, $u$ can be interpreted as the
(now finite) area of a holomorphic disk ending on the Harvey-Lawson brane
indicated by the dashed line in Fig.\ \ref{vertex}. In terms of this parameter,
the length of the trajectory \eqref{ncD4brane} up to $|x_0|$ displays a 
$t\sim u^2/2$ divergence naturally associated to a D4-brane.

\begin{figure}[tbp]
\centering
\includegraphics[scale=1]{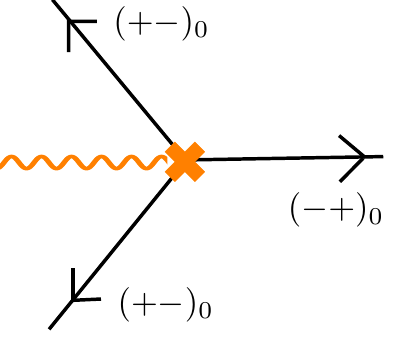}
\caption{Local structure of BPS trajectories near a $+-$ branch point.} 
\label{fig:localexp}
\end{figure}

Around the branch points, the analysis is basically analogous to the 
gauge theory case, with three BPS trajectories emanating from each ordinary
double point, leading to the local structure in Fig.\ \ref{fig:localexp}.
A slight inconvenience in following these trajectories around $C$ is the presence 
of the logarithmic branch cut running between $x=0$ and $\infty$: The
$(+-)_0$ strand begins/ends at the branch point, but not its image
after a non-trivial monodromy around $C$. As a potential remedy, we have 
included some partial notes on our numerical implementation in Appendix 
\ref{sec:coding}.

The most consequential novelty, which we will observe in our first example
in section \ref{conifold}, is the need to allow for ``stacked'' BPS trajectories,
in other words, trajectories carrying multiplicity $k>1$, with peculiar
interaction rules. These interactions of higher multiplicity are necessary to 
account for ``tachyon condensation'' that occurs at the intersection of 
(elementary) BPS trajectories away from the branch points but that does 
not separate the stacked trajectories.

\begin{figure}[t]
\begin{center}
\includegraphics[width=0.6\textwidth]{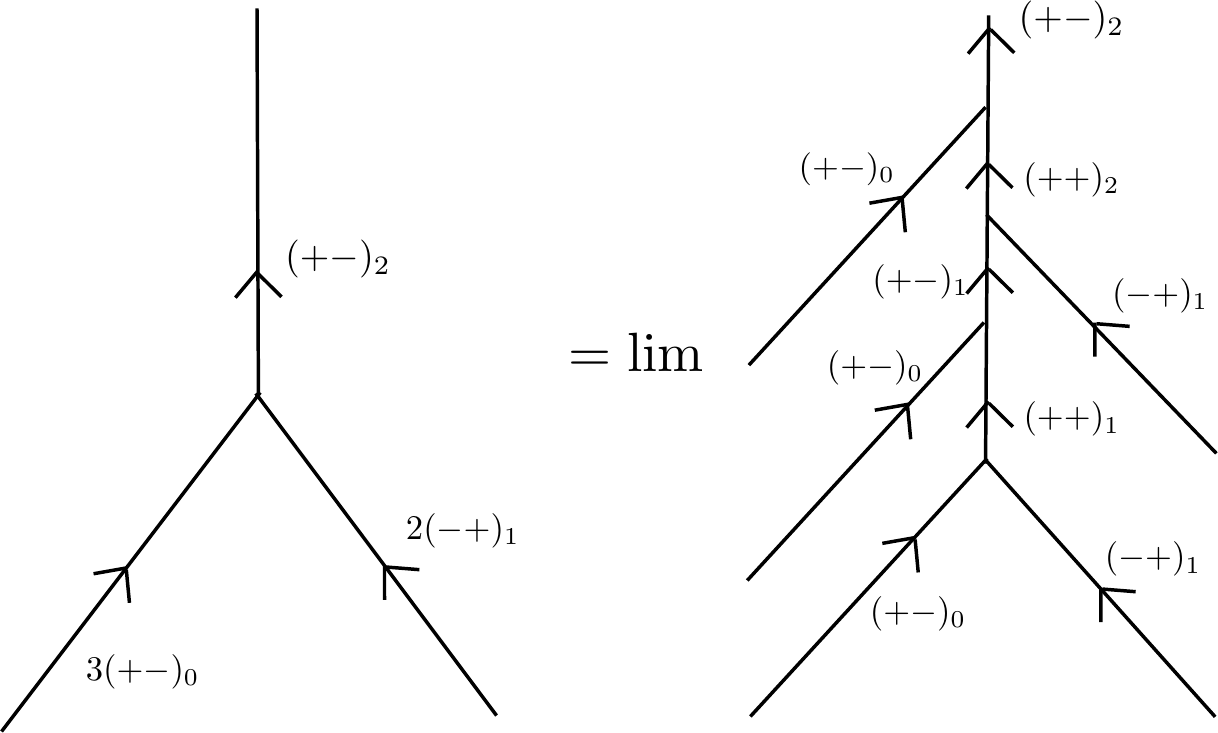}
\caption{Resolution of junctions with multiplicity.}
\label{multi23inter}
\end{center}
\end{figure}

To be more specific, we adopt the notation $k(ij)_n$ to indicate multiplicity 
$k$. Taking into account that the charges coming into a vertex must add up to
zero in the latticeoid of charges, we find that we need to allow for the following
interactions of BPS trajectories for every $k\ge 1$:
\begin{equation}
\label{eq:reaction1}
k(ij)_{n_1}+k(ji)_{n_2} \to (ii)_{(n_1+n_2)k}
\end{equation}
\begin{equation}
\label{eq:reaction2}
(k+1)(ij)_{n_1} + k(ji)_{n_2} \to (ij)_{n_1(k+1)+n_2k}
\end{equation}
Pictorially, one might think of these multiple junctions as a sequence of elementary
junctions of the type $(+-)_0 + (-+)_1\to (++)_1$ followed by 
$(++)_1+(+-)_0\to (+-)_1$, in the limit in which the interaction vertices sit on 
top of each other. We have depicted the interaction \eqref{eq:reaction2} for $k=2$ 
in Fig.\ \ref{multi23inter}.

\subsection{\ldots for old Geometries}

\begin{figure}[tbp]
  \begin{subfigure}[b]{0.3\textwidth}
  \centering
    \includegraphics[width=0.8\textwidth]{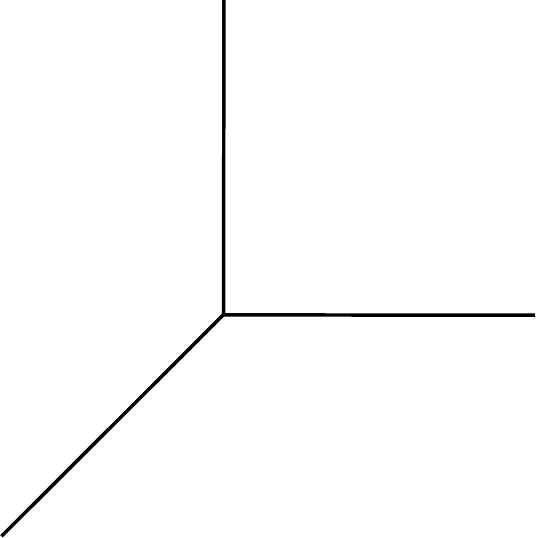}
    \caption{$\C^3$}
    \label{fig:toricc3}
  \end{subfigure}
  \hfill
  \begin{subfigure}[b]{0.3\textwidth}
    \includegraphics[width=\textwidth]{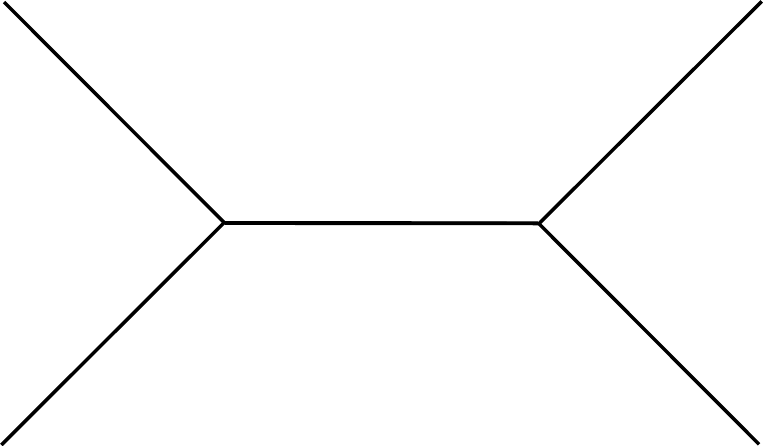}
    \caption{Resolved conifold}
    \label{fig:toriccon}
  \end{subfigure}
    \hfill
  \begin{subfigure}[b]{0.3\textwidth}
  \centering
    \includegraphics[width=0.62\textwidth]{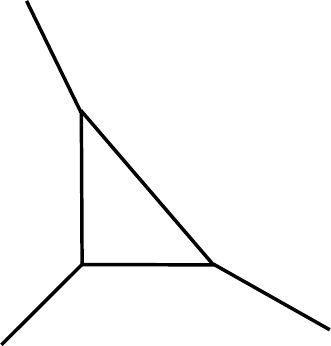}
    \caption{local $\CP^2$}
    \label{fig:toriccp2}
  \end{subfigure}
  \caption{Toric diagrams of some local Calabi-Yaus.}
  \label{fig:toricd}
\end{figure}

In the rest of the paper we investigate the three simplest toric Calabi-Yau manifolds: 
$\C^3$, the resolved conifold (``local $\CP^1$"), and local $\CP^2$. The toric diagrams 
are shown in Figure \ref{fig:toricd}. In Table \ref{tab:curves} we collect the (mirror 
to the) D-term equations, which are used to solve for the variables $y_i$ in terms of 
two of them that we call $x$ and $y$ (recall that one of the variables is set to 1). 
The choice of which variables are kept can be interpreted as choosing the leg of the 
toric diagram on which the probing brane sits (see discussion around Figure \ref{vertex}); 
this choice is immaterial as far as the BPS spectrum is concerned. We adjust the 
curve $W=\sum y_i=0$  with the framing operation 
$$y\to y, x\to (-y)^fx$$ 
so that the resulting expression is quadratic in $y$ (this is not always possible for 
more complicated examples). Note that the sign in the framing rule is completely 
innocuous at the level of pictures and merely implements a reflection of the $x$-plane.

\begin{table}[hp]
\begin{center}
\begin{tabular}{l|c|c}
Geometry & (mirror to the) D-term equations & framed curve $W=\sum y_i$ \\
 \hline
 \hline
$\C^3$ & $\emptyset$ & $-x + y + y^2$\\
\hline
Resolved conifold & $y_1y_2y_3^{-1}y_4^{-1}=Q$& $-1 + y + x y - Q x y^2$\\
\hline
Local $\CP^2$ &$y_1y_2y_3y_4^{-3}=Q$ &$-Q x^3 - y + x y + y^2$
\end{tabular}
\caption{Framed mirror curves.}
\label{tab:curves}
\end{center}
\end{table}

\section{Resolved Conifold}
\label{conifold}

The conifold is the singular geometry $X = \{x_1 x_2 - x_3 x_4 = 0\} \subseteq 
\C^4$. It has two small resolutions $\hat X^{\pm}$ obtained by blowing up the ideals 
$(x_1, x_3)$ and $(x_2, x_4),$ respectively.  The resolutions contain the curves 
$C^{\pm} \cong \CP^1$.

The moduli space is given by the complexified volume of the compact $\CP^1$, denoted 
by $\log Q$ below. The two large volume regions $|Q| >> 1$ and $|Q| << 1$ correspond to 
the two resolutions $\hat X^{\pm}$ of the conifold and are connected by a birational 
transformation known as a ``flop''.  The two resolutions are in fact isomorphic. We 
will see below that the flop transformation acts in an interesting way on spectral 
networks, providing the first motivation for the junction rules. 

\subsection{Webs and quiver representations}

In large volume terminology, the compact branes are classified by their charges $Q_2$ 
and $Q_0,$ where $Q_2$ is the D2-brane charge wrapping the compact $\CP^1$ and $Q_0$ the 
D2-charge. The central charge is
\begin{equation}
Z_Q(Q_2,Q_0)=\frac{Q_2}{2\pi i}\log Q+Q_0
\end{equation}
and the stable branes have charge $\pm Q_2 + n Q_0, n \in \Z.$ 

\begin{figure}[t]
\begin{center}
\begin{tikzpicture}[scale=1] 
\tikzset{ 
my loop/.style={->,to path={ 
.. controls +(-45:2) and +(45:2) .. (\tikztotarget) \tikztonodes}}
} 
\tikzset{ 
left loop/.style={->,to path={ 
.. controls +(225:2) and +(135:2) .. (\tikztotarget) \tikztonodes}}
} 

\path (0,0) node[draw,shape=circle,color=Mblue] (v0) {$v_1$}; 
\path (4,0) node[draw,shape=circle,color=Morange] (v1) {$v_2$}; 
\path[->, thick, bend left = 20] (v0) edge node[above, yshift = -0.3ex] {$a_1$} (v1);
\path[->, thick, bend left = 40] (v0) edge node[above] {$a_2$} (v1);
\path[->, thick, bend left = 20] (v1) edge node[below,yshift=0.7ex] {$b_1$} (v0);
\path[->, thick, bend left = 40] (v1) edge node[below] {$b_2$} (v0);
\end{tikzpicture}
\caption{Quiver corresponding to the conifold geometry.} 
\label{fig:conifoldquiver}
\end{center}
\end{figure}
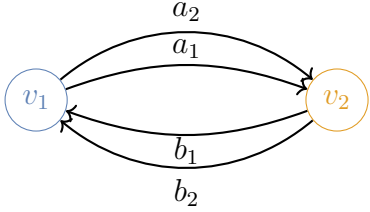
\begin{figure}[h]
\begin{center}
\begin{subfigure}[b]{0.4\textwidth}
\includegraphics[width=\textwidth]{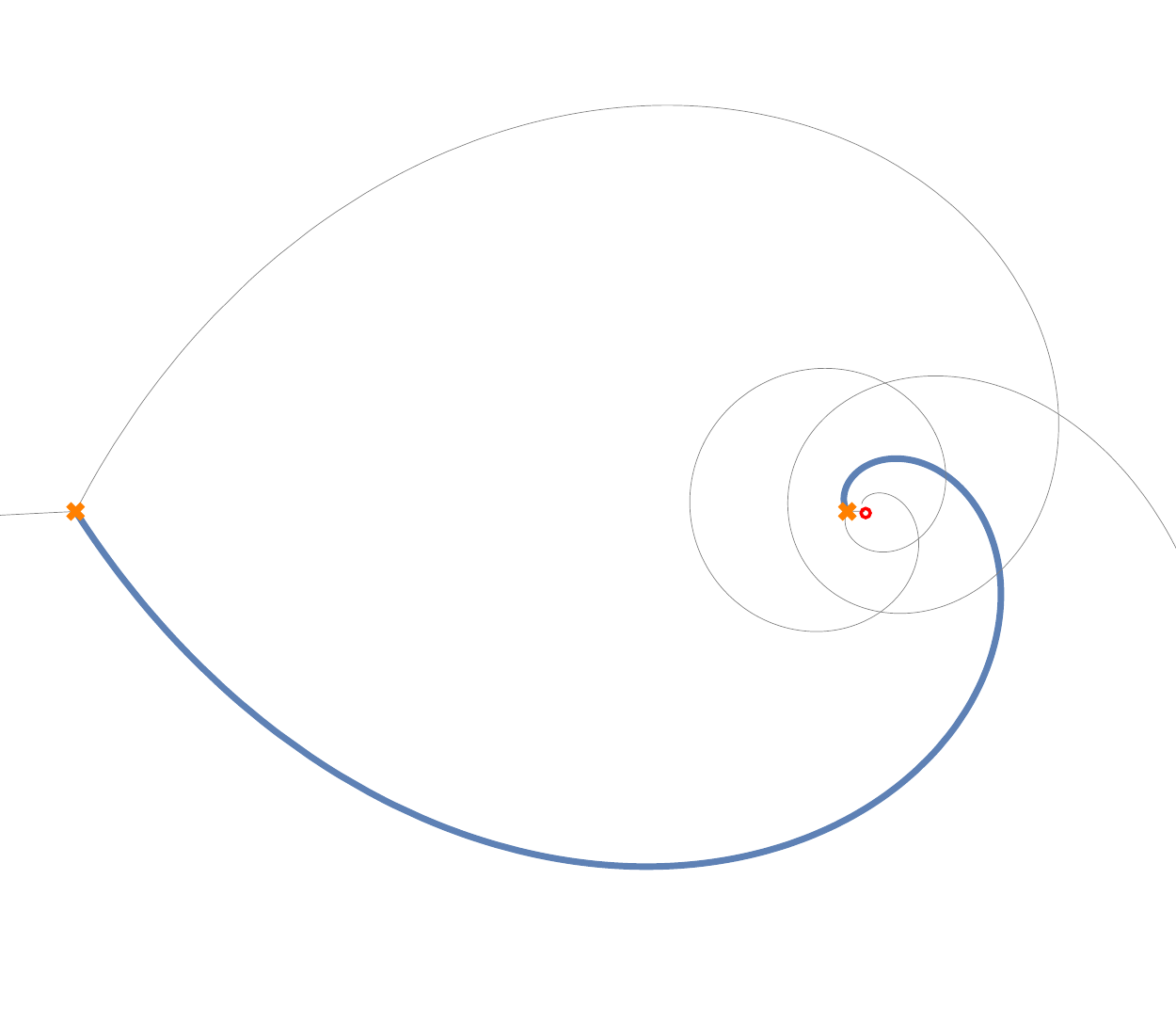}
\caption{$\Dtwobar$ + D0 brane.}
\label{fig:conifoldS1}
\end{subfigure}
  \hfill
\begin{subfigure}[b]{0.4\textwidth}
\includegraphics[width=\textwidth]{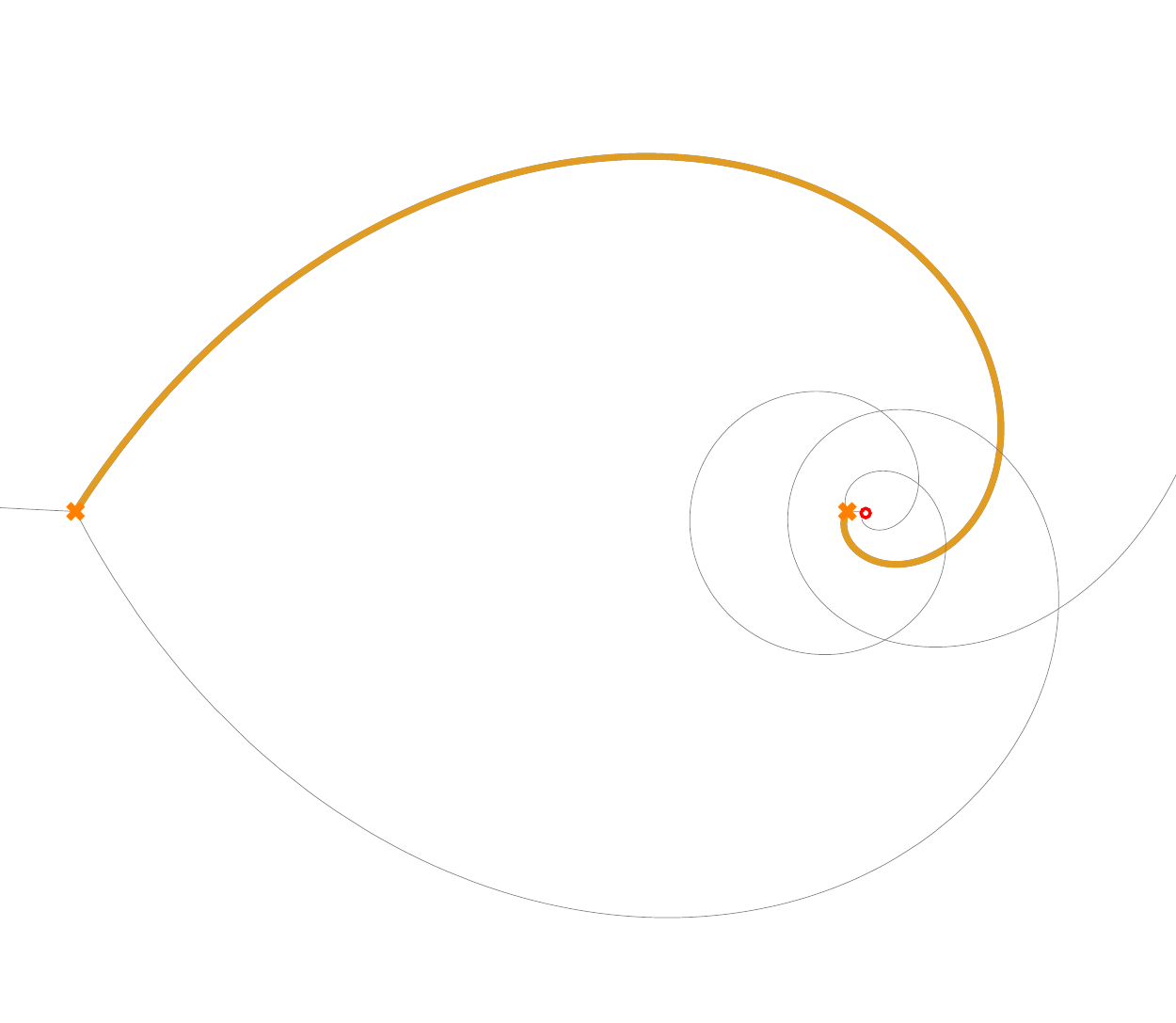}
\caption{D2 brane.}
\label{fig:conifoldS2}
\end{subfigure}
\caption{The two basic states for the conifold.}

\includegraphics[width=8cm]{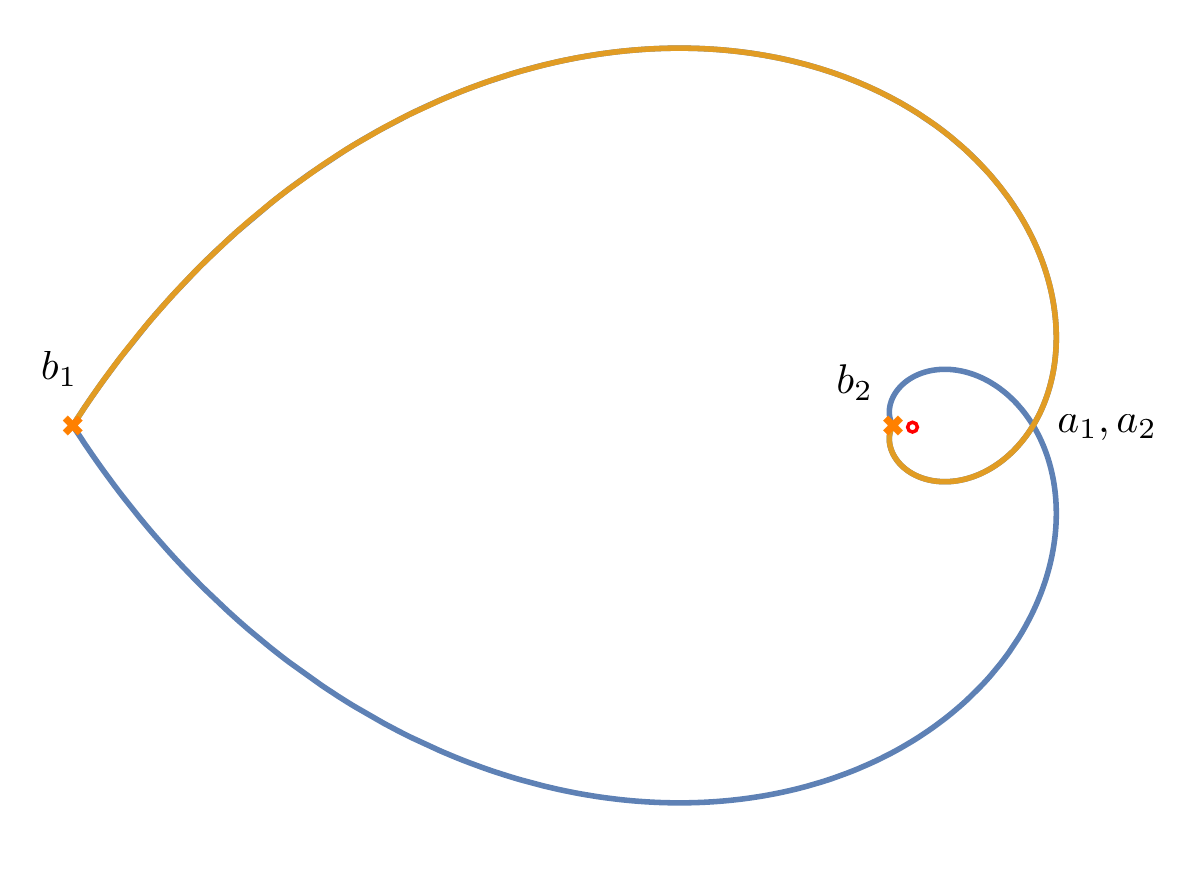}
\caption{Superimposing the two basic finite webs on the conifold, showing the four 
bifundamental fields arising at intersection points. Note that the two basic states 
do not occur at the same phase. }
\label{fig:superimpose}
\end{center}
\end{figure}

The D0 brane is a bound state of the basic states (``fractional branes") D2 
and $\Dtwobar + {\textrm D}0$ respectively.  These two basic states are 
represented in the quiver quantum mechanics by the two nodes of the quiver in Figure 
\ref{fig:conifoldquiver} and have dimension vectors $(1,0)$ and $(0,1)$ respectively.
The two basic states are realized by finite webs shown in figures \ref{fig:conifoldS1} 
and \ref{fig:conifoldS2} and have constant topology throughout the moduli space.  
The two basic states intersect in four points, which give rise to the four bifundamental 
fields in the conifold quiver.  On the curve $C$ the four intersection points are 
depicted in Figure \ref{fig:superimpose}.

\begin{figure}[tbh]
\centering
\includegraphics[width=8cm]{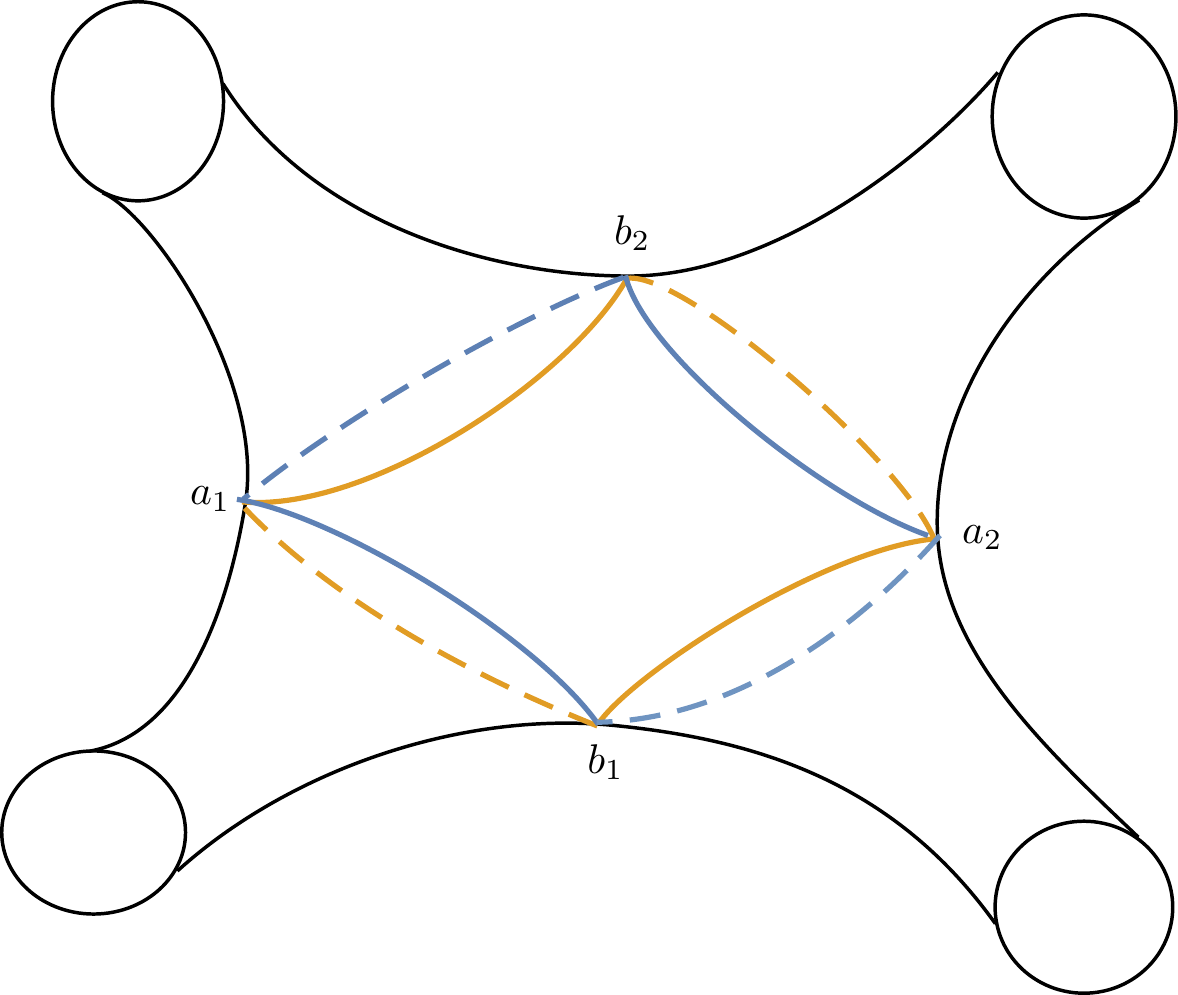}
\caption{Rendering on the mirror curve of the two basic states on the resolved conifold.}
\label{fig:conifoldbasic}
\end{figure}

Two intersections correspond to the fields $a_1$ and $a_2$ at the two branch points.  
The third intersection point, away from the two branch points corresponds to two fields 
$b_1, b_2$, since $\pi:\Sigma\to C$ is a double covering away from the branch points.  
The four intersection points on $\Sigma$ are shown in Figure \ref{fig:conifoldbasic}. 
Taking orientations into account, $a_1$ and $a_2$ transform in the dual representation 
to $b_1$ and $b_2.$  Moreover, the two holomorphic disks from the top and bottom of 
the ``pillowcase'' in Figure \ref{fig:conifoldbasic} contribute the two terms in the 
Klebanov-Witten superpotential
\begin{equation}
\cW=\Tr (a_1b_1a_2b_2-a_1b_2a_2b_1).
\end{equation}
The resulting quiver for the conifold is shown in Figure \ref{fig:conifoldquiver}.

\paragraph{A tale of two phases} 
\begin{figure}[tbp]
\centering
\includegraphics[width=0.8\textwidth]{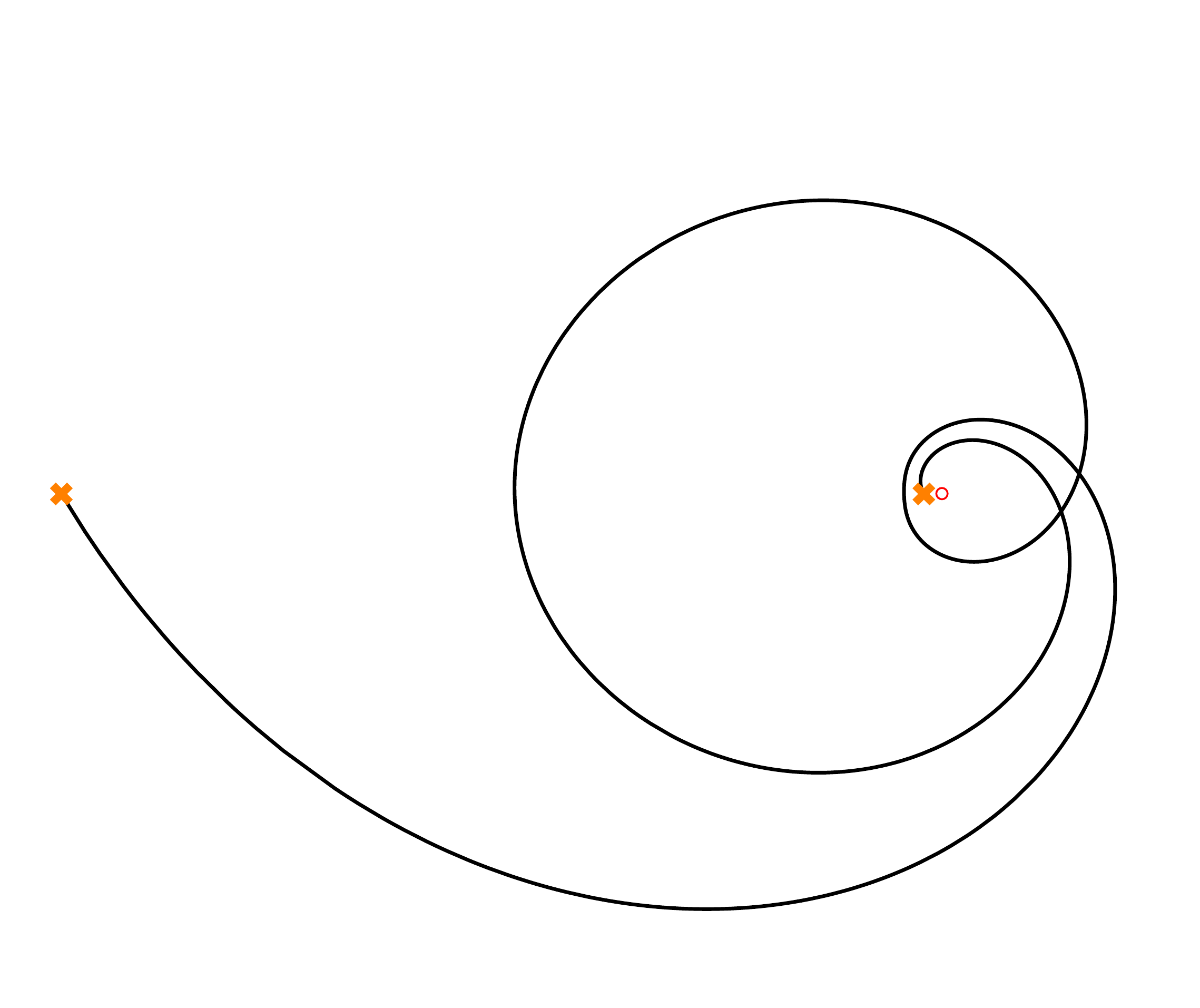}
\caption{$\Dtwobar + 2{\rm D}0$ or brane $(2,1)$ at $|Q|>1$.}
\label{fig:conifoldB}
\includegraphics[width=0.8\textwidth]{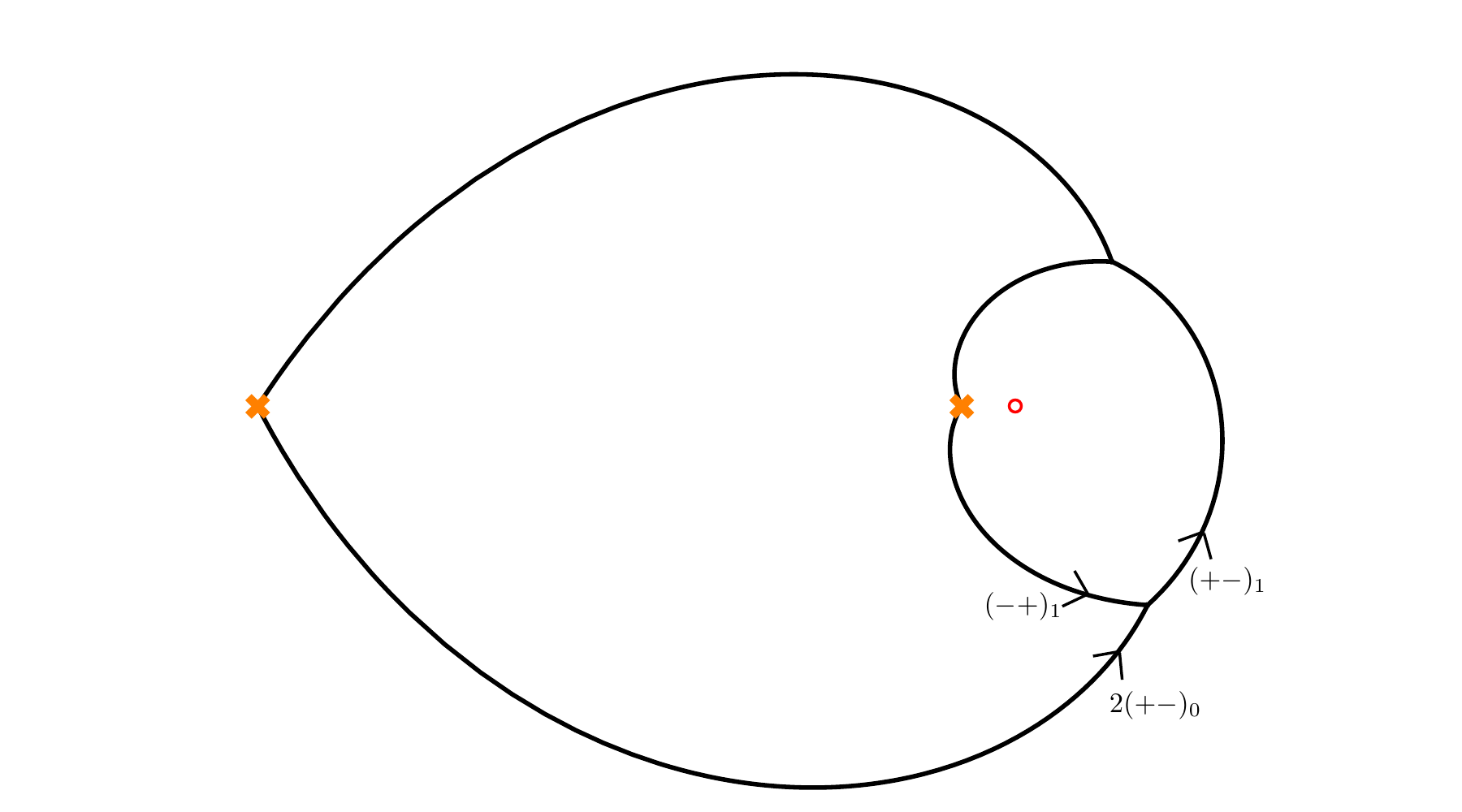}
\caption{Brane $(2,1)$ at $|Q|<1$. }
\label{fig:conifoldC}
\end{figure}
Bound states of the two basic branes are realized differently in the two phases. 
Recall the heuristic picture developed in section \ref{BPSQ} where bound states are 
formed by resolving intersections.  In the quiver quantum mechanics, the bound states 
arise from tachyonic fields condensing and the matter fields acquiring a non-zero 
vacuum expectation value (VEV).  In terms of quiver representations, this corresponds 
to arrows associated to resolved intersections taking non-zero values. 

For $|Q|>1$, the bound states are made of concatenated copies of the basic finite webs 
that have detached from the branch point anchors. This precisely mimics the situation in 
$SU(2)$ gauge theory shown in Figure \ref{fig:finitewebs}. Indeed, in terms of the quiver 
in Figure \ref{fig:conifoldquiver}, this corresponds to only the $b$-arrows taking on 
non-zero values, effectively reducing to the Kronecker-2 quiver of the $SU(2)$ gauge 
theory.  This is in agreement with the representation theory of the conifold quiver as explained in Appendix \ref{appendixC}.
An example bound state corresponding to the representation with dimension 
vector $(2,1)$ in this phase is shown in Figure \ref{fig:conifoldB}.

\paragraph{Down the rabbit hole} 

In the other phase $|Q|<1$, naturally, the situation is the opposite. As can be expected, 
the bound states form by resolving intersections associated to the $a$-arrows, namely the 
collision points of the BPS trajectories. This makes use of the junction rules introduced 
in section \ref{executive}.  A representative bound state in the $(k,k+1)$ or $(k+1,k)$ 
family is shown in Figure \ref{fig:conifoldC}.  The $a$-intersection points have been 
resolved by a $(+-)_k$  
strand.  This observation will be useful later to visualize these representations as 
string modules.
Namely, by symmetry of the two phases, we learn that we should also associate the $(k,k+1)$ or 
$(k+1,k)$ representation of the Kronecker-2 quiver to the resolution of a
multiple intersection of $(k+1)(+-)_0$ with $k(-+)_1$ by insertion of a (single) $(+-)_k$
stub.

\begin{figure}[tp]
\centering
\includegraphics[width=0.8\textwidth]{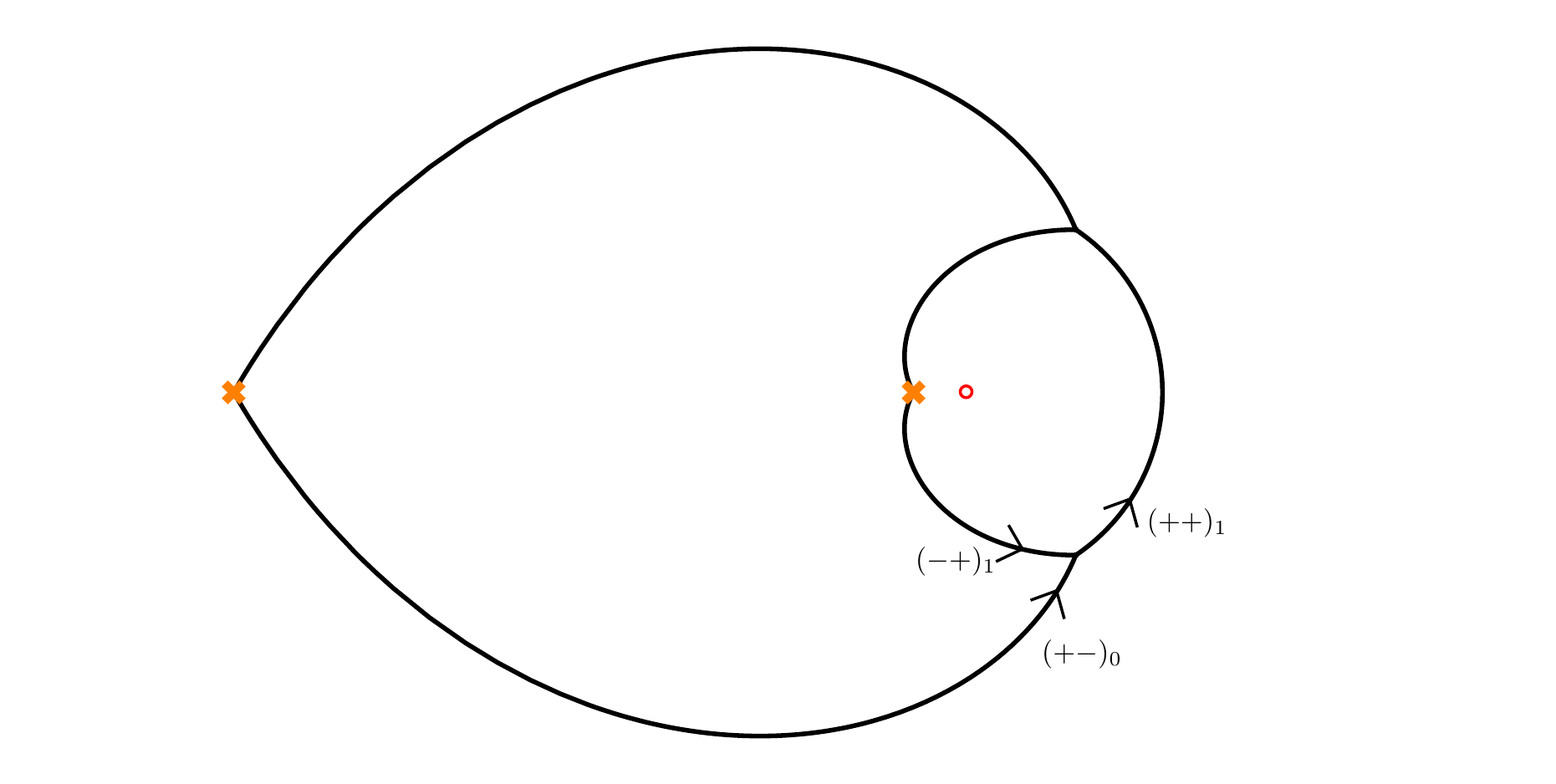}
\caption{The D0 brane $(1,1)$ at $|Q|<1$.}
\label{fig:conifoldD0}
\end{figure}

\begin{figure}[tbp]
\begin{center}
\includegraphics[width=8cm]{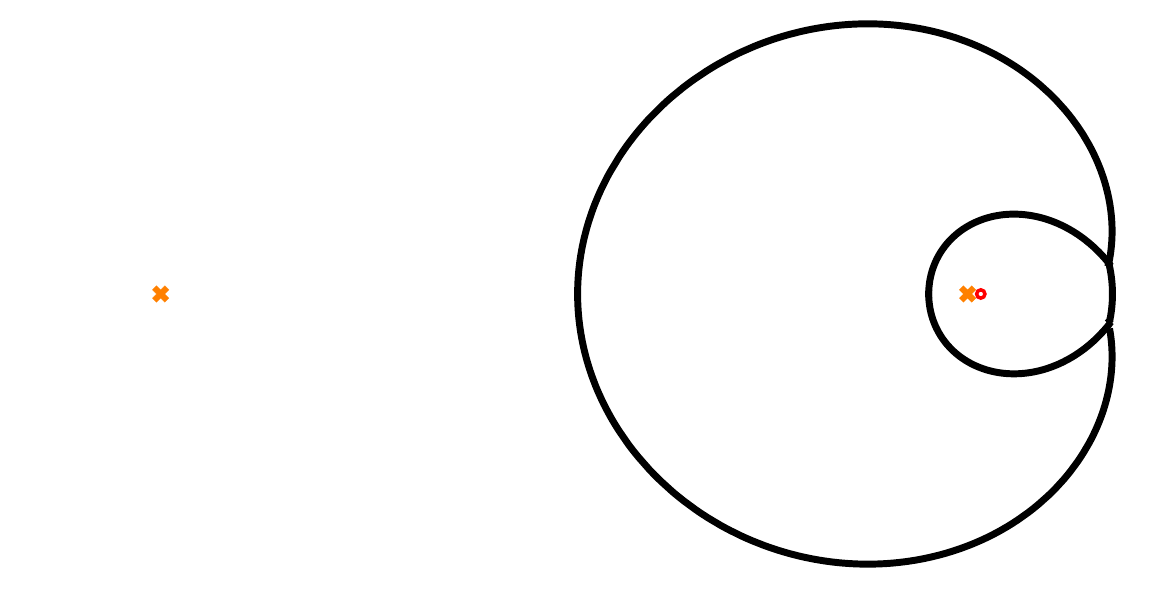}
\caption{Generic finite web for the D0-brane on the resolved conifold.}
\label{fig:genericcon}
\end{center}
\end{figure}

\paragraph{A SLAG's point of view of stringy geometry}
A special D0-brane corresponding to dimension vector $(1,1)$ is shown in Figure 
\ref{fig:conifoldD0}.  A generic D0-brane with both $a$ and $b$ bifundamentals 
turned on has a $(++)$ or $(--)$ and is detached from the branch points, as shown 
in Figure \ref{fig:genericcon}.

\begin{figure}[tbp]
\centering
\includegraphics[width=0.9\textwidth]{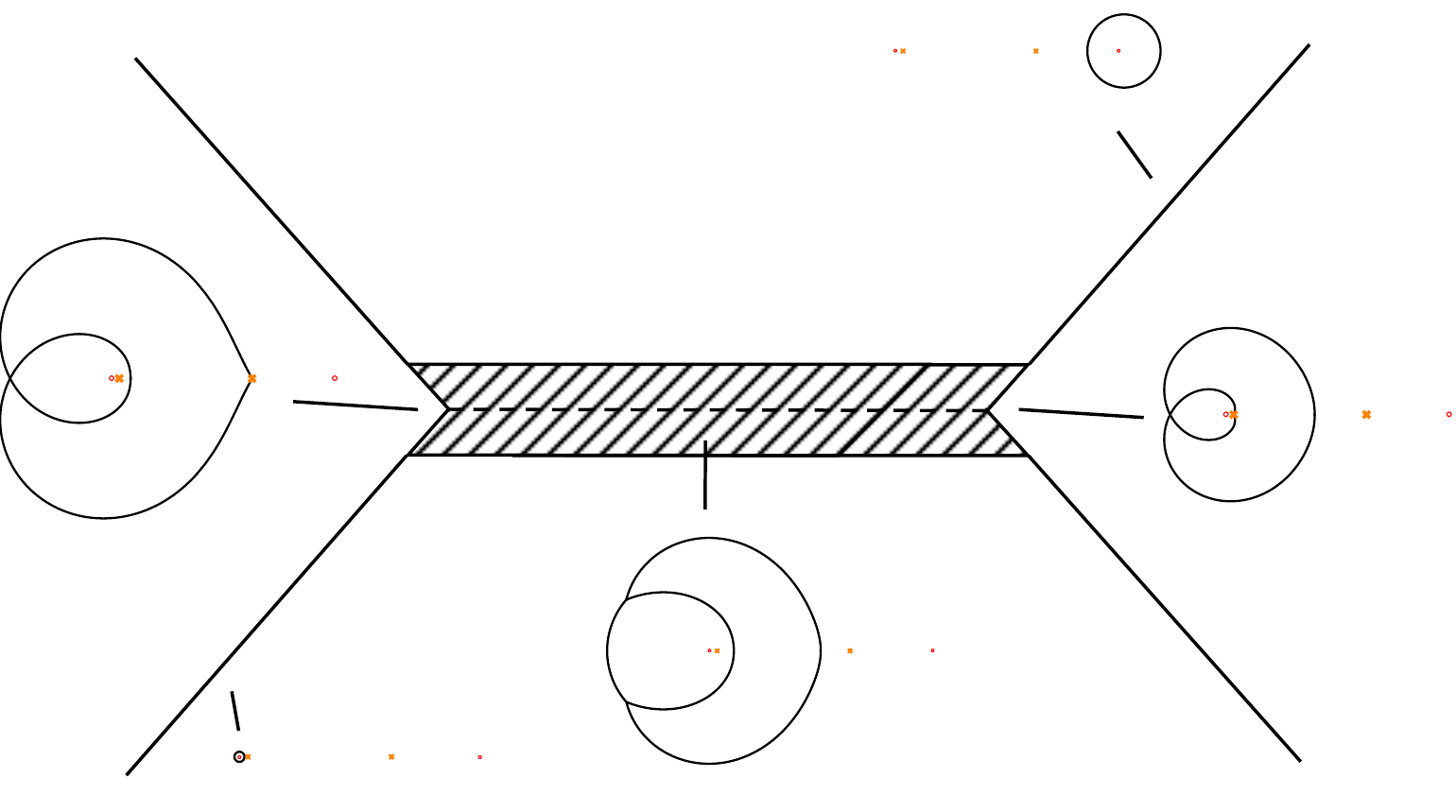}
\caption{Moduli of D$0$-brane on the resolved conifold. Networks plotted in a variable 
such that the $x=\infty$ puncture lies at finite distance.}
\label{fig:conmoduli}
\end{figure}

\begin{figure}[tbp]
\centering
\includegraphics[width=0.6\textwidth]{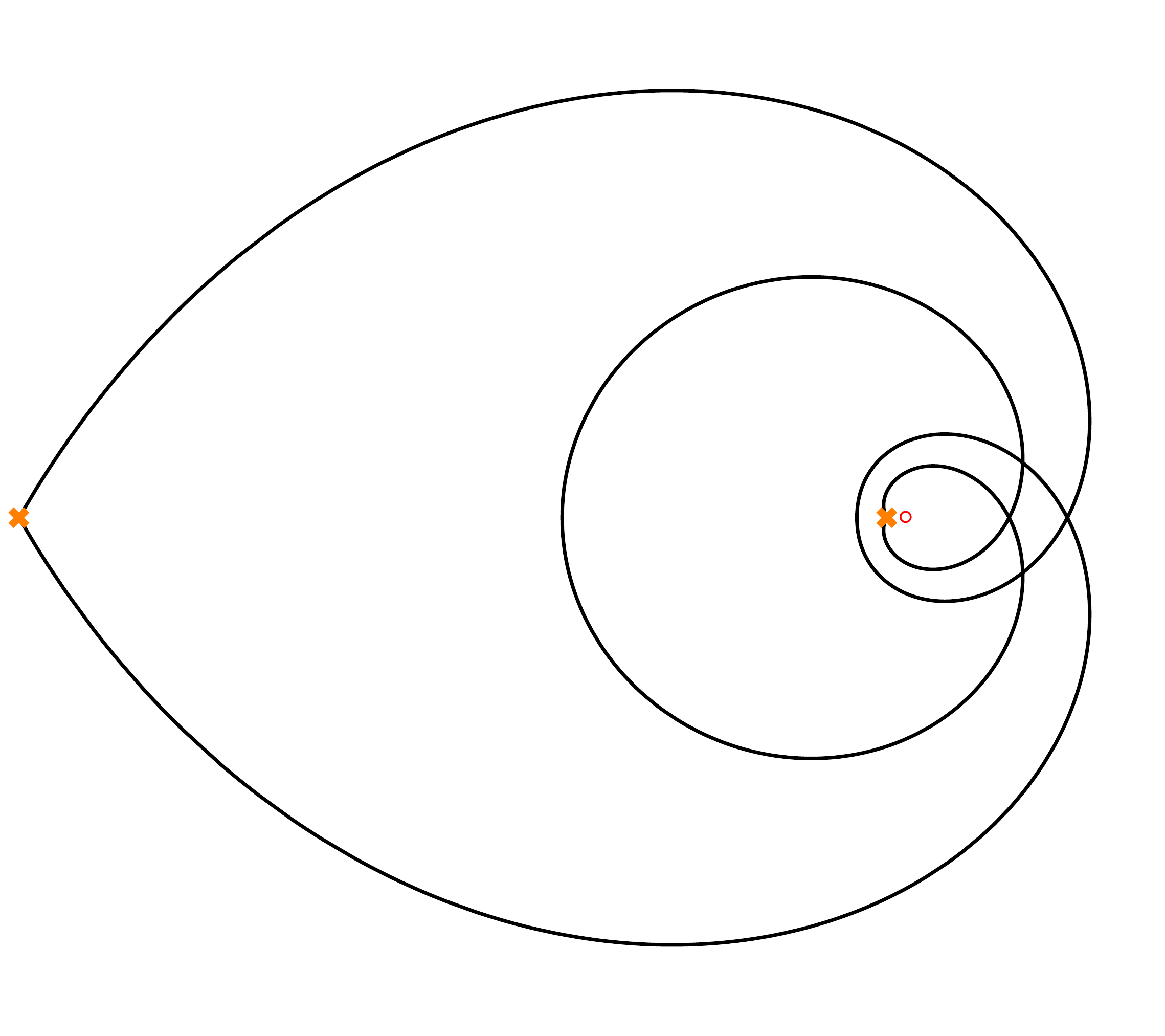}
\caption{Two fixed points for the D0 brane $(1,1)$ at $|Q|>1$.}
\label{fig:conifoldD0A}
\end{figure}

Unlike the $(k,k+1)$ and $(k+1,k)$ bound states, the representation theory of the 
D0-brane does not reduce to that of the Kronecker-2 quiver.  Indeed, all four 
fields can gain expectation values.  For dimension vector $(1,1)$ the F-term constraints 
are vacuous, and the resolved conifold geometry is recovered from the quiver quantum 
mechanics as the GIT theory quotient of four fields with charges $(1,1,-1,-1)$ by a 
$U(1)$ gauge group.  A map of the moduli space of the D0 is shown in Figure 
\ref{fig:conmoduli}, where the vertical reflection symmetry exchanges $(++)$ 
and $(--)$ strands.  The compact part of the D0-brane moduli space is $\CP^1$ and the two corresponding fixed point are shown in Figure \ref{fig:conifoldD0A}.

\begin{figure}[htbp]
\begin{center}
\includegraphics[width=9cm]{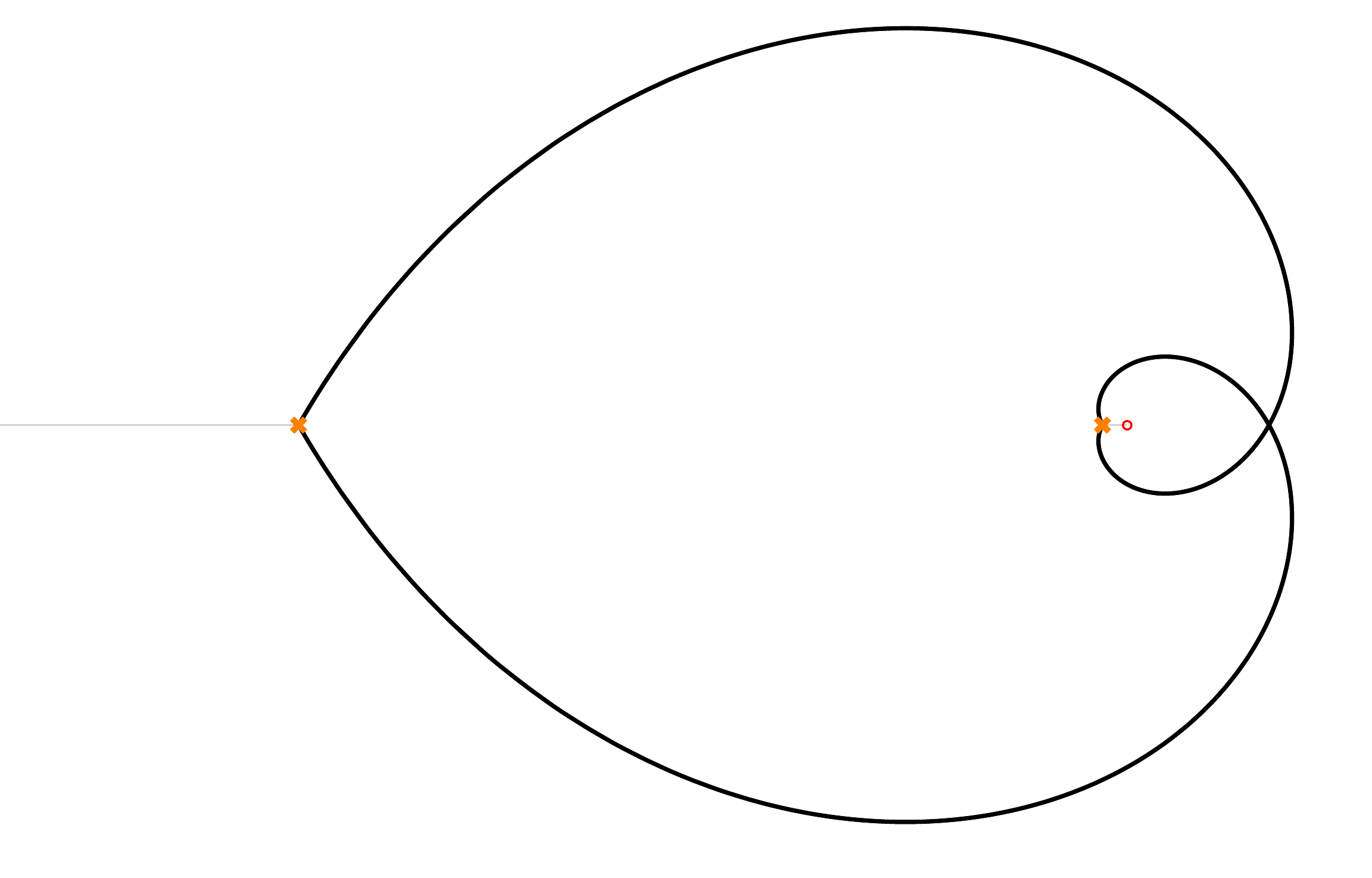}
\caption{Spectral network at $Q=-1$, on the line of marginal stability.}
\label{fig:conifoldmarginal}
\end{center}
\end{figure}

In summary, the flop transformation preserves the spectrum and interchanges the 
two realizations of the Kronecker-2 quiver. At $|Q|=1$, the two basic branes have 
aligned central charges and they come to coexist with all of their bound states 
in the same spectral network, as depicted in Figure \ref{fig:conifoldmarginal}. All the 
resolutions contract to zero length, nicely interpolating between the $|Q| > 1$ phase and the 
$|Q| < 1$ phase.

\section{A Local Calabi-Yau}
\label{localP2}
Local $\mathbb{P}^2$ is the total space of the bundle $\cO(-3) \rightarrow \mathbb{P}^2,$ 
which can be obtained by resolving the orbifold $\C^3/\Z_3.$  The smooth geometry and 
the orbifold correspond to the large volume and orbifold points in the complexified 
K\"ahler moduli space \cite{agm}.
\begin{figure}[tp]
\centering
\begin{tikzpicture}[scale=0.8] 
\path (0,0) node[draw,shape=circle,color=Mblue] (v0) {$v_0$}; 
\path (0:4cm) node[draw,shape=circle,color=Morange] (v1) {$v_1$}; 
\path (60:4cm) node[draw,shape=circle,color=Mgreen] (v2) {$v_2$}; 
\path[->,thick, bend right = 20] (v0) edge node[below] {$a_1,a_2,a_3$} (v1);
\path[->,thick, bend right = 0] (v0) edge node[] {$$} (v1);
\path[->,thick, bend right = -20] (v0) edge node[above] {$$} (v1);
\path[->, thick, bend right = 20] (v1) edge node[right] {$b_1,b_2,b_3$} (v2);
\path[->, thick, bend right = 0] (v1) edge node[right] {$$} (v2);
\path[->, thick, bend right = -20] (v1) edge node[left] {$$} (v2);
\path[->, thick, bend right = 20] (v2) edge node[left] {$c_1,c_2,c_3$} (v0);
\path[->, thick, bend right = 0] (v2) edge node[left] {} (v0);
\path[->, thick, bend right = -20] (v2) edge node[right] {} (v0);
\end{tikzpicture}
\caption{Quiver for local $\projective^2$.} 
\label{fig:localP2quiver}
\end{figure}
The quiver is shown in Figure \ref{fig:localP2quiver}
and has superpotential
\begin{equation}
\cW = \Tr \left( \epsilon_{ijk} a^i b^j c^k \right).
\end{equation}
The spectrum of BPS branes is much richer than in the previous example and displays an 
intricate wall-crossing structure that was studied in \cite{dfr}. At large volume, the 
stable branes are sheaves on $\CP^2$ while near the orbifold point they are in 
correspondence with quiver representations \cite{aspinwallPi}. In this section we 
explore the relation between quiver representations and spectral networks near the 
orbifold point, show an example of D-brane decay, and identify massless branes at 
the conifold point.
The central charge in various bases, as well as the conversion between them, are given 
in appendix \ref{sec:centralcharge}.

\subsection{Orbifold point}
\begin{figure}[tbp]
\begin{center}
\includegraphics[width=9cm]{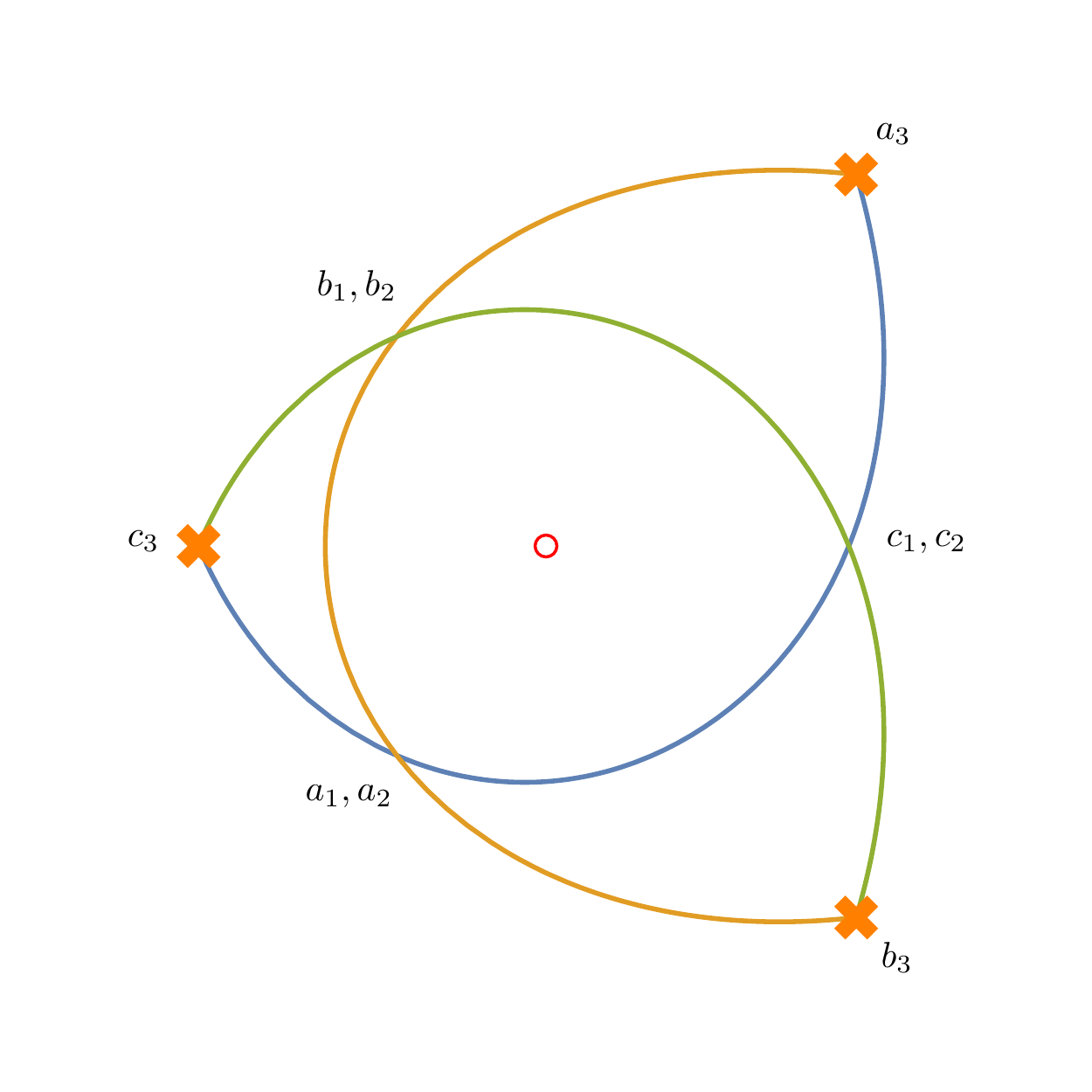}
\caption{The three fractional branes near the orbifold point and the 
bifundamental fields corresponding to their intersection points. Note that the fractional branes all 
occur at different phases.}
\label{fig:localp2}
\end{center}
\centering
\begin{tikzpicture}[scale=10]
\draw[help lines, color=gray!30, dashed,step=.1cm] (-.49,-.19) grid (.49,.19);
\draw[ultra thick] (-.5,0)--(.5,0) node[right]{$\Re Z$};
\draw[ultra thick] (0,-.2)--(0,.2) node[above]{$\Im Z$};
\draw[->,very thick, Mblue] (0,0)--(0.298695,-0.0562444) node[right, black]{$F_0$};
\draw[->,very thick, Morange] (0,0)--(0.40261, 0) node[above right, black]{$F_1$};
\draw[->,very thick, Mgreen] (0,0)--(0.298695, 0.0562444) node[above right, black]{$F_2$};
\end{tikzpicture}
\caption{Central charges of the fractional branes at 
$\psi = -\frac{1}{6}.$}
\label{Fig:localp2Zs}
\end{figure}

The fractional branes near the orbifold point are represented by the orange, green 
and blue finite webs respectively in Figure \ref{fig:localp2}.   The fractional branes 
$F_0, F_1,$ and $F_2$ correspond to the simple representation with dimension vectors 
$(1,0,0)$, $(0,1,0)$ and $(0,0,1).$  The central 
charges of the basic fractional branes are shown in Figure \ref{Fig:localp2Zs}.  For 
clarity, we have chosen a point in the complex structure moduli space where the central 
charges of the two fractional branes nearly align. This ensures that the resolutions of 
Lagrangian intersections will be localized close to the original intersection point.  The central charges of these branes as a function 
of the complex structure modulus is relegated to Appendix \ref{sec:centralcharge}. 
Each pair of fractional branes intersect in three points.  For each pair, one 
intersection point is at a branch point and two intersection points on $\Sigma$ 
come from the two distinct lifts of an intersection point on $C.$
The resulting quiver is shown in Figure \ref{fig:localP2quiver}.

\paragraph{Kronecker-3 Quiver}
We first consider bound states of two of the two fractional branes $F_1$ and $F_2$.  The quiver quantum mechanics reduces to the Kronecker-3 subquiver consisting of
three arrows $b_1, b_2, b_3$ between the nodes $v_1$ and $v_2.$
We label the intersection at the 
bottom branch point by $B$ and label the two intersection points near the top 
center by `$+$' and `$-$'.  This is the same labelling scheme for the Kronecker-3 quiver used in Figure \ref{fig:K3}.
As explained in section \ref{BPSQ}, resolving 
intersections corresponds to giving a VEV to the corresponding fields in the quiver 
quantum mechanics.  The notation in the figures to come is explained in the next subsection.

\begin{figure}[tbp]
  \begin{subfigure}[b]{0.45\textwidth}
    \includegraphics[width=\textwidth]{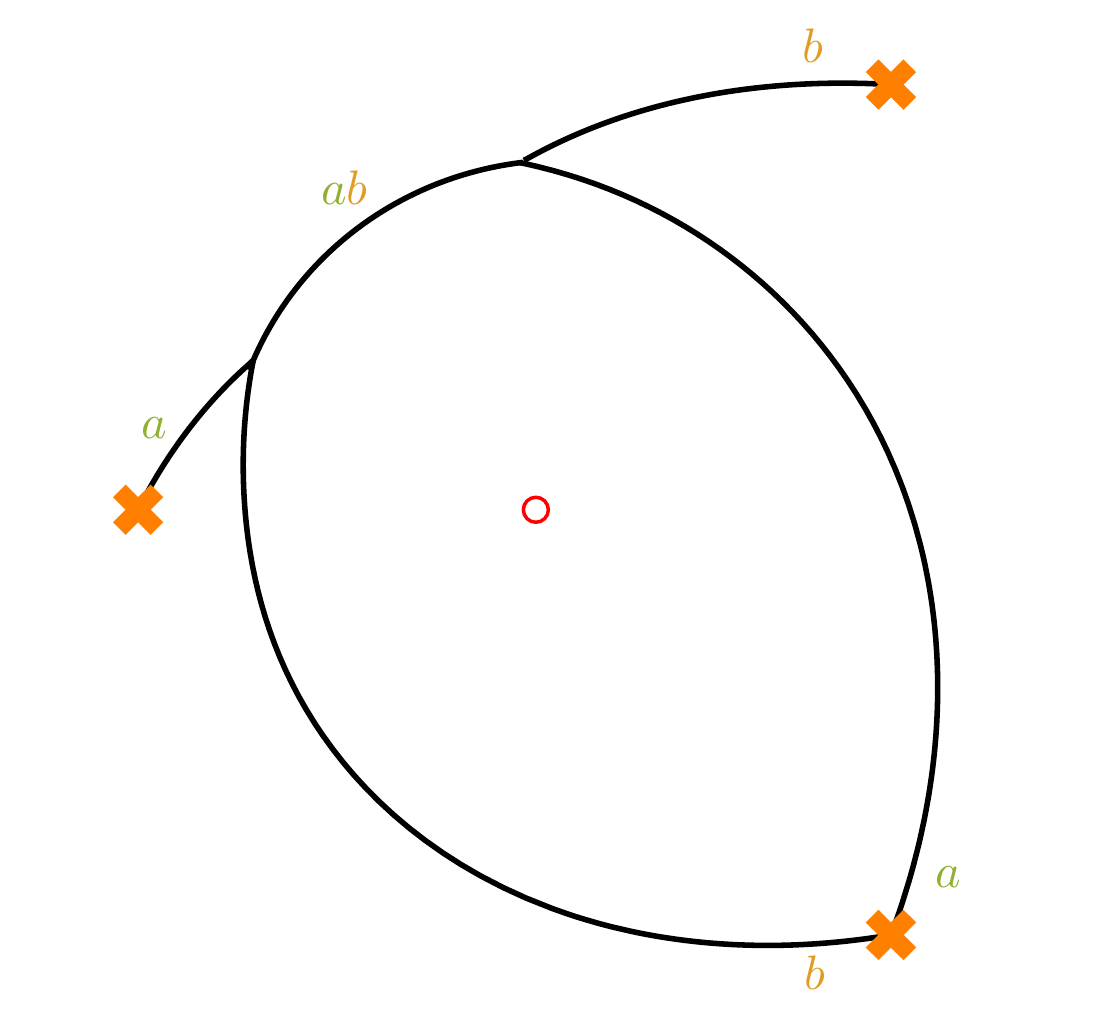}
    \caption{Resolve by a 1:1 strand ($++$ or $--$).}
    \label{fig:f1}
  \end{subfigure}
  \hfill
  \begin{subfigure}[b]{0.45\textwidth}
\raisebox{.25cm}{\includegraphics[width=\textwidth]{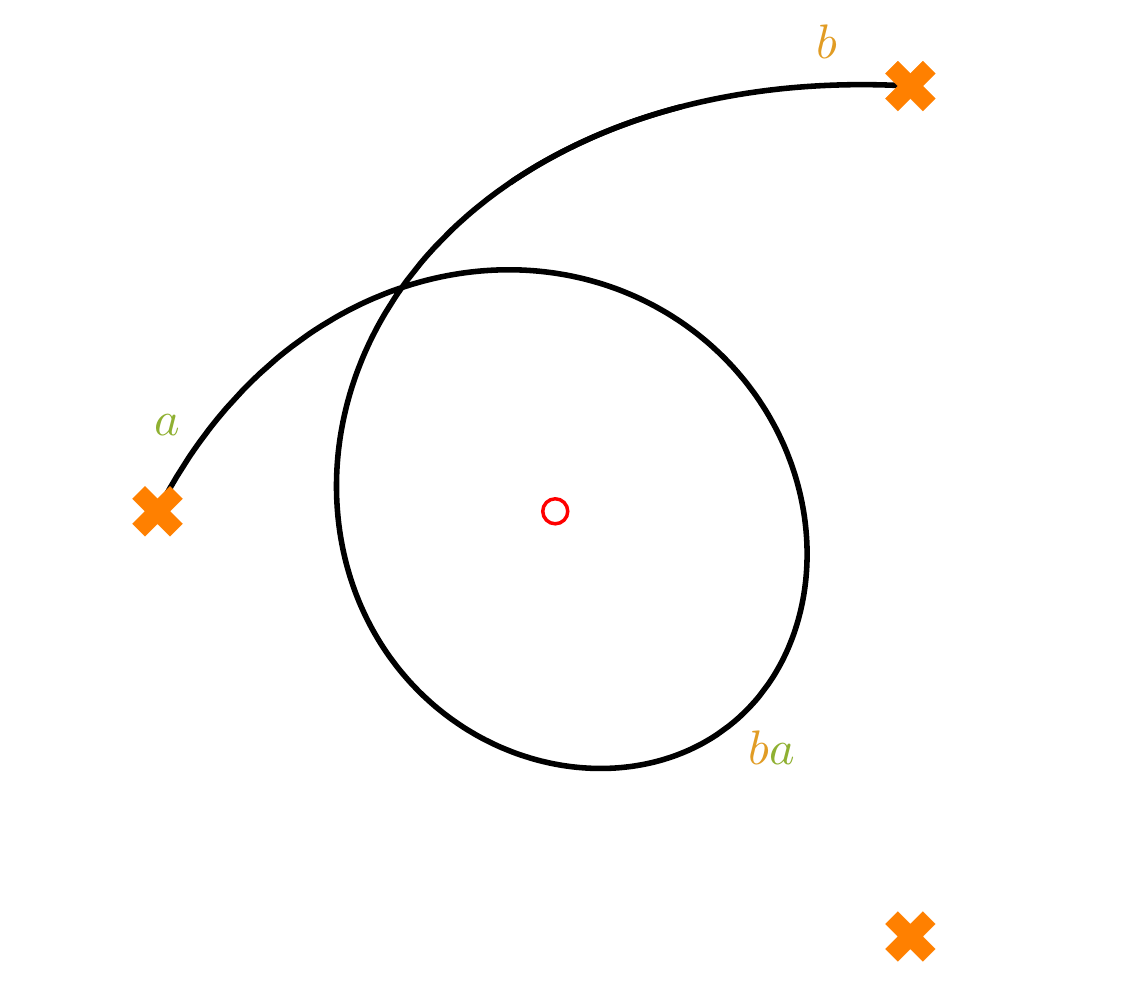}}
    \caption{Detach from bottom branch point.}
    \label{fig:f2}
  \end{subfigure}
  \caption{Three torus fixed points from $K_3(1,1)$}
\end{figure}

We start by considering bound states corresponding to representations with dimension 
vector $(1,1)$. The moduli space of representations is $K_3(1,1) \cong 
\{\C^3-0\}/\C^\times\cong\CP^2$.  This moduli space has three torus fixed points and 
accordingly the Euler characteristic is three.
For each of the three torus fixed points there is a corresponding finite web. The finite 
webs are shown in figures \ref{fig:f1} and \ref{fig:f2}. The finite web shown in Figure 
\ref{fig:f1}, resolves the intersection in the top center by either a $(++)$ or a $(--)$ 
strand.  The two possible resolutions correspond to two distinct finite webs 
contributing two to the Euler characteristic.  Depending on the choice of resolution 
a VEV is given to the `$+$'- or `$-$'-arrow in the Kronecker-3 quiver corresponding to 
the $(++)$ or $(--)$ strand. The second finite web, Figure \ref{fig:f2}, detaches a 
strand from the bottom intersection point, giving a VEV to the third B-arrow in the 
Kronecker quiver. Note that there is a continuous family of finite webs interpolating 
between the three distinguished members: it is possible to gradually shorten the $(++)$ 
link to zero size, simultaneously detaching the strands at the bottom branch point. The 
link can then be regrown with a $(--)$ strand. This does not quite match the $\CP^2$ 
moduli space, though it is reminiscent of its toric diagram. 

\begin{figure}[htb]
\centering
  \begin{subfigure}[b]{0.45\textwidth}
    \includegraphics[width=\textwidth]{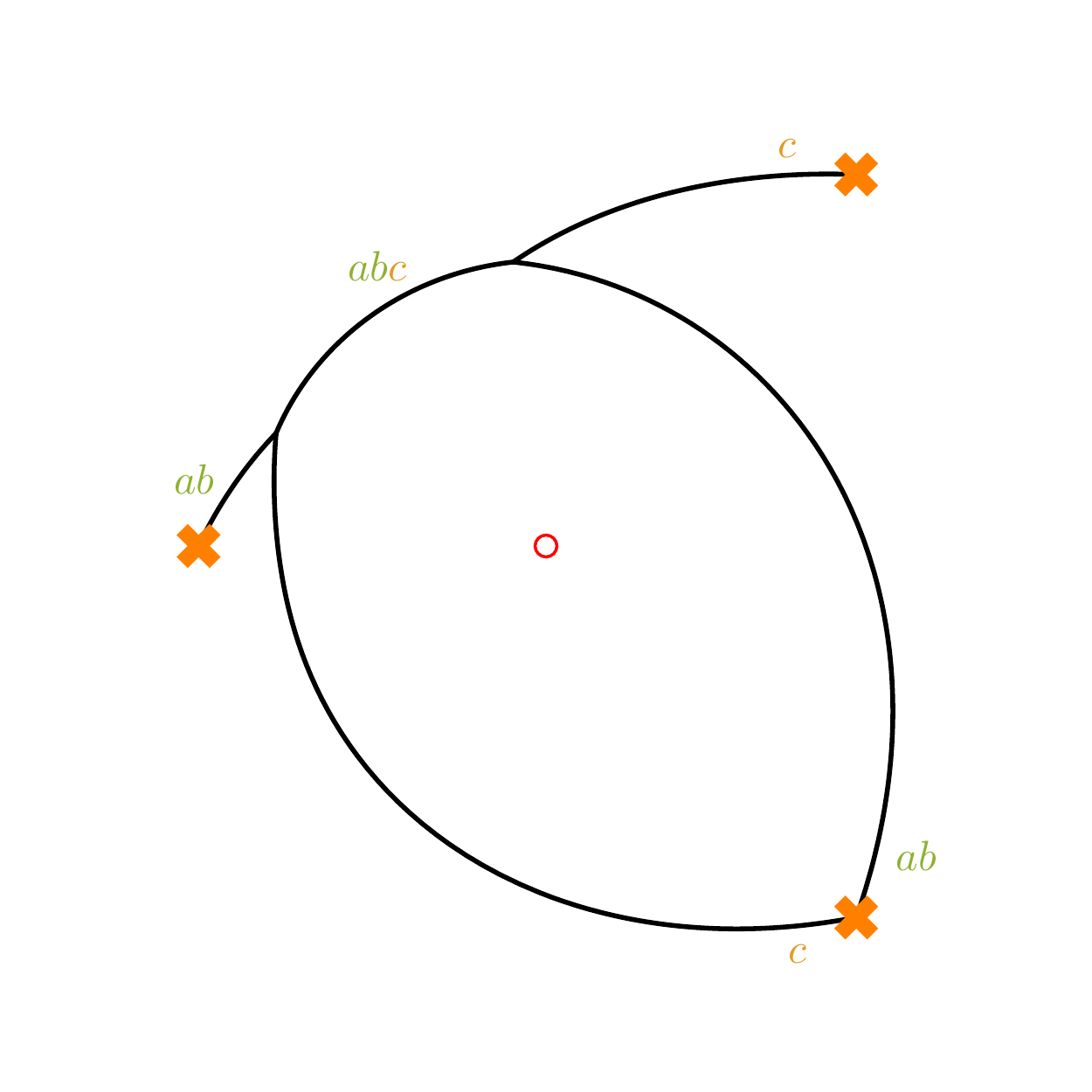}
    \caption{Resolve by a 2:1 strand.}
    \label{fig:ff1}
  \end{subfigure}
  \hfill
  \begin{subfigure}[b]{0.45\textwidth}
    \includegraphics[width=\textwidth]{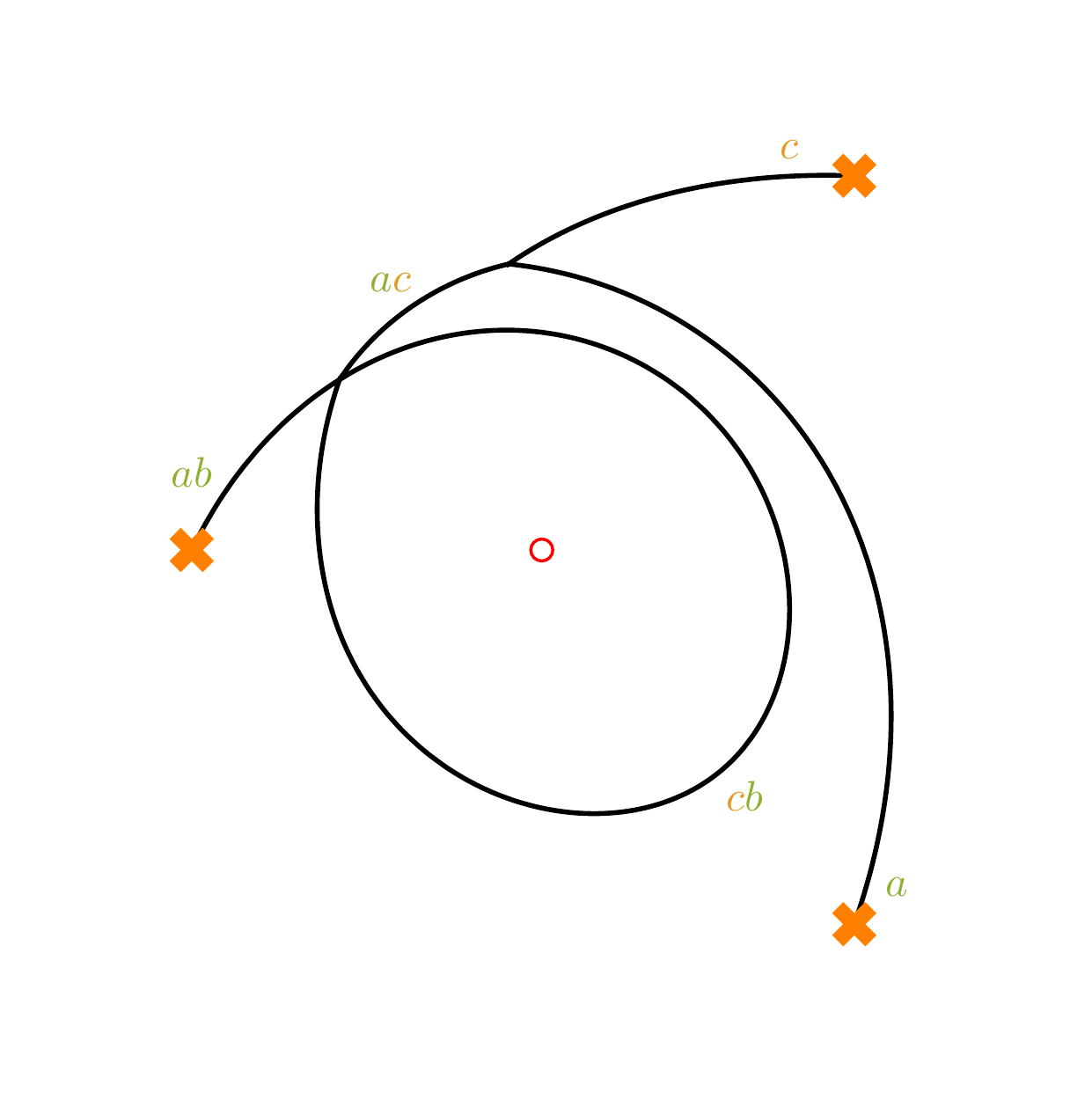}
    \caption{Resolve by a ($++$) or ($--$) strand and detach from bottom branch point.}
    \label{fig:ff2}
  \end{subfigure}
  \caption{Three torus fixed points from $K_3(1,2)$.}
\end{figure}

The moduli space for the dimension vector $(1,2)$ is again $\CP^2$. We again find 
three finite webs corresponding to the three torus fixed points.  The first finite 
web shown in Figure \ref{fig:ff1} resolves the top intersection by a $(+-)_1$ strand.  
The corresponding ``abelianized''
quiver representation is shown in Figure \ref{fig:k12a}.  The other 
two finite webs shown in Figure \ref{fig:ff2} resolve the top intersection point by 
a $(++)$ or $(--)$ strand. Note that of the two strands starting from the top left 
branch point in figure \ref{fig:ff2}, only one of them goes around the loop, and 
collides with the one that didn't get to make a $(++)$ strand offspring.  The quiver 
representation corresponding to resolving by a $(++)$ strand is shown in Figure 
\ref{fig:k12b}.
Less obvious than in the previous case, there is also a family of finite webs 
interpolating the two pictures, obtained by resolving the 4-way junction in figure 
\ref{fig:ff2}.

\begin{figure}[tbp]
\centering
  \begin{tikzpicture}
	\node[Mgreen] (pm) at (30:2cm) {$\C_b$};
	\node[Mgreen] (pp) at  (150:2cm) {$\C_a$};
	\node[Morange] (z) at (0,0) {$\C_c$};
	\draw[->, very thick] (z) -- node[above] {$-$} (pm);
	\draw[->, very thick] (z) -- node[above] {$+$}  (pp);
	\end{tikzpicture}

\caption{Representation for resolving by a 2:1 strand.}
\label{fig:k12a}
\bigskip
    \begin{tikzpicture}

	\node[Mgreen] (pp) at  (150:2cm) {$\C_a$};
	\node[Mgreen] (pb) at  (-90:2cm) {$\C_b$};
	\node[Morange] (z) at (0,0) {$\C_c$};

	\draw[->, very thick] (z) -- node[above] {$+$}  (pp);
	\draw[->, very thick] (z) -- node[right] {B} (pb);
	\end{tikzpicture}
\caption{Representation for resolving by a 1:1 strand ($++$ according to the direction 
of forking).}
\label{fig:k12b}
\end{figure}

\paragraph{From tree modules to networks}
We now explain how to obtain representations of the Kronecker-3 quiver from exponential 
networks.  In section 
\ref{quiverReps} we described a special family of quiver representations for dimension 
vector $(F_n, F_{n+2})$ of the Kronecker-3 quiver in terms of tree modules, where $F_n$ 
is the $n$-th Fibonacci number. We now wish to exhibit exponential networks 
corresponding to these representations.

\begin{figure}[t]
\begin{subfigure}[b]{0.45\textwidth}
\includegraphics[width=\textwidth]{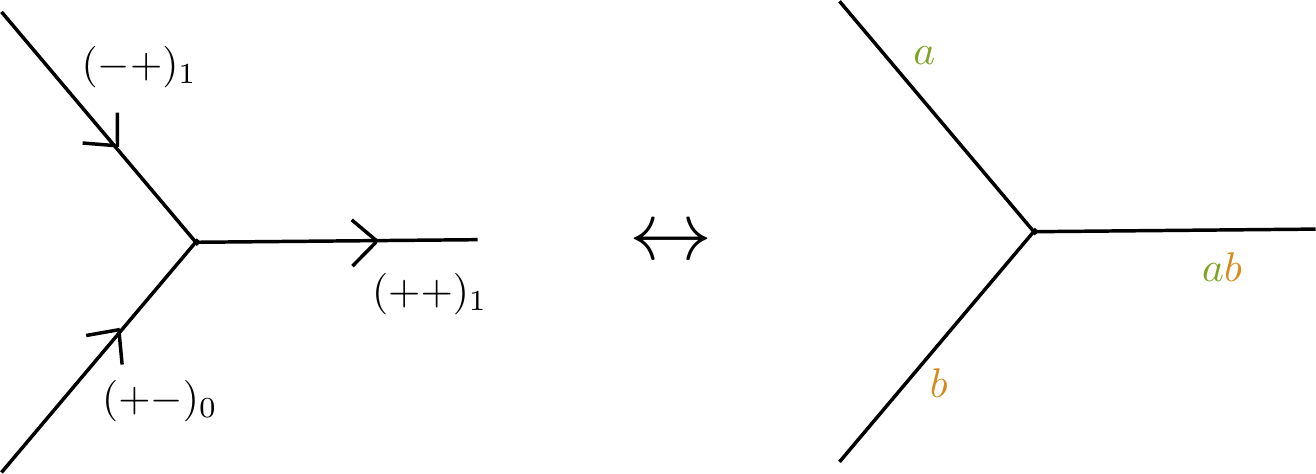}
\caption{Notations for 1:1 collision.}
\label{fig:conversion1}
\end{subfigure}
  \hfill
\begin{subfigure}[b]{0.45\textwidth}
\includegraphics[width=\textwidth]{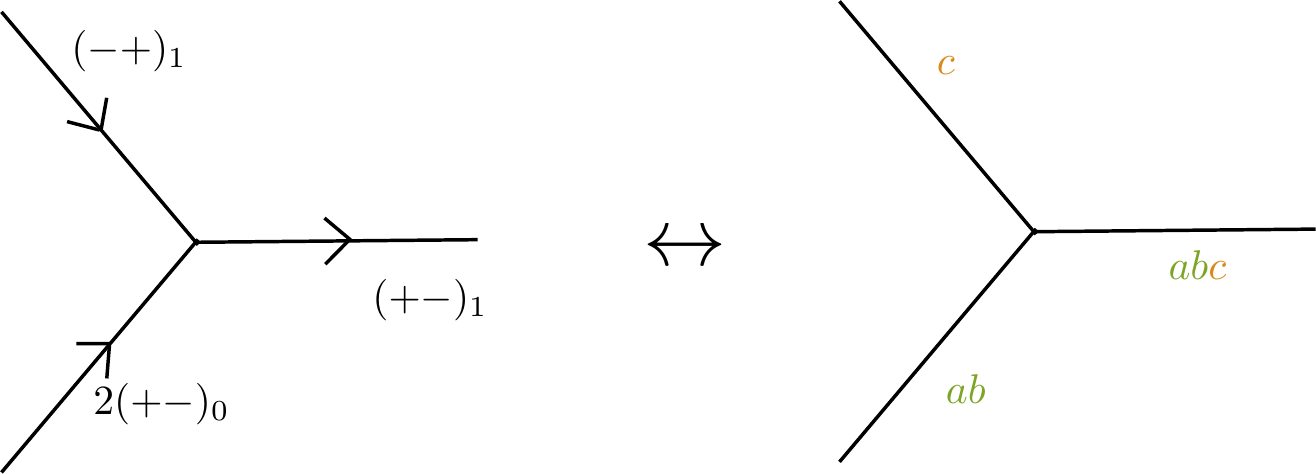}
\caption{Notations for 2:1 collision.}
\label{fig:conversion2}
\end{subfigure}
\caption{Converting between $(ij)_n$ and tree module notation}
\end{figure}

\begin{figure}[htbp]
\centering
\includegraphics[width=\textwidth]{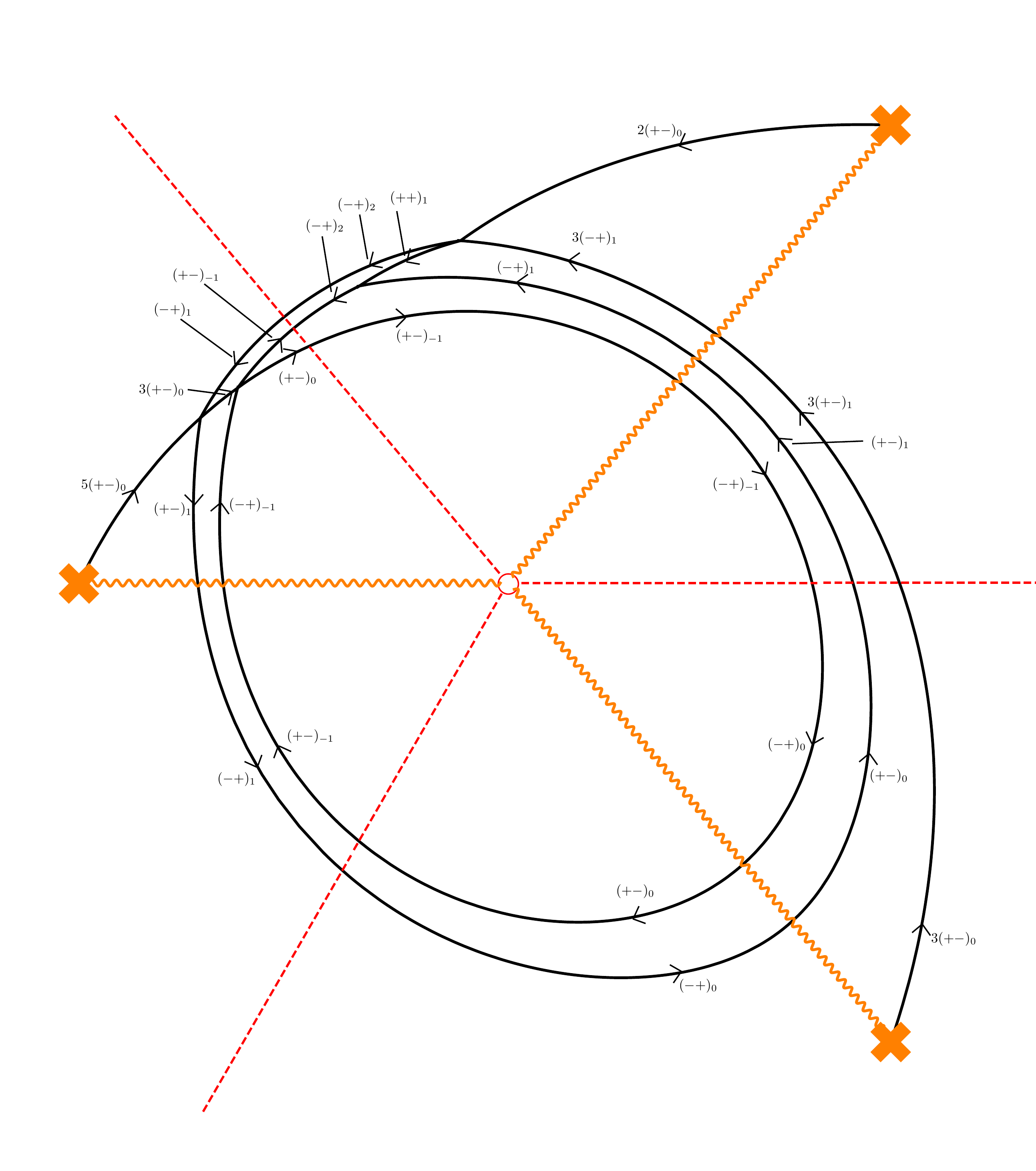}
\caption{$(ij)_n$ notation illustrated on the finite web for the (2,5) representation 
of the Kronecker-3 quiver.} 
\label{fig:sample25}
\end{figure}

\begin{figure}[htbp]
\centering
\includegraphics[width=\textwidth]{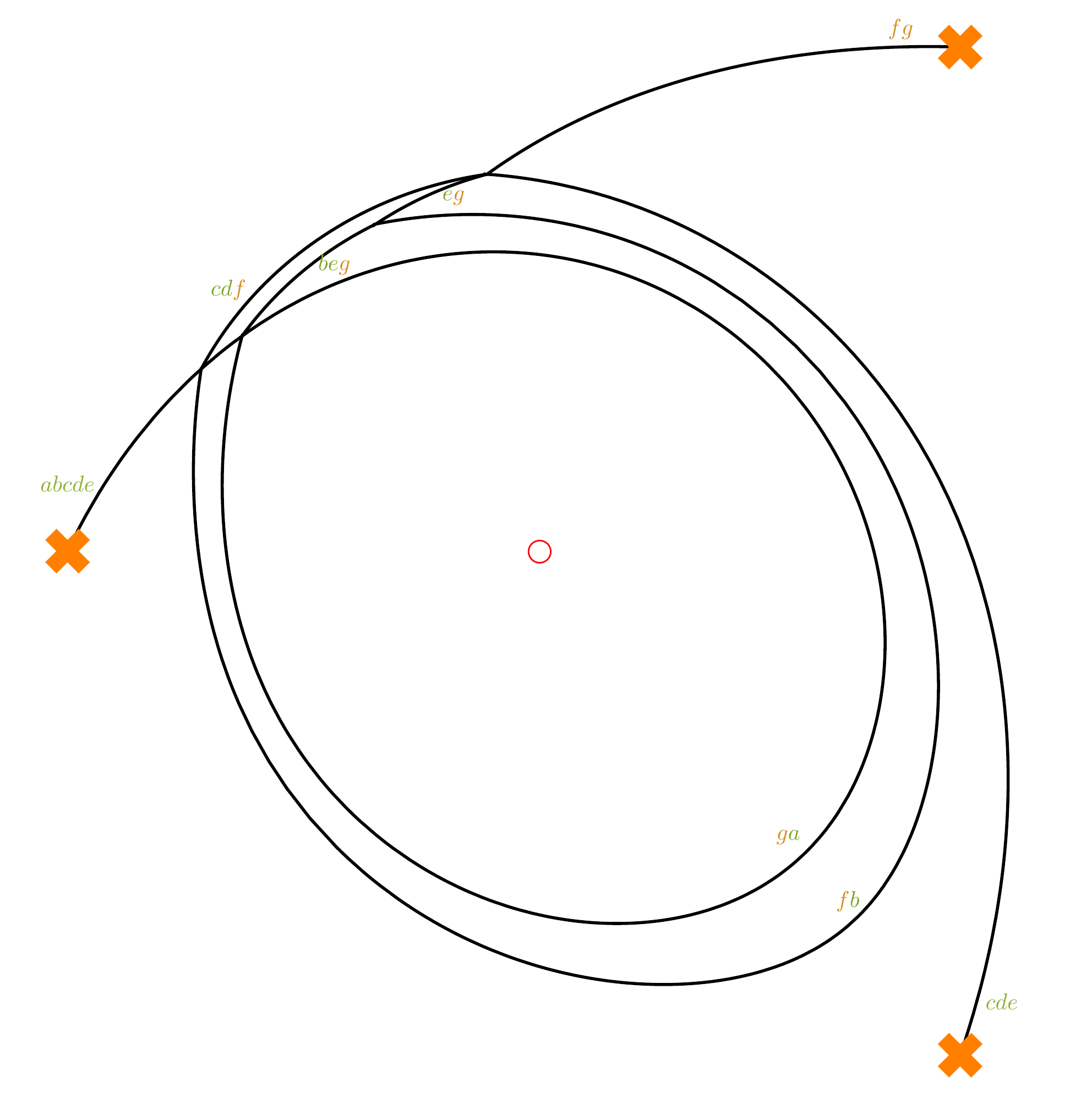}
\caption{Tree module notation illustrated on the finite web for the (2,5) representation 
of the Kronecker-3 quiver.} 
\label{fig:samplealternative}
\end{figure}

To simplify the translation, and reduce the clutter, 
we will use an alternative notation that first appeared in Figure \ref{fig:boundq}.
In the new notation, there is one label for each copy of the original basic states
that make up the bound state. Every label follows a strand that starts and ends 
where the corresponding basic state did, possibly traveling through resolutions 
and possibly ending detached from the branch point. Note that the number of labels 
on a strand is not necessarily equal to its multiplicity. Rather, it can be recovered 
from the conversions shown in Figures \ref{fig:conversion1} and \ref{fig:conversion2}. 
We will from now on refer to strands born according to reactions \ref{eq:reaction1} 
and \ref{eq:reaction2} as $k:k$ and $k+1:k$ strands respectively.  
The conversion rules might be best illustrated on a sample network of sufficient
complexity. To this end, we show the fully labelled $(2,5)$ representation
of the Kronecker-3 quiver realized on local $\CP^2$, in the ``old'' 
$(ij)_n$-notation in Fig.\ \ref{fig:sample25}, and in the new notation in Fig.\ 
\ref{fig:samplealternative}.

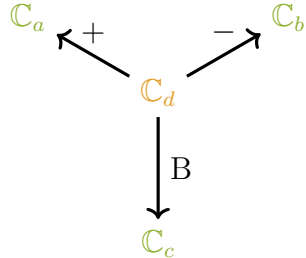
\begin{figure}[htbp]
\centering
\begin{tikzpicture}
\node[Mgreen] (pm) at (30:2cm) {$\C_b$};
\node[Mgreen] (pp) at  (150:2cm) {$\C_a$};
\node[Mgreen] (pb) at  (-90:2cm) {$\C_c$};
\node[Morange] (z) at (0,0) {$\C_d$};
\draw[->, very thick] (z) -- node[above] {$-$} (pm);
\draw[->, very thick] (z) -- node[above] {$+$}  (pp);
\draw[->, very thick] (z) -- node[right] {B} (pb);
\end{tikzpicture}
\caption{Conventions for the the covering quiver.  Also, the tree module for dimension vector $(1,3).$}
\label{fig:k13}
\end{figure}

Tree modules can be described in terms of a covering quiver of the original quiver 
representation.  They bear a striking resemblance to quantum Hall halos \cite{Denef}.  
In the covering quiver of the Kronecker-3 quiver we will always orient the arrows such 
that $B$ is in the vertical direction and `$+$'/`$-$' are at 120 degrees to B.  
These conventions are easily illustrated in Figure \ref{fig:k13}.
Giving a covering quiver representation, we slice it horizontally into string modules 
by forgetting the vertical `B' arrows.  The resulting collection of quivers will 
correspond to representations of the Kronecker-2 quiver with the two arrows corresponding 
to  `$+$' and `$-$'. Each time there is a $(k+1,k)$ representation of the Kronecker-2 
quiver with $k$ copies of $\C$ meeting $k+1$ copies of $\C$ we will associate a $k+1:k$ 
strand to the resolution of the corresponding intersection.  This translation agrees
with the representation that we gave in Figure \ref{fig:k12a} for the finite web
shown in Figure \ref{fig:ff1}.  In the odd Fibonacci case the covering quiver has a left-right asymmetry
and there will be additional $(++)$ or $(--)$ strands to resolve the ambiguities.

\begin{figure}[tbp]
\centering
\includegraphics[width=8cm]{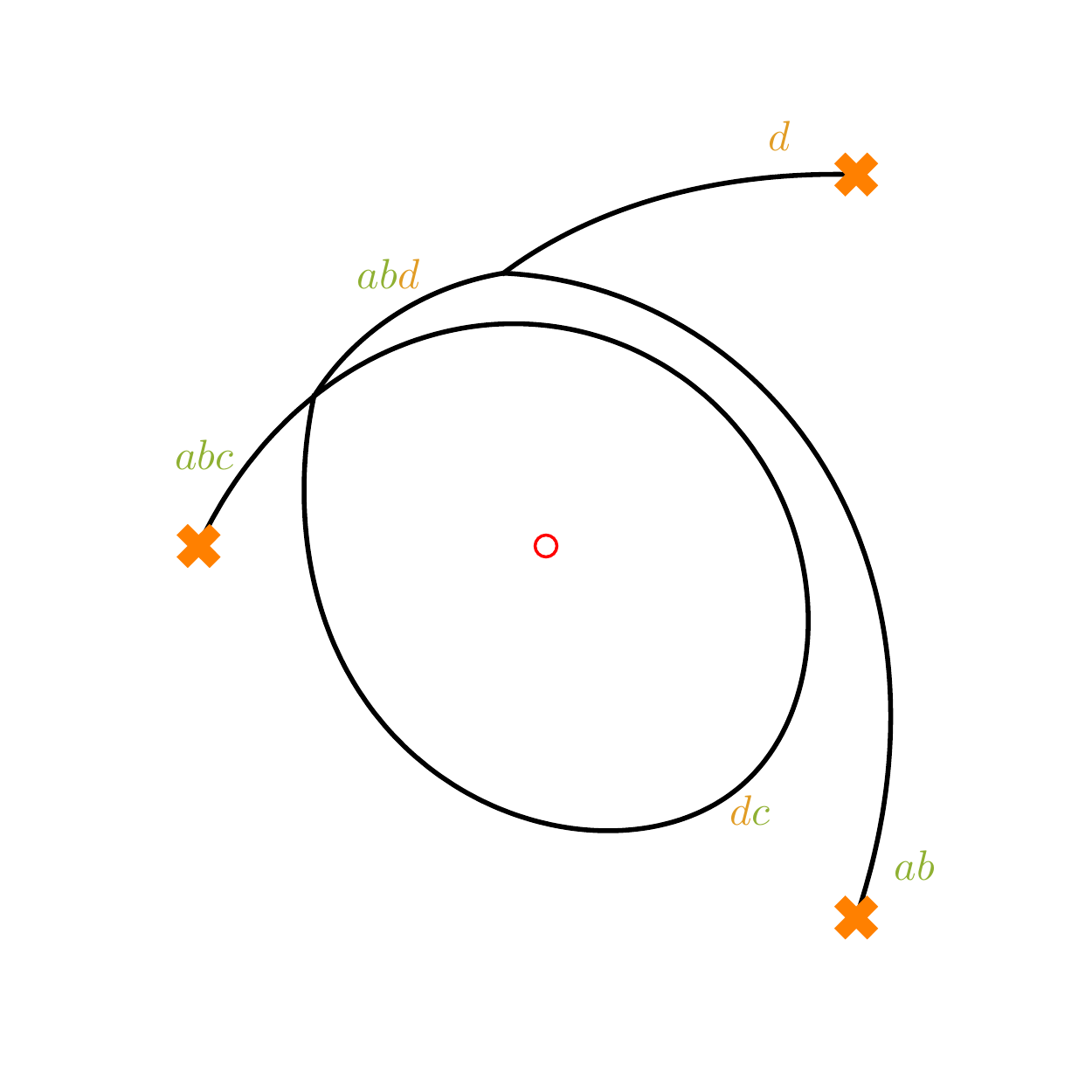}
\caption{The finite web for the unique (fixed) point in $K_3(1,3)$.}
\label{fig:013bound}
\end{figure}
The representation for dimension vector $(1,3)$ has $K_3(1,3) \cong pt$ as its moduli 
space. Therefore there is a single torus fixed point which is represented by a single 
finite web shown in Figure \ref{fig:013bound}.  There is a single 2:1 strand resolving 
the top intersection point and one of the strands detaches from the bottom branch point. 
The corresponding quiver representation appears in equation \eqref{13rep} and
Figure \ref{fig:k13} illustrates the rules for converting between a finite web and its
associated tree module.

\begin{figure}[tbp]
\centering
\begin{tikzpicture}
\node[Morange] (pm) at (30:2cm) {$\C_g$};
\node[Morange] (pp) at  (150:2cm) {$\C_f$};
\node[Mgreen] (pmb) at ($(pm) +(90:2cm)$) {$\C_e$};
\node[Mgreen] (pmp) at ($(pm) +(-30:2cm)$) {$\C_c$};
\node[Mgreen] (ppb) at ($(pp) +(90:2cm)$) {$\C_d$};
\node[Mgreen] (ppm) at ($(pp) +(210:2cm)$) {$\C_a$};
\node[Mgreen] (z) at (0,0) {$\C_b$};
\draw[<-, very thick] (z) -- node[above] {$-$} (pm);
\draw[<-, very thick] (z) -- node[above] {$+$}  (pp);
\draw[->, very thick] (pm) -- node[right] {$B$} (pmb);
\draw[->, very thick] (pm) -- node[above] {$+$} (pmp);
\draw[->, very thick] (pp) -- node[right] {$B$} (ppb);
\draw[->, very thick] (pp) -- node[above] {$-$} (ppm);

\draw[Mblue, thick] (5,0.4) 
             arc [start angle=0,   end angle=360,
                  x radius=5cm, y radius=1cm]
                  node [above right, pos=.10] {$(+-)_2$} ;
\end{tikzpicture}
\caption{A tree module for a dimension $(2,5)$ representation.}
\label{fig:kcover25a}

\includegraphics[width=8cm]{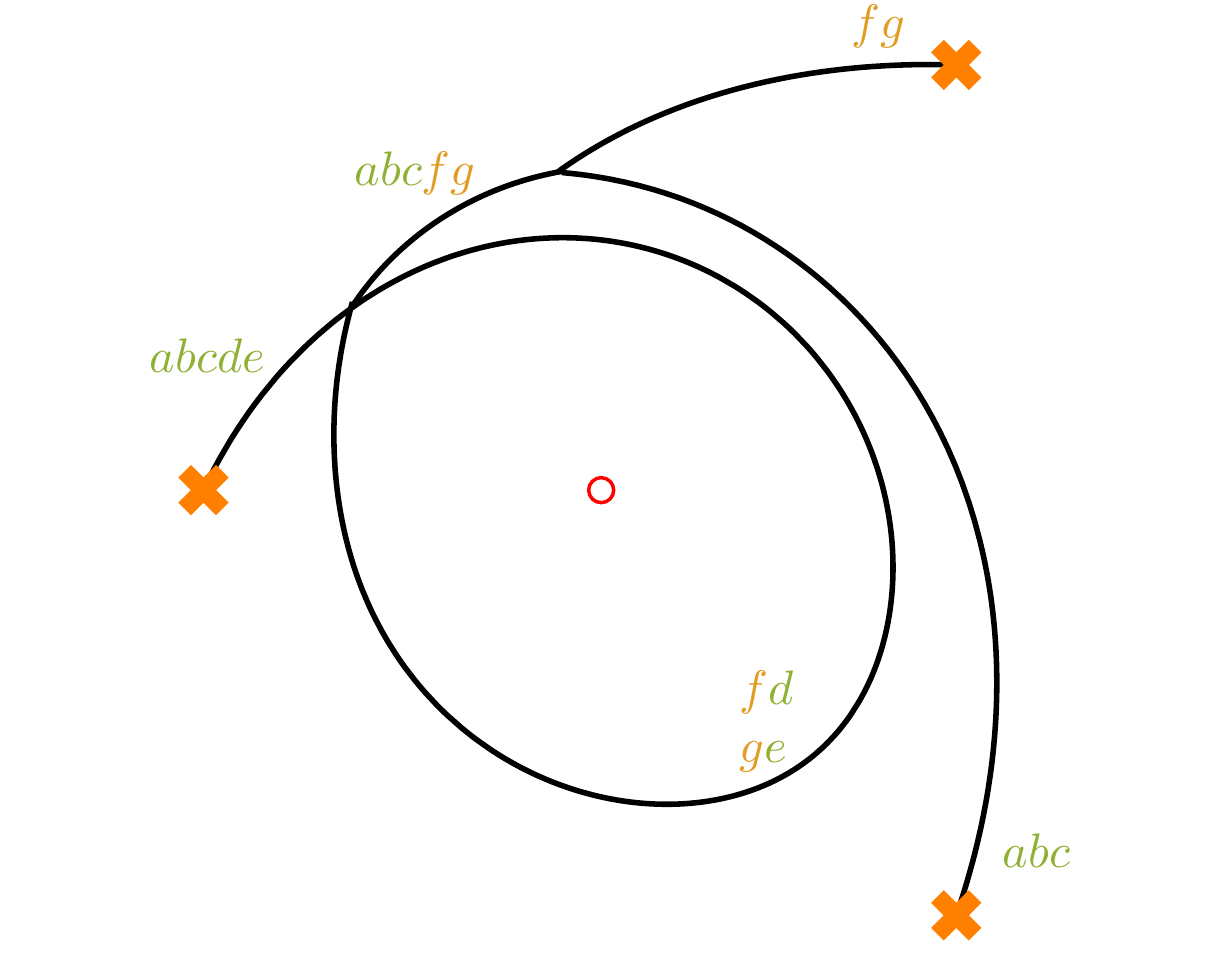}
\caption{A finite web for dimension $(2,5)$ representation.}
\label{fig:025a}

\begin{tikzpicture}
\node[Morange] (pp) at  (150:2cm) {$\C_g$};
\node[Morange] (pb) at  (-90:2cm) {$\C_f$};
\node[Mgreen] (ppb) at ($(pp) +(90:2cm)$) {$\C_a$};
\node[Mgreen] (ppm) at ($(pp) +(210:2cm)$) {$\C_e$};
\node[Mgreen] (pbp) at ($(pb) +(-30:2cm)$) {$\C_d$};
\node[Mgreen] (pbm) at ($(pb) +(-150:2cm)$) {$\C_c$};
\node[Mgreen] (z) at (0,0) {$\C_b$};
\draw[<-, very thick] (z) -- node[above] {$+$}  (pp);
\draw[<-, very thick] (z) -- node[right] {B} (pb);
\draw[->, very thick] (pp) -- node[right] {$B$} (ppb);
\draw[->, very thick] (pp) -- node[above] {$-$} (ppm);
\draw[->, very thick] (pb) -- node[above] {$+$} (pbp);
\draw[->, very thick] (pb) -- node[above] {$-$} (pbm);

\draw[Mblue, thick] (3,-2.6) 
             arc [start angle=0,   end angle=360,
                  x radius=3cm, y radius=.85cm]
                  node [above right, pos=.10] {$(+-)_1$} ;
                  
\draw[Mblue, thick] (1.2,0.43) 
          arc [start angle=0,   end angle=360,
                 x radius=3cm, y radius=0.85cm]
                 node [above right, pos=.10] {$(+-)_1$} ;

\end{tikzpicture}
\caption{A different tree module for a dimension $(2,5)$ representation.}
\label{fig:kcover25b}
\end{figure}

\paragraph{More exotic Fibonacci representations}
We now turn to the $(2,5)$ and $(3,8)$ Fibonacci representations.
There are three families of tree modules corresponding to the three torus fixed points 
for the representation vector $(2,5)$.  The representation shown in Figure 
\ref{fig:kcover25a} corresponds to a finite web with a resolution by a 3:2 strand and 
two strands detaching from the bottom branch point.  There indeed is such a finite 
web and it is shown in Figure \ref{fig:025a}.

There is a second type of representation shown in Figure \ref{fig:kcover25b}.  The 
corresponding finite web is show in Figure \ref{fig:samplealternative}
(as well as in Fig.\ \ref{fig:sample25} in the ``old'' notation).  An interesting feature is the $(++)$ or $(--)$ strand that determines if the arrow between the nodes `e' and `g' is a '+' or a '-'. 

\begin{figure}[tbp]
\includegraphics[width=\textwidth]{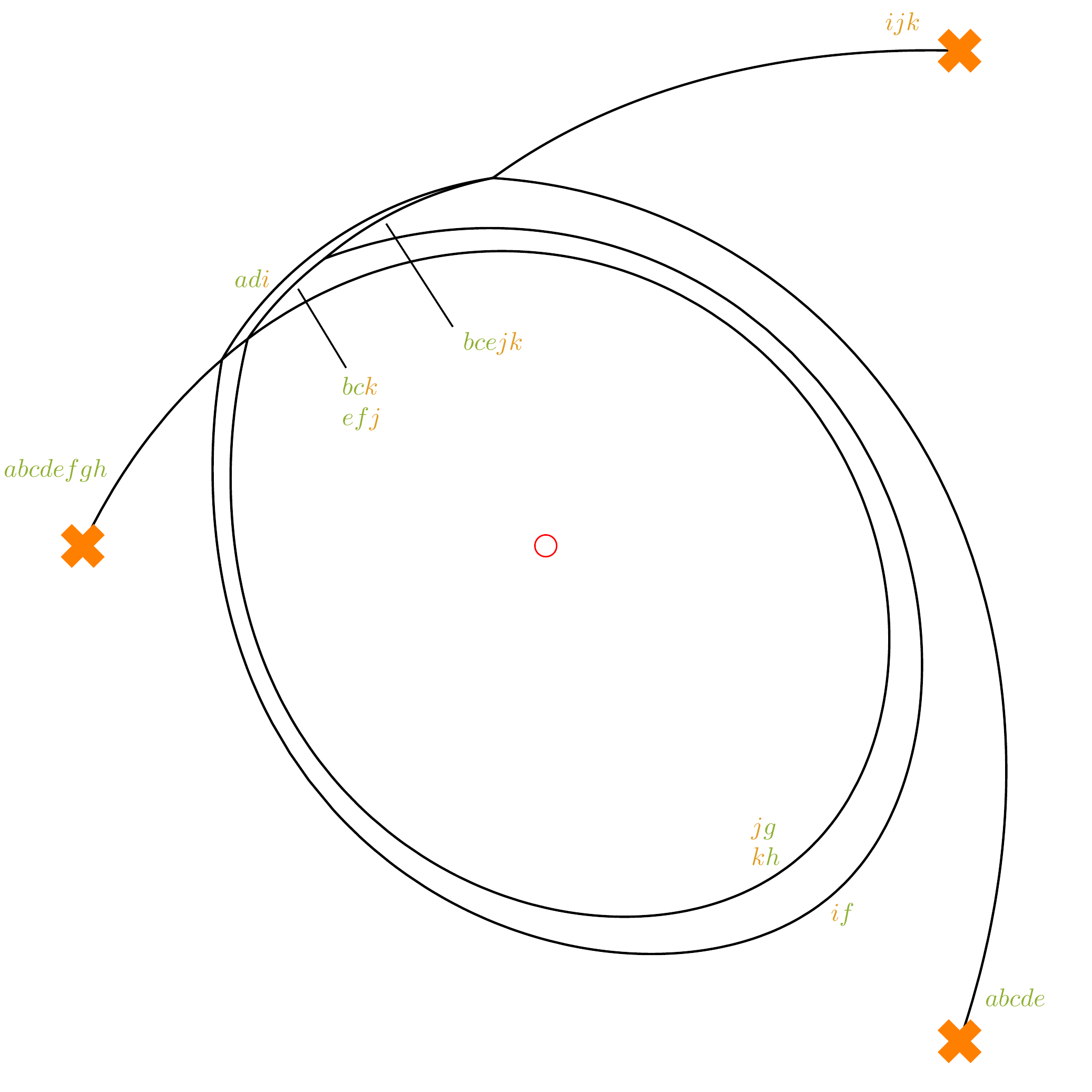}
\caption{The finite web for the dimension $(3,8)$ representation.}
\label{fig:038}
\end{figure}

Finally we consider the $(3,8)$ representation shown in Figure \ref{fig:kcover38}.  
The moduli space is again a single point and there is one finite web.  The network 
is shown in Figure \ref{fig:038}.

\begin{figure}[htbp]
\centering
\begin{tikzpicture}
\node[Morange] (pm) at (30:2cm) {$\C_j$};
\node[Morange] (pp) at  (150:2cm) {$\C_k$};
\node[Morange] (pb) at  (-90:2cm) {$\C_i$};
\node[Mgreen] (pmb) at ($(pm) +(90:2cm)$) {$\C_g$};
\node[Mgreen] (pmp) at ($(pm) +(-30:2cm)$) {$\C_e$};
\node[Mgreen] (ppb) at ($(pp) +(90:2cm)$) {$\C_h$};
\node[Mgreen] (ppm) at ($(pp) +(210:2cm)$) {$\C_c$};
\node[Mgreen] (pbp) at ($(pb) +(-30:2cm)$) {$\C_d$};
\node[Mgreen] (pbm) at ($(pb) +(-150:2cm)$) {$\C_a$};
\node[Mgreen] (z) at (0,0) {$\C^2_{b,f}$};
\draw[<-, very thick] (z) -- node[above] {$-$} (pm);
\draw[<-, very thick] (z) -- node[above] {$+$}  (pp);
\draw[<-, very thick] (z) -- node[right] {B} (pb);
\draw[->, very thick] (pm) -- node[right] {$B$} (pmb);
\draw[->, very thick] (pm) -- node[above] {$+$} (pmp);
\draw[->, very thick] (pp) -- node[right] {$B$} (ppb);
\draw[->, very thick] (pp) -- node[above] {$-$} (ppm);
\draw[->, very thick] (pb) -- node[above] {$+$} (pbp);
\draw[->, very thick] (pb) -- node[above] {$-$} (pbm);
\draw[Mblue, thick] (5,0.5) 
             arc [start angle=0,   end angle=360,
                  x radius=5cm, y radius=1cm]
                  node [above right, pos=.10] {$(+-)_2$} ;
\draw[Mblue, thick] (3,-2.6) 
             arc [start angle=0,   end angle=360,
                  x radius=3cm, y radius=.85cm]
                  node [above right, pos=.10] {$(+-)_1$} ;
                  
\draw[Mblue, thick] (1.4,0.4) 
          arc [start angle=0,   end angle=360,
                 x radius=3cm, y radius=0.85cm]
                 node [above, pos=-.25] {$(+-)_1$} ;
\draw[Mblue, thick] (4.8,0.4) 
          arc [start angle=0,   end angle=360,
                 x radius=3cm, y radius=0.85cm]
                 node [above, pos=-.25] {$(+-)_1$} ;
\end{tikzpicture}
\caption{Covering quiver for the dimension $(3,8)$ representation.}
\label{fig:kcover38}
\end{figure}

\subsection{Large volume}
\begin{figure}[htbp]
\centering
\includegraphics[width=0.8\textwidth]{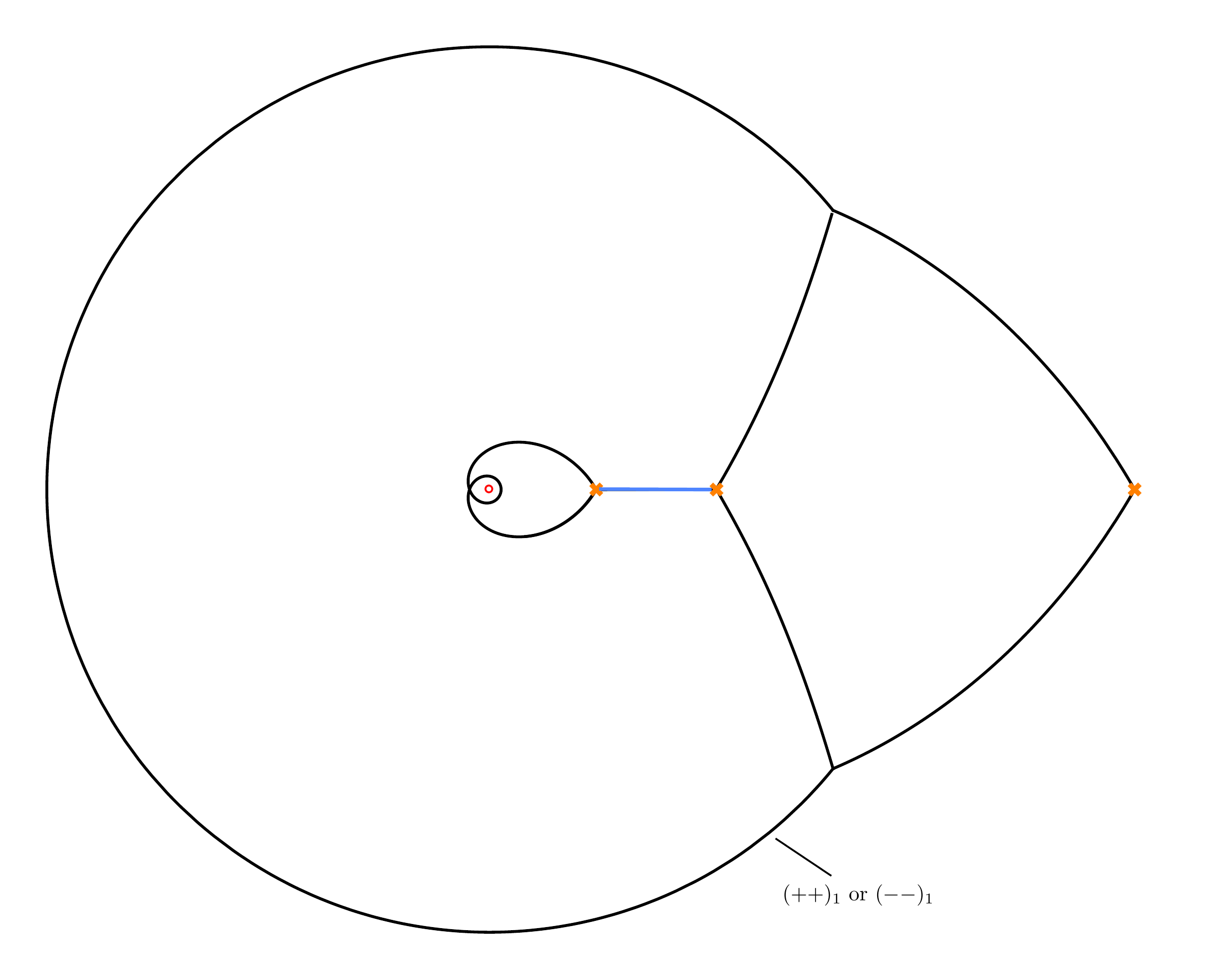}
\caption{Network at $\vartheta=0$ at a point in moduli space where the D4- and 
D0-brane coexist. The blue segment is the D4-brane. The inner loop, along with the outer finite web for either choice of $(++)$ or $(--)$, constitute the three fixed points of the D0-brane.}
\label{fig:P2LargeVolume}
\end{figure}

At large volume the brane charges are linear combinations of the D0, D2 and D4
brane charges.  The branes with compact support can be mathematically described as 
sheaves on $\CP^2$. 
The compact part of the moduli space of the D0 brane is $\CP^2$.  There are three 
finite webs corresponding to the fixed points. They appear in Figure 
\ref{fig:P2LargeVolume}; one is attached to the leftmost branch point, while the other 
two arise from the piece of network with a $(++)$ or $(--)$ strand.

The webs corresponding to the D0- and D4- branes are shown in Figure \ref{fig:P2LargeVolume}.  The figure is drawn at a point in moduli space where the central charges of the D0- and D4-branes align.  The D4-brane corresponds to the network consisting of a single strand connecting two branch points.  The D4-brane becomes massless at the conifold point, and grows to infinite size towards large volume.

\begin{figure}[tbp]
  \begin{subfigure}[b]{0.45\textwidth}
    \includegraphics[width=\textwidth]{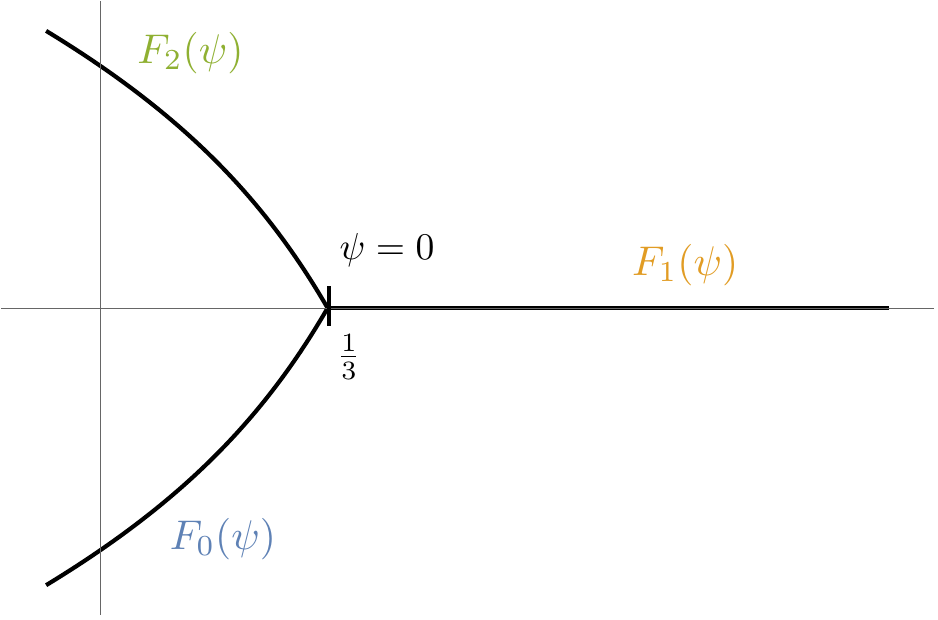}
    \caption{Parametric dependence of the periods along the negative $\psi$-axis. 
    }
    \label{fig:evolution}
  \end{subfigure}
  \hfill
  \begin{subfigure}[b]{0.45\textwidth}
    \includegraphics[width=\textwidth]{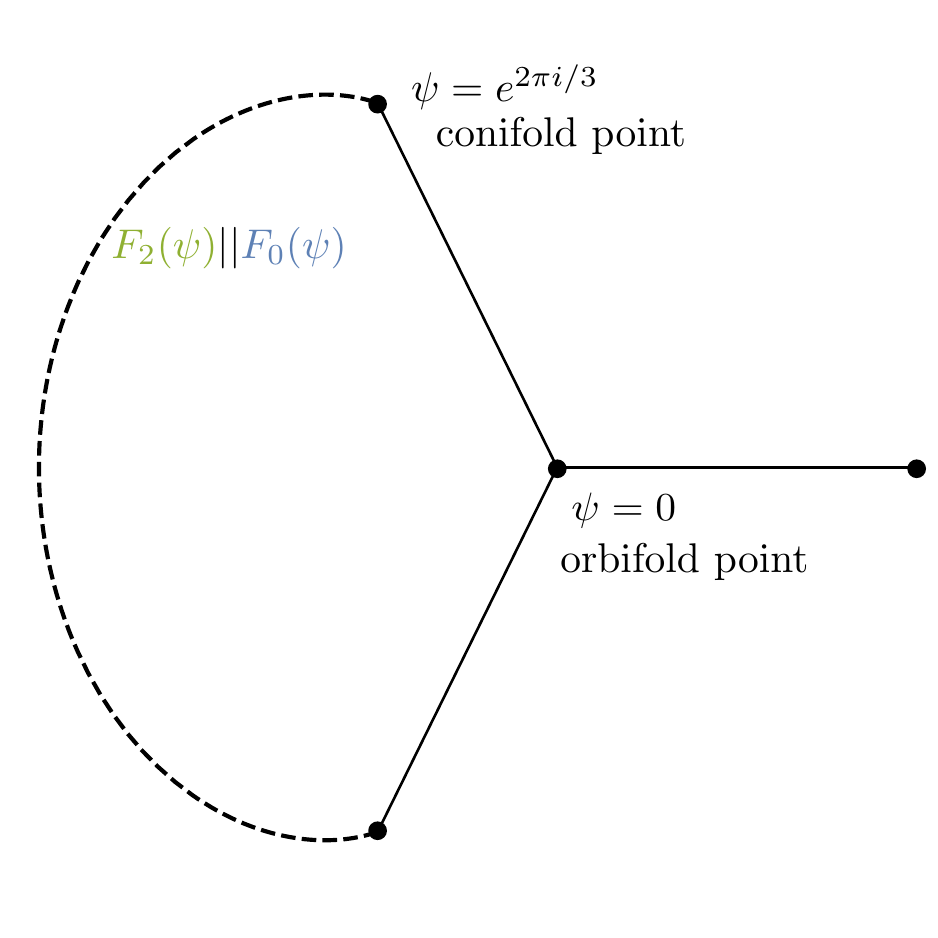}
    \caption{Wall of marginal stability}
    \label{fig:d2wall}
  \end{subfigure}
  \caption{Stability of the large volume D2-brane}
  \label{fig:d2stab}
\end{figure}

\begin{figure}[!htbp]
  \begin{subfigure}[b]{0.3\textwidth}
    \includegraphics[width=\textwidth]{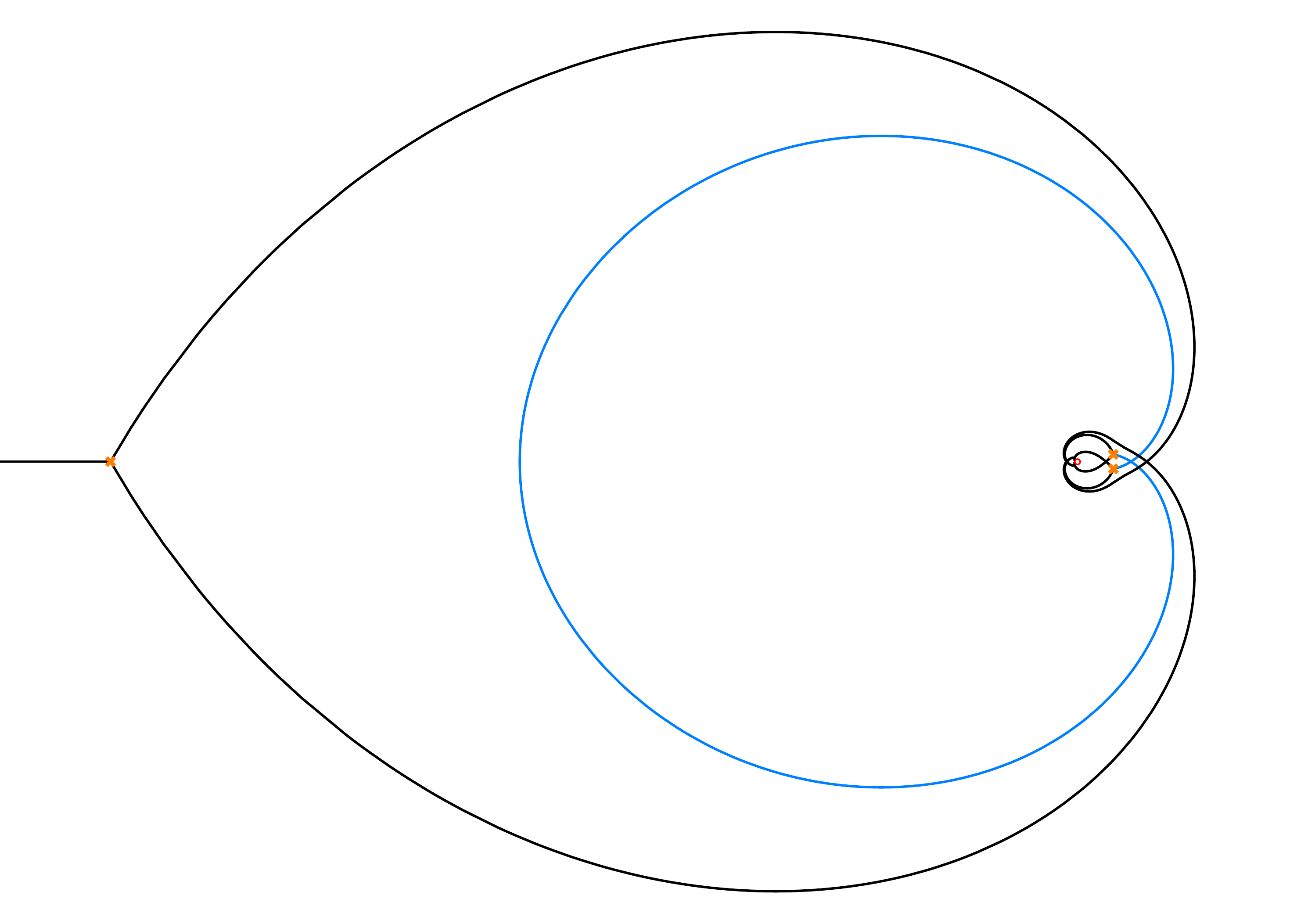}
    \caption{$\psi<\psi_{critical}$}
  \end{subfigure} 
  \hfill
  \begin{subfigure}[b]{0.3\textwidth}
    \includegraphics[width=\textwidth]{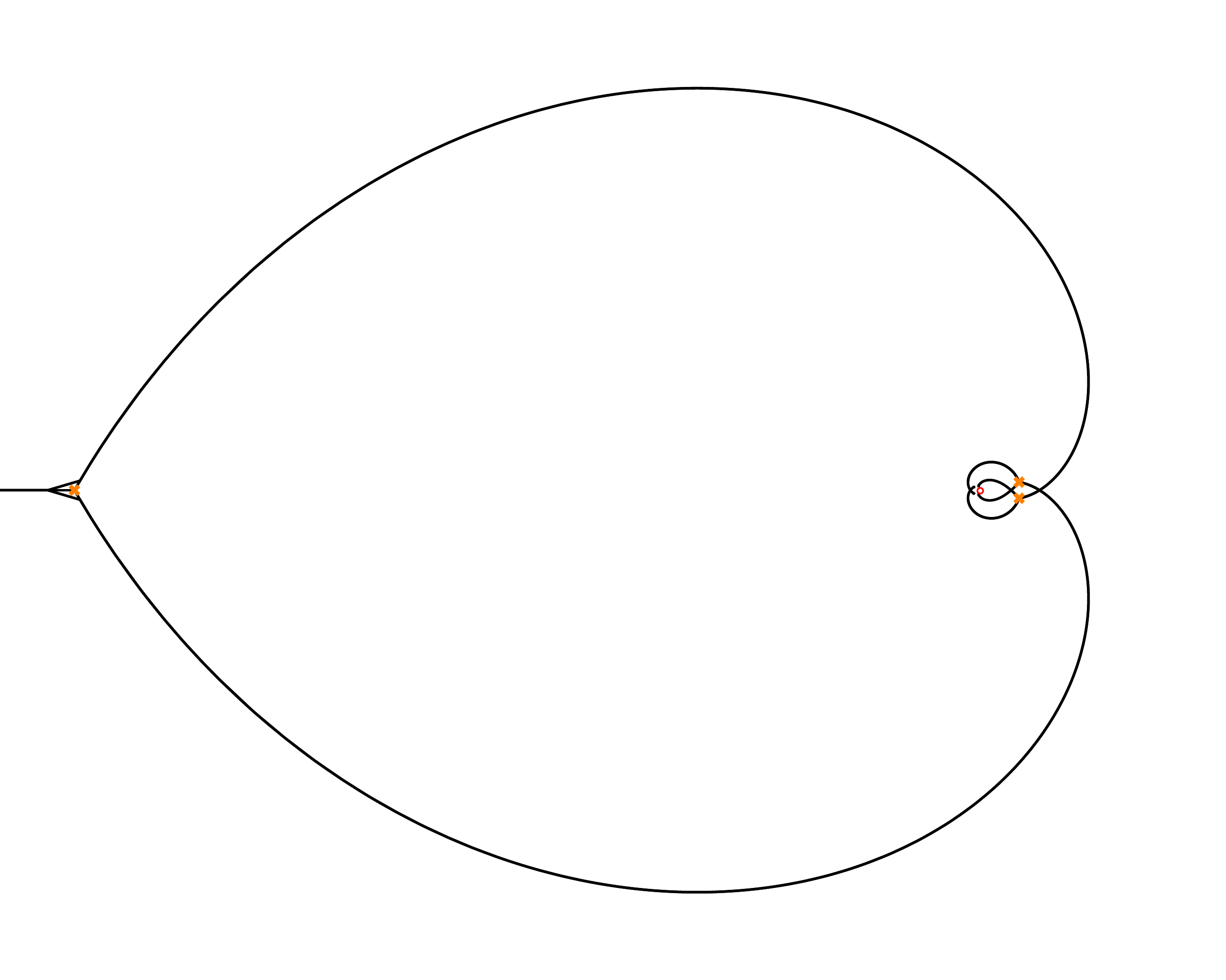}
    \caption{$\psi=\psi_{critical}$}
  \end{subfigure}
  \hfill
 \begin{subfigure}[b]{0.3\textwidth}
    \includegraphics[width=\textwidth]{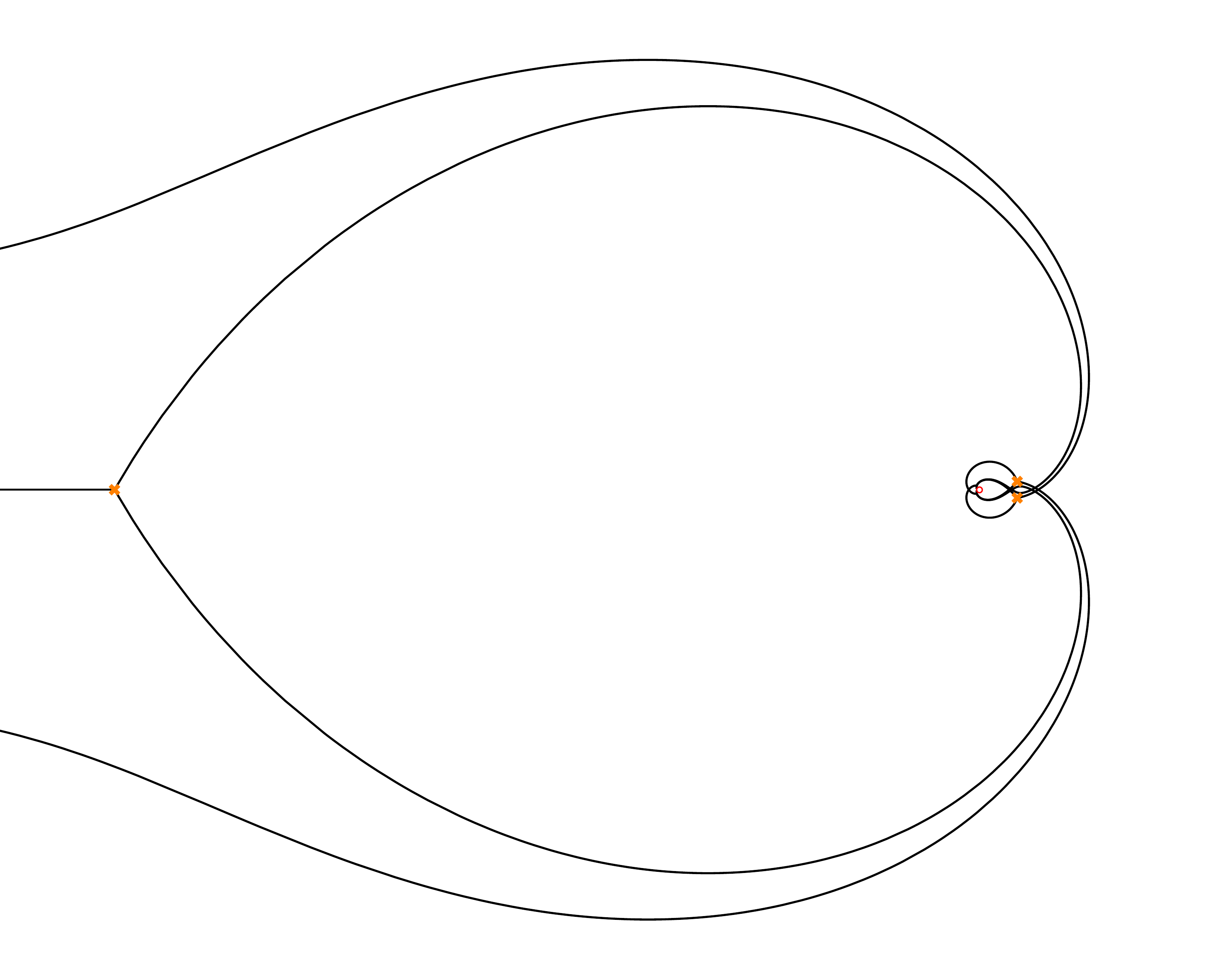}
    \caption{$\psi>\psi_{critical}$}
  \end{subfigure}
  \caption{Decay of the large volume D2-brane along the negative $\psi$-axis.}
  \label{fig:d2p2}
\end{figure}

Finally we consider a D2-brane brane near the large volume point. In the orbifold basis it has charge 
$(1,0,-1)$. From Figure \ref{fig:evolution}, we see that it becomes massless at the 
orbifold point. However the CFT is non-singular there so we expect that the D2-brane 
decays somewhere on the way from large volume. Figure \ref{fig:d2stab} also provides a 
natural suggestion for the location and mechanism of the decay, namely that the D2-brane 
decays to objects with charges $(1,0,0)$ and $(0,0,-1)$ on the locus where the periods 
$F_2$ and $F_0$ anti-align \cite{dfr}. We get a very nice visual corroboration of this fact by 
plotting the networks, as shown in Figure \ref{fig:d2p2}. 

A natural avenue for further study is transporting the Fibonacci representations from 
near the orbifold point to the large volume point.  These should correspond to the 
mirrors of Fibonacci bundles \cite{bram} on the mirror of local $\CP^2.$

\section{Flat Space}
\label{C3}

\subsection{Quiver}

\begin{figure}[bp]
\centering
\includegraphics[width=0.6\textwidth]{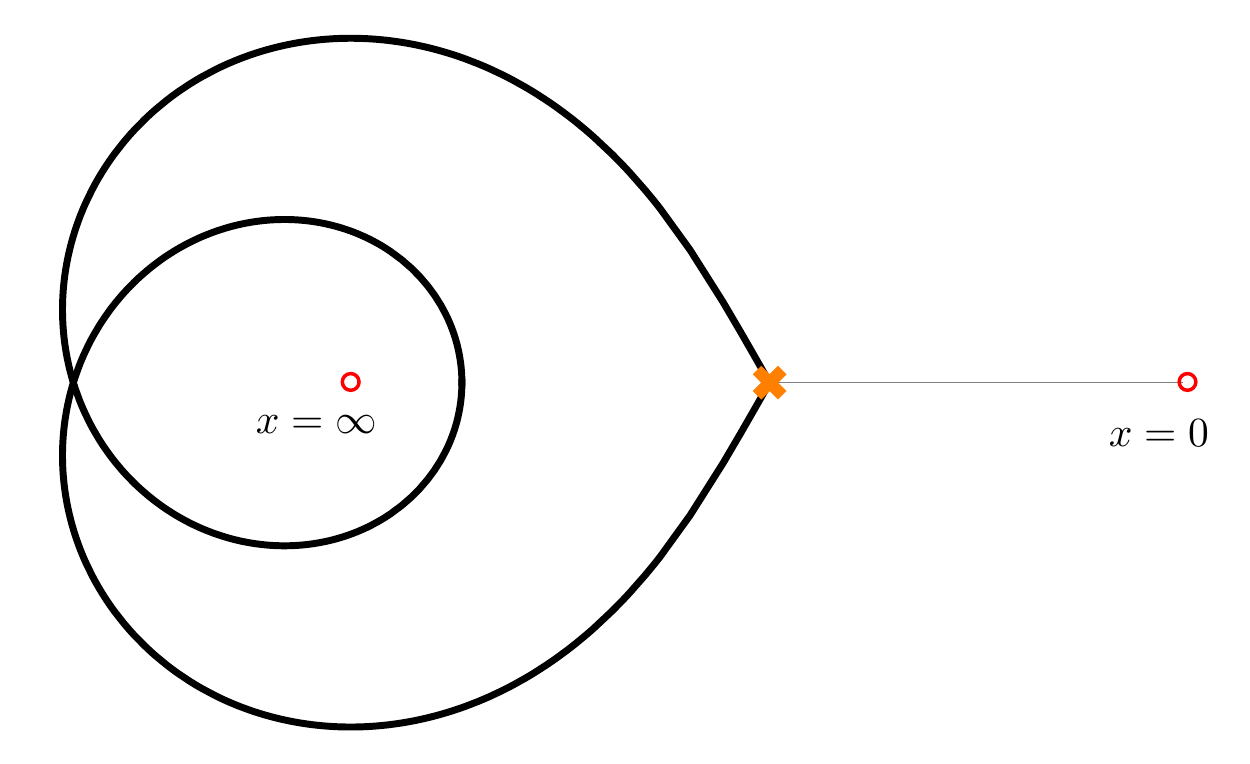}
\caption{Network at $\vartheta=0$ on $\complex^3$. Compare with Figure \ref{fig:P2LargeVolume}.}
\label{fig:c3phase0}
\end{figure}

\begin{figure}[tp]
\centering
\includegraphics[width=\textwidth]{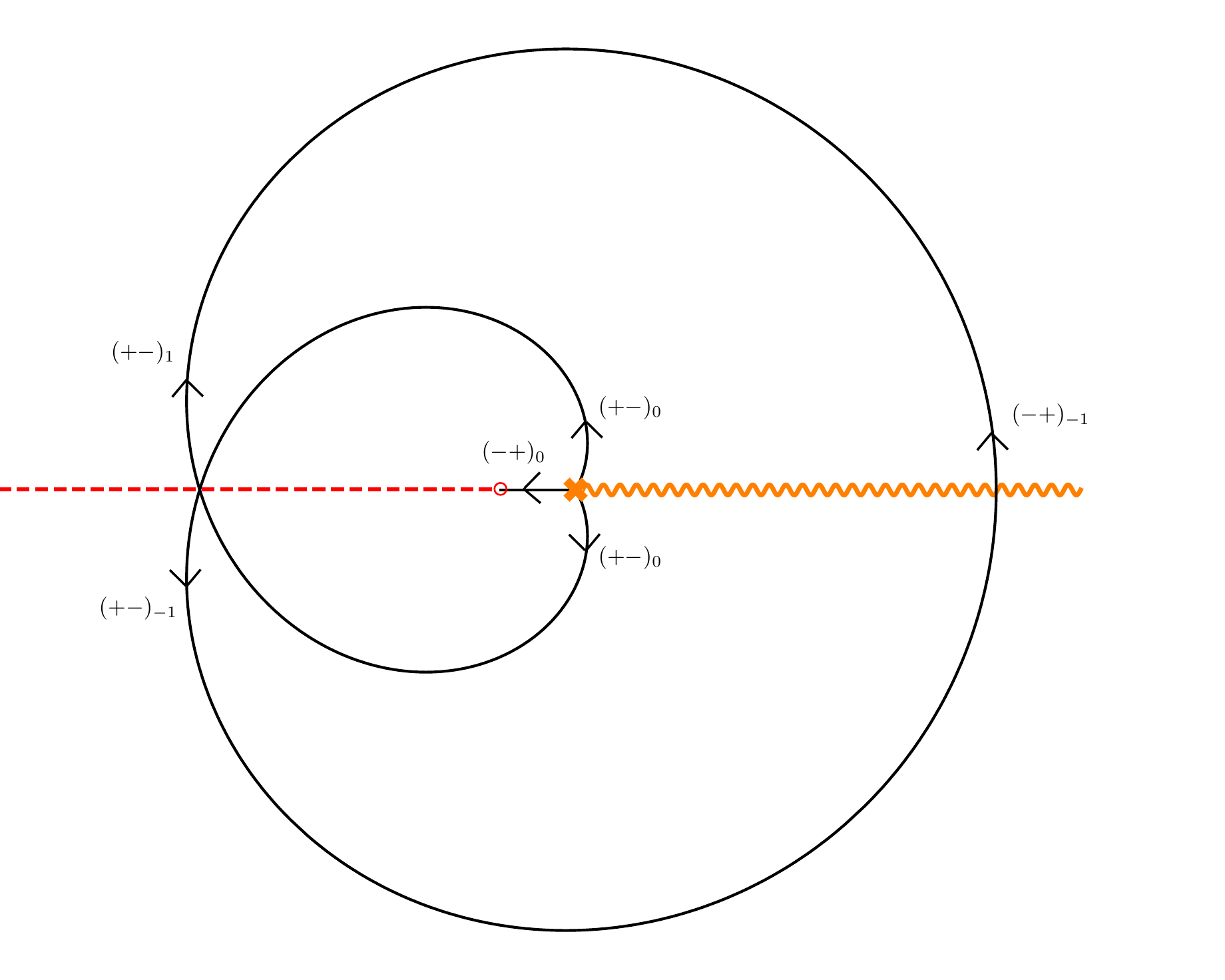}
\caption{Sample network for the $\C^3$ geometry.} 
\label{fig:samplec3}
\end{figure}

\begin{figure}[tbp]
\centering
\includegraphics[scale=1]{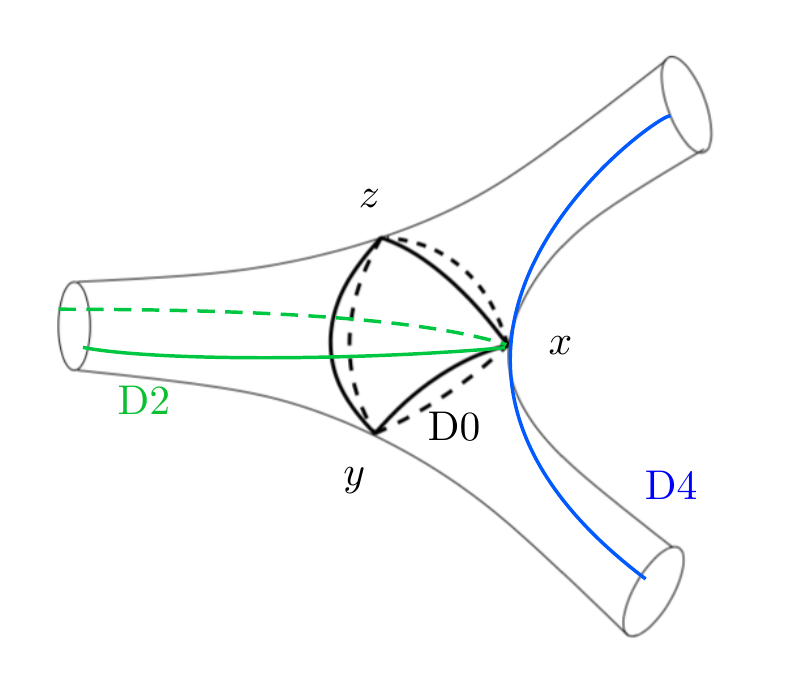}
\caption{Rendering of the branes on the mirror curve for $\complex^3$.}
\label{fig:d0upstairs}
\end{figure}

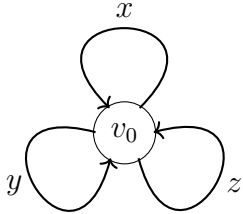
\begin{figure}[tbh]
\begin{center}
\begin{tikzpicture}[scale=1] 
\tikzset{ 
Lloop/.style={->,to path={ 
.. controls +(285:2) and +(375:2) .. (\tikztotarget) \tikztonodes}}
} 
\tikzset{ 
Rloop/.style={->,to path={ 
.. controls +(165:2) and +(255:2) .. (\tikztotarget) \tikztonodes}}
} 
\tikzset{ 
above loop/.style={->,to path={ 
.. controls +(45:2) and +(135:2) .. (\tikztotarget) \tikztonodes}}
} 
\path (0,0) node[draw,shape=circle] (v0) {$v_0$}; 
\draw[thick] (.2,.33) edge[above loop] node[above] {$x$} (-.2,.33) ;
\draw[thick, rotate=120] (.2,.33) edge[above loop] node[left] {$y$} (-.2,.33);
\draw[thick, rotate=240] (.2,.33) edge[above loop] node[right] {$z$} (-.2,.33);
\end{tikzpicture}
\caption{Quiver corresponding to $\C^3$.} 
\label{fig:C3quiver}
\end{center}
\end{figure}

The compact spectrum of $\complex^3$ consists only of the D0-brane. The network is 
shown in Figure \ref{fig:c3phase0}, more fully decorated in Figure 
\ref{fig:samplec3},
\footnote{In this section we plot in a variable 
$w=\frac{x}{1/4-x}$ such that the puncture at $x=\infty$ lies at finite distance. As 
a visual benefit Figure \ref{fig:c3phase0} is easily recognized as a subset of Figure 
\ref{fig:P2LargeVolume}.}
and rendered on $\Sigma$ in Figure \ref{fig:d0upstairs}.  
There are three self-intersection points in the network.  One intersection is at the 
branch point.  The intersection point at the left in Figure \ref{fig:c3phase0} lifts 
to two intersection points on $\Sigma.$ These three intersection points have the same 
orientation and are the matter fields in the quiver with a single vertex and three loops shown in Figure \ref{fig:C3quiver}.
The superpotential $\cW = \Tr \left(xyz - xzy \right)$ arises from the two holomorphic 
disks with opposite orientation shown in Figure \ref{fig:d0upstairs}.

\paragraph{Higher framing}

\begin{figure}[htbp]
\centering
\includegraphics[width=0.6\textwidth]{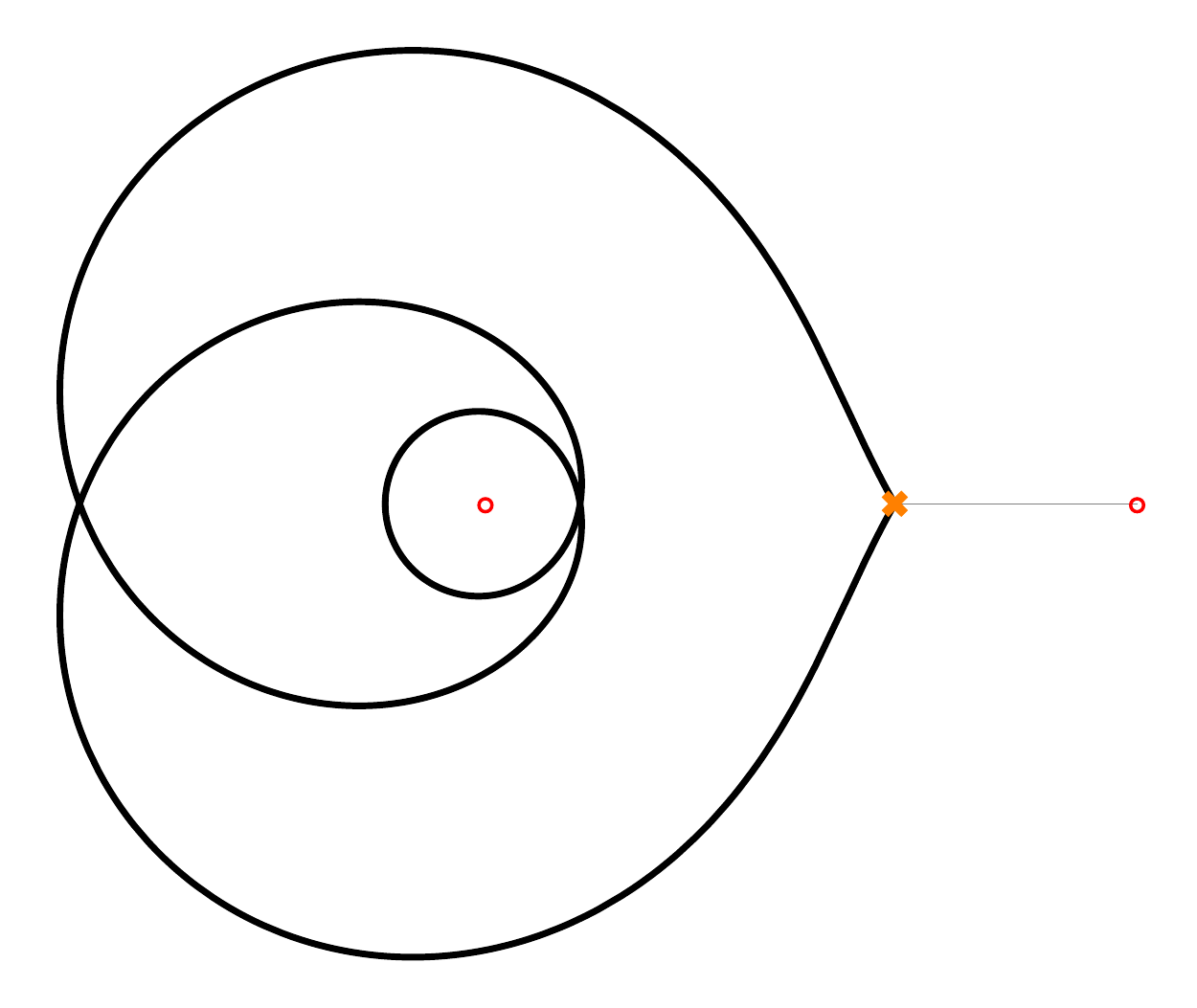}
\caption{The D0 brane at framing where the mirror curve is cubic in $y$.}
\label{fig:d0framing3}
\end{figure}

As a brief consistency check we verify that the D0 brane exists at higher framing and that 
its mass is independent of the framing. Figure \ref{fig:d0framing3} shows the D0 brane at 
a framing such that the mirror curve is cubic in $y$. The additional self-intersection 
points on $C$ are absent on $\Sigma$ because the strands lift to different sheets. 

\paragraph{Moduli of the D0-brane}
\begin{figure}[tp]
\centering
\includegraphics[width=0.6\textwidth]{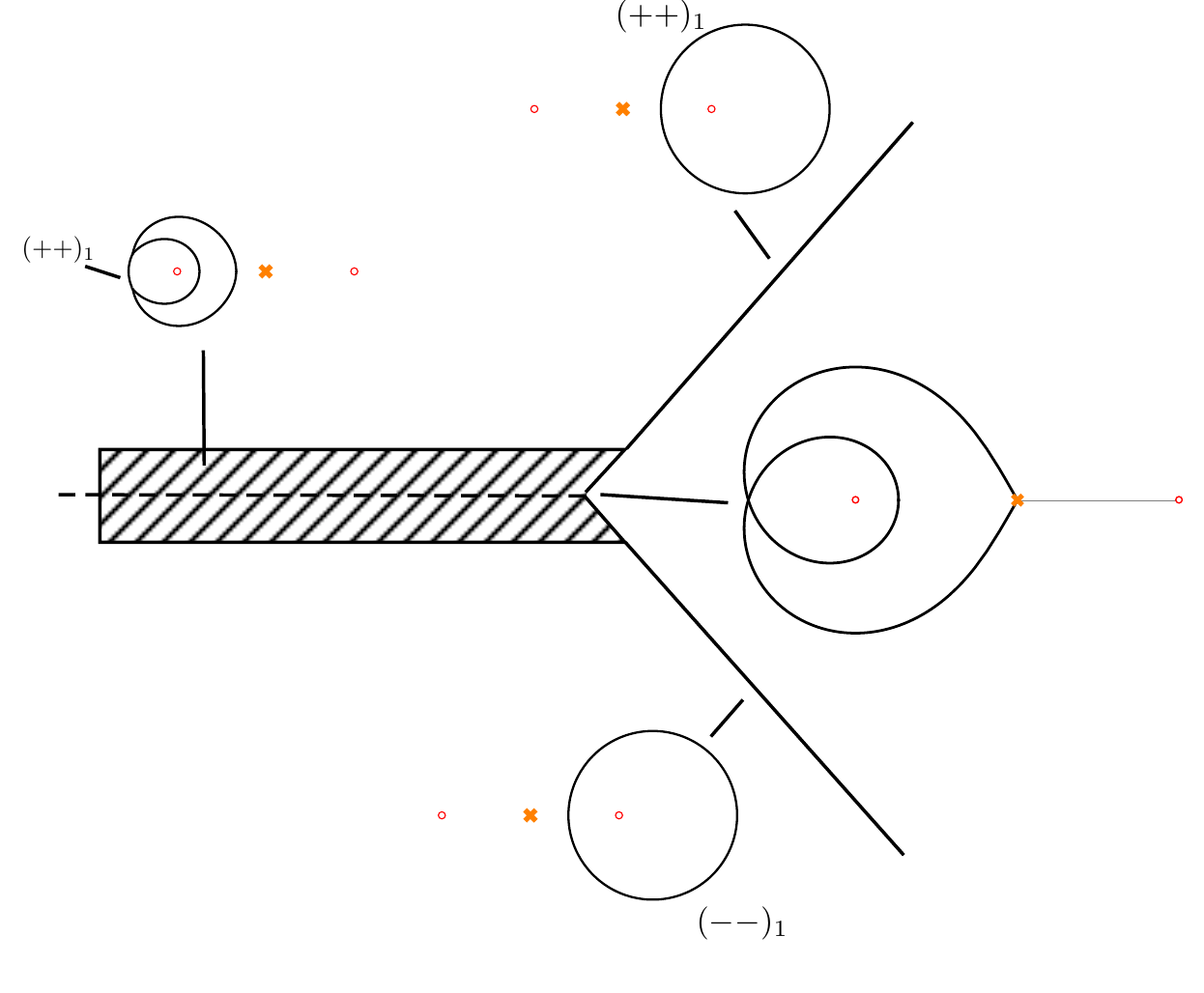}
\caption{Moduli of D$0$-brane on $\C^3$.}
\label{fig:d0moduli}
\end{figure}
The finite web corresponding to the fixed point D0 brane has the following moduli available 
for deformation. The first modulus is detaching the finite web from its branch point 
anchor, i.e. moving it towards the waist in the pair of pants shown in Figure 
\ref{fig:d0upstairs}. The second modulus is to resolve the left intersection point by 
opening a $(++)_1/(- -)_1$ strand according to the junction 
rule. The resulting moduli space is drawn in Figure \ref{fig:d0moduli}. A generic web 
with both moduli turned on is shown as a member of the fat strip in Figure 
\ref{fig:d0moduli}. The size of that strand is arbitrary and can be grown until it eats 
up the entire finite web. If the finite web is attached to the branch point, the bubble can 
detach on the other side as shown along both edges in Figure \ref{fig:d0moduli}, which 
corresponds to moving towards on one of the legs.  Note that the vertical reflection 
symmetry implements the interchange of$(++)_1$ and $(- -)_1$ in both figures.

\bigskip

\hfill
\parbox[b]{8cm}{{\it
Alice: How long is forever? \\ White Rabbit: Sometimes, just one second.} }

\ssubsection{Mirror ADHM moduli spaces}
\label{mirrorADHM}

\paragraph{Non-compact branes}
In this section, we describe non-compact D2- and D4-branes.  The D4-brane is 
represented by the strand starting from the branch point running into the puncture 
$x=0$ at $\vartheta=0$. We regularize its central charge by cutting the strand at some 
large finite mass.  Non-compact D4-branes can be used to geometrically engineer framed 
BPS states \cite{fbps}. 

Evidence for this identification includes the divergence of the central charge as 
$\log(x)^2$ (see \eqref{ncD4brane}), two oppositely oriented intersections with the 
D0-brane giving rise to the ADHM quiver in Figure \ref{fig:ADHM}
\footnote{The two extra intersections give rise to the fields $i$ and $j$ in the ADHM quiver.  The extra holomorphic disk modifies the potential to $\cW = xyz - xzy + z ij.$  The F-term relations for the $z$ field reduce to the ADHM relations after replacing $x$ and $y$ by $B_1$ and $B_2$.}, 
and a compact model in the large volume region of local 
$\mathbb{P}^2$, see Figure \ref{fig:P2LargeVolume}.  In a similar way, we identify the 
strand starting at the branch point and going into the puncture $x=\infty$ as a non-compact 
D2-brane: the central charge diverges as $\log x$ and we can also recognize it as a 
decompactified limit of the D2 brane on the resolved conifold (see also
\eqref{ncD2brane}). 

\begin{figure}[htbp]
\centering
\includegraphics[width=0.6\textwidth]{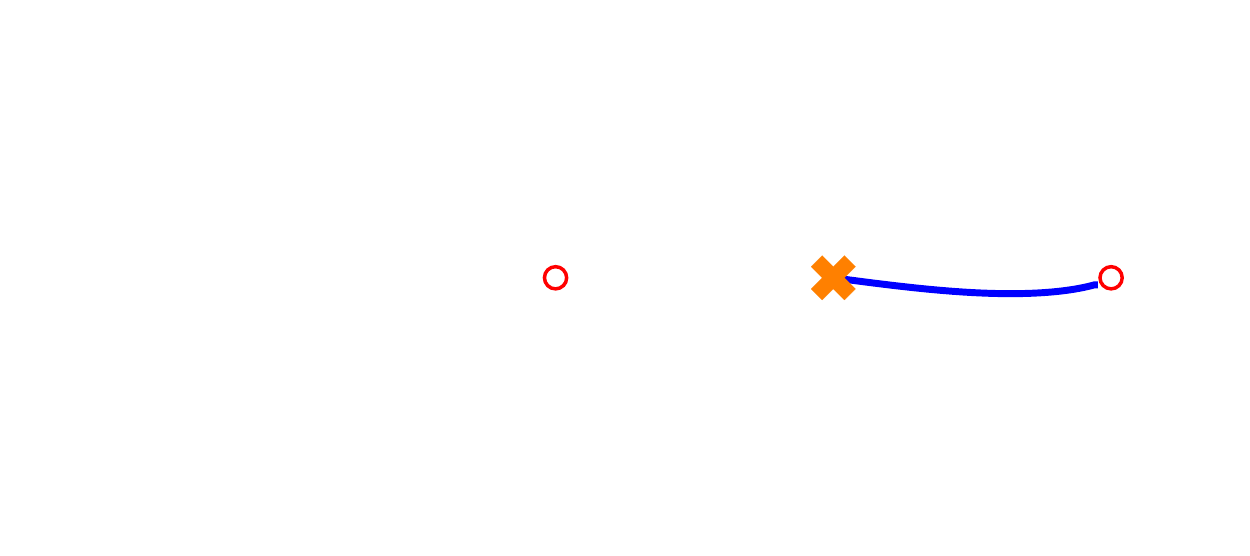}
\caption{D4 brane at non-zero B-field.}
\label{fig:c3loneD4}
\end{figure}

In addition to complexified K\"ahler moduli, on local Calabi-Yau one should also keep 
track of the B-field in non-compact directions. If B-fields $B_{12}, B_{34}$ are turned 
on in the directions parallel to the D4 brane, then to leading order in the B-field 
the phase its central charge will be given by $i(B_{12}+B_{34})$. We use this to 
identify the B-field with the phase $\vartheta_{D4}$ at which the D4 brane exists, 
see Figure \ref{fig:c3loneD4}. 

\paragraph{D0-D4 bound states}
\begin{figure}[htbp]
\centering
\includegraphics[width=0.6\textwidth]{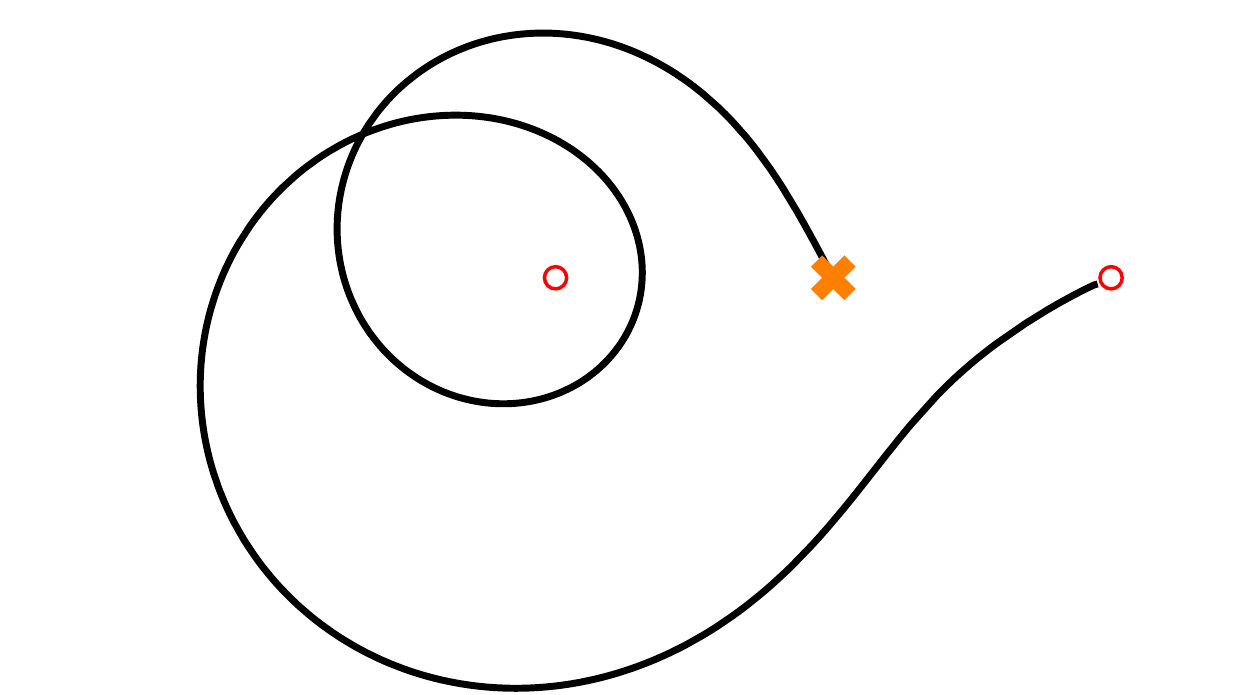}
\caption{Bound state of the D4- with the D0 brane at non-zero B-field.}
\label{fig:D4plusD0}
\end{figure}

Now we consider bound states of the non-compact D4-brane with D0-branes.
The central charge of the D0 brane is independent of the B-field, so at nonzero 
B-field the bound states of the D0 and the D4 have distinct phases and are 
nicely separated. The bound state of the D4 brane with a D0 brane is shown in 
Figure \ref{fig:D4plusD0}. The regularization is enforced by imposing that the diverging 
strand ends at precisely the same point as the pure D4 (in practice, we simply ``shoot'' 
backwards from that point). The regularized central charge obeys the usual 
additivity\footnote{This additivity of the central charge holds for open segments in 
general, whether they are near $x=0$ or not. In general an open segment could be interpreted 
as an open brane or a ``soliton'' in the sense of \cite{gmn12}, and can also form 
bound states with D0 branes. This is a natural avenue for future research.}. 

The moduli space of $k$ D0-branes sitting inside $N$ coincident D4 branes is 
isomorphic to the moduli space of $k$-instantons in $U(N)$ Yang-Mills \cite{wittenADHM, 
douglasD0D4}. However this holds at zero B-field, so the bound states all occur at 
the same phase hence are not readily studied. For $N=1$ there are no instantons in 
pure Yang-Mills, but as explained in \cite{SWnc} the inclusion of a B-field into 
the problem maps to a non-commutative deformation of the gauge theory, which does 
admit instantons precisely for $N=1$. This problem is also related to the representations 
of the Nakajima quiver described in section \ref{quiverReps}, and we now make the 
connection with spectral networks.

According to Nakajima's theorem
, and following our identification of finite webs with toric fixed points in moduli space, we should be able to associate a finite web corresponding to a bound state of the D4 and $k$ D0 branes to each partition of $k$. As a guiding principle in this quest we postulate the row-column duality on Young diagrams is implemented on finite webs by exchanging all $(++)$ and $(--)$ strands.

\begin{figure}[htbp]
\hfill\parbox[b]{8.5cm}{{\it
``and what is the use of a book," thought Alice,\\
 ``without pictures or conversations?"\\ } } 
\centering
\includegraphics[width=0.8\textwidth]{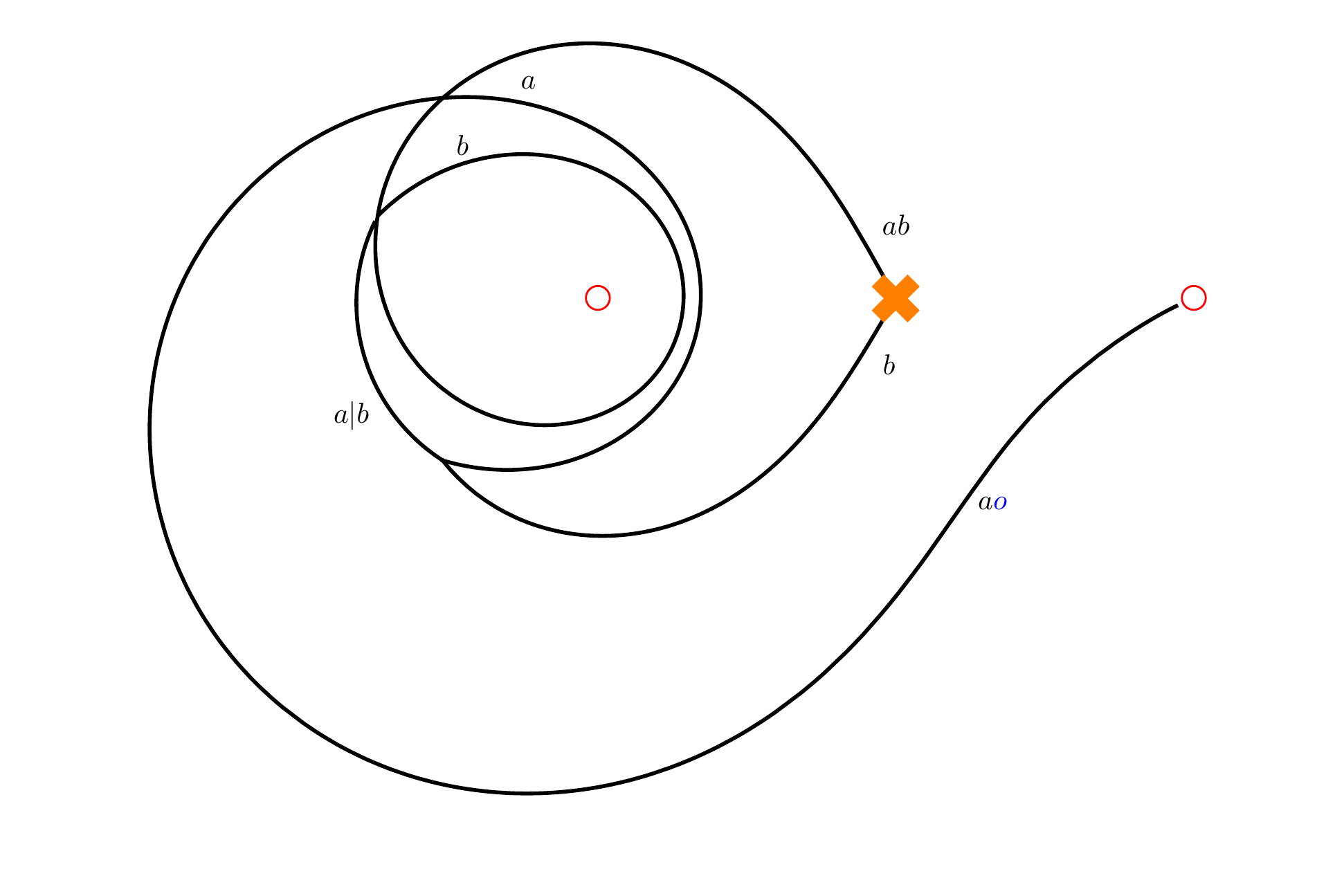}
\caption{Bound state of the D4 brane with two D0 branes.}
\label{fig:D4plus2D0}
\end{figure}

\begin{figure}[htbp]
\centering
\includegraphics[width=0.8\textwidth]{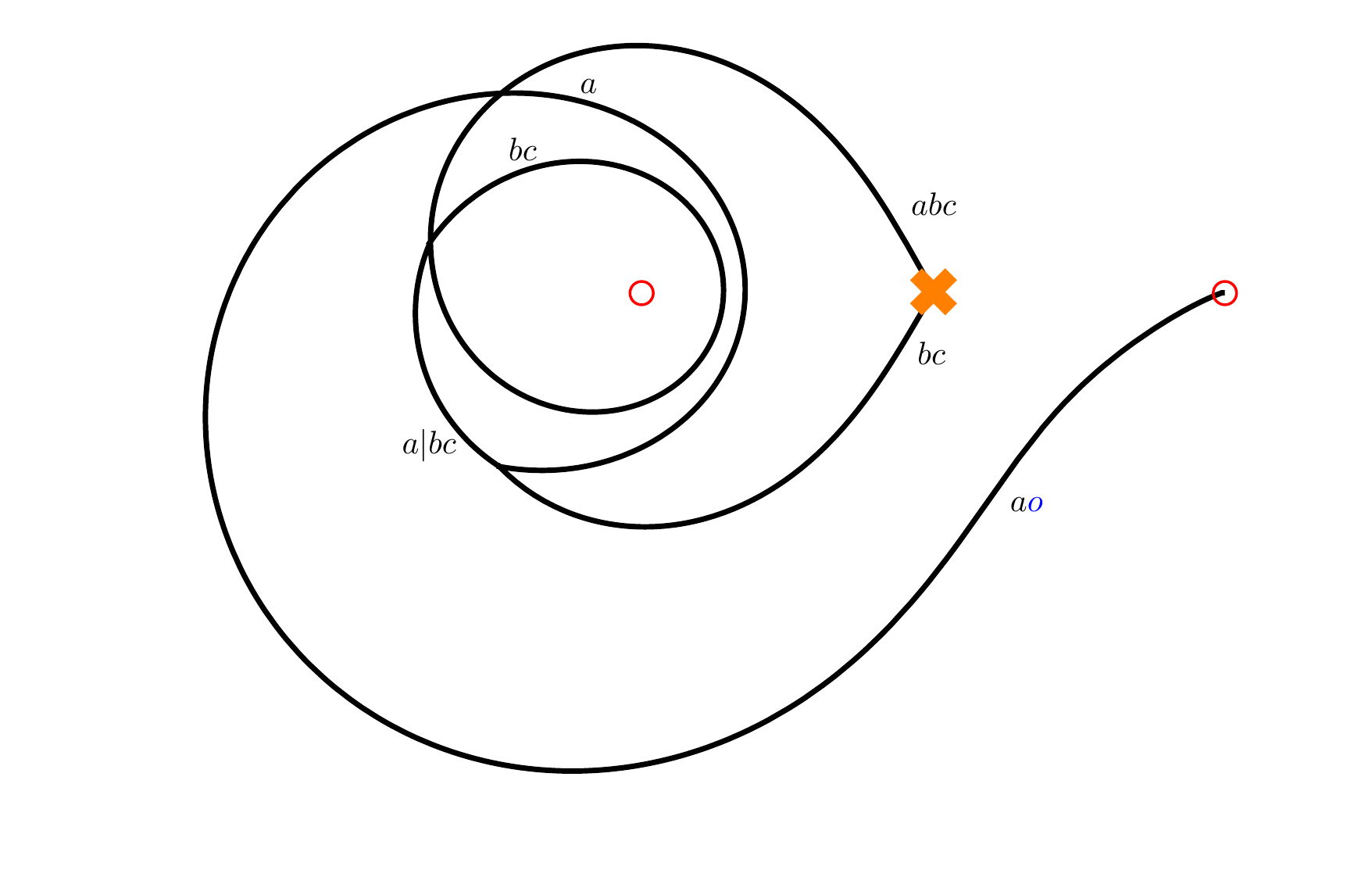}
\caption{D4 + $3$D0, one $(+-)_1$ street corresponding to the partition (2,1).}
\label{fig:D4plus3D0two}
\end{figure}

\begin{figure}[htbp]
\centering
\includegraphics[width=0.8\textwidth]{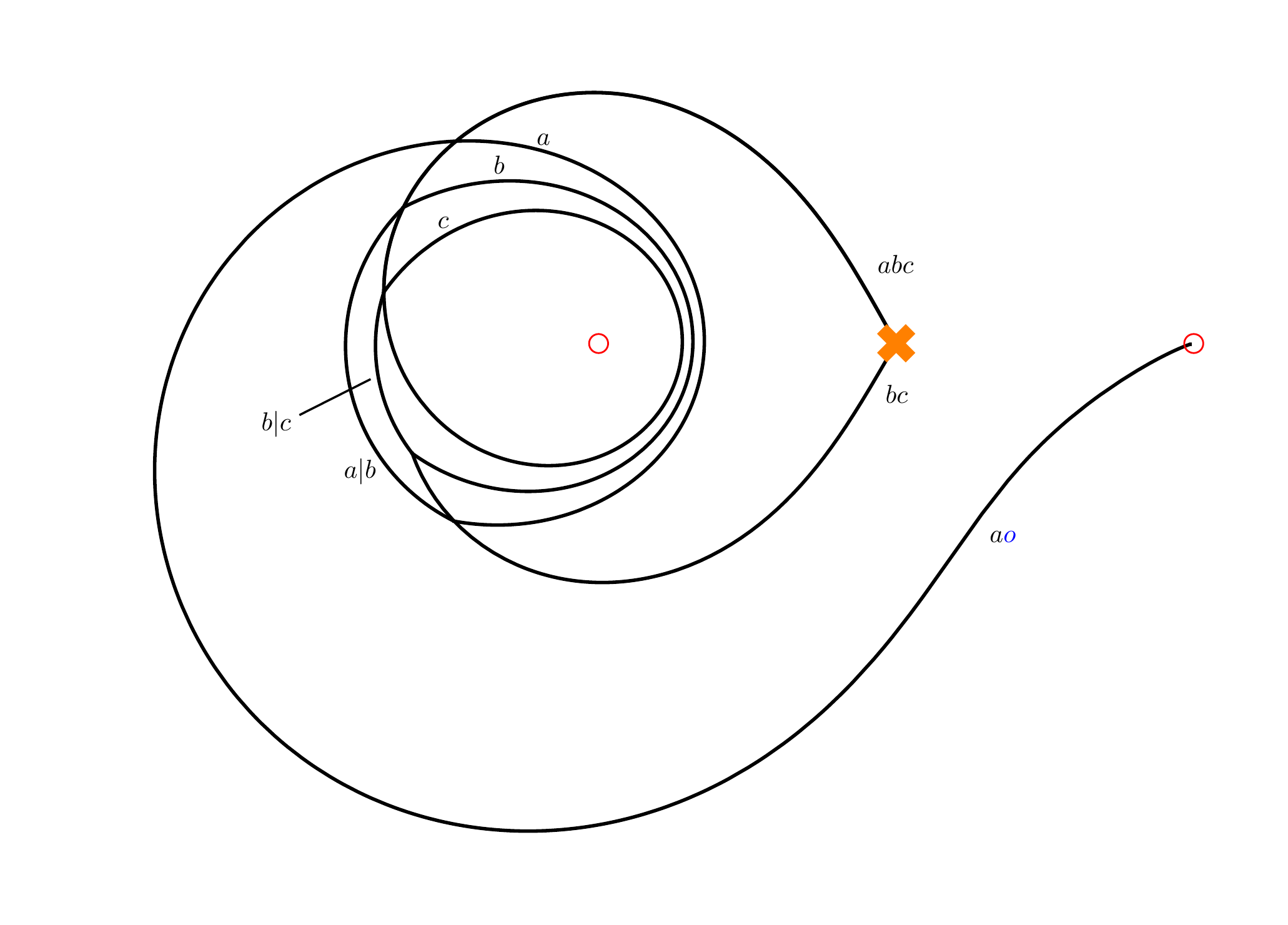}
\caption{D4 $+$ $3$D0, two $(++)/(--)$ streets corresponding to the partitions (3) and (1,1,1).}
\label{fig:D4plus3D0one}
\end{figure}

\begin{figure}[htbp]
\centering
\includegraphics[width=0.8\textwidth]{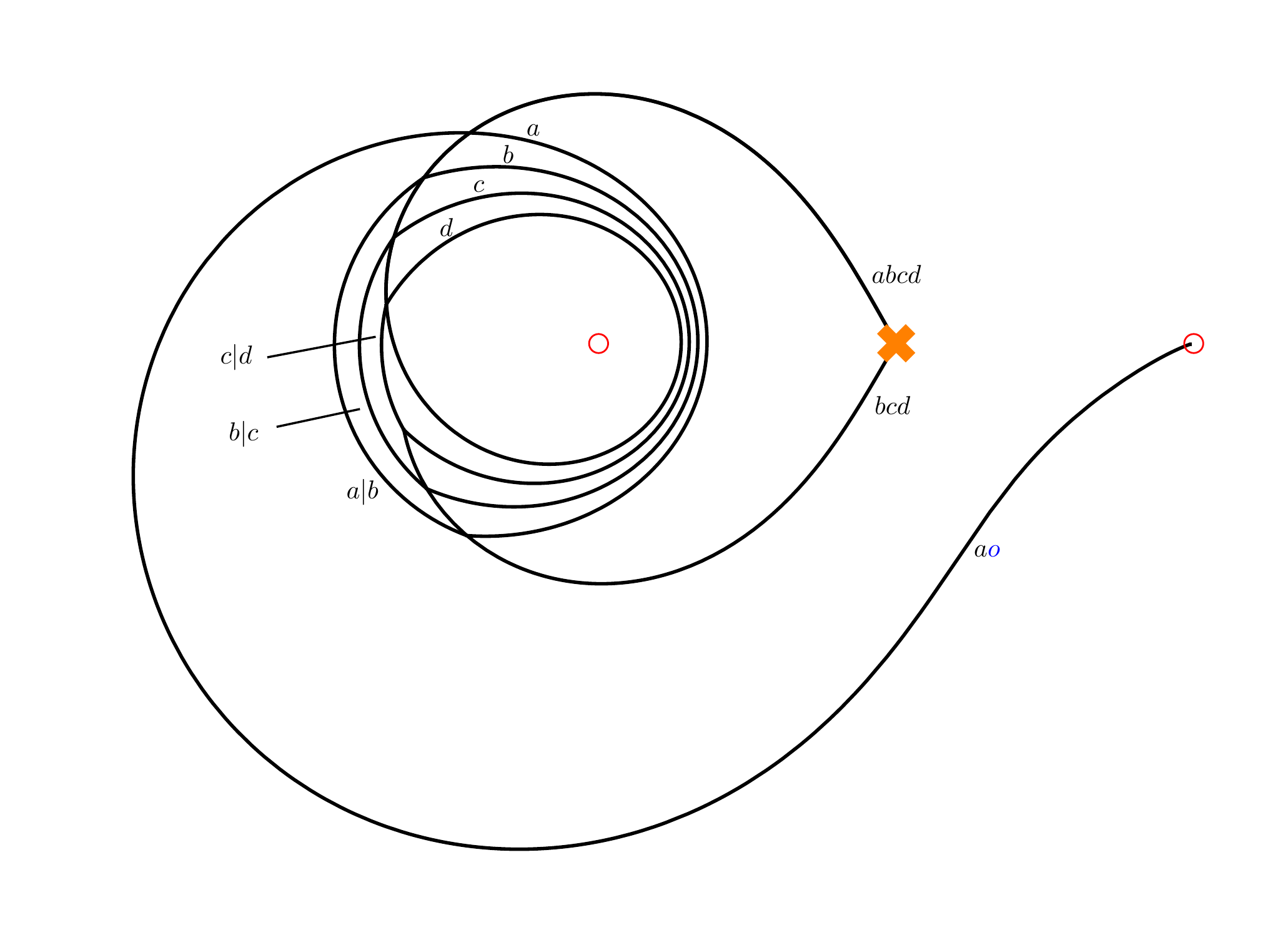}
\caption{D4 + $4$D0, corresponding to the partitions (4) and (1,1,1,1).}
\label{fig:D4plus4D0column}
\end{figure}

Figures \ref{fig:D4plus2D0} through \ref{fig:D4plus4D0column} show finite webs 
corresponding to bound states of the D4 brane with a small number of D0 branes. 
The connection to covering quivers appears to work in a similar way as for the Kronecker-3 quiver 
described earlier. Nonzero entries of the $B_1$ and $B_2$ matrices correspond to a connection 
between basic constituents of the bound state (here, the D4 brane and each separate 
D0 brane). We identify a left-right symmetry breaking arrow in the covering quiver with a connection through a $(++)$ or $(--)$ strand. 
The labels in the diagrams indicate the connection structure. In this section 
instead of colors we use the label $a|bc$ for a strand born of strands with labels $a$ 
and $bc$.

The bound state of the D4 with two D0's is shown in Figure \ref{fig:D4plus2D0}. There is a single $(++)$ or $(--)$ strand that resolves the intersection corresponding to $B_1$ or $B_2.$  The choice determines whether the corresponding partition is (2) and (1,1) which are transposes of each other.

\begin{equation}
\begin{tikzcd}
             \C_b &    \\
                  & \C_a \arrow[lu, "B_1",swap] \\
	            &         {\color[rgb]{0.000000,0.000000,1.000000}\C_o}    \arrow[u, "i"]  
\end{tikzcd}
\end{equation}
The finite web for the (2,1) partition of 3 is shown in Figure \ref{fig:D4plus3D0two}. The covering quiver is left-right symmetric and accordingly there is no $(++)$ or $(--)$ strand, only a $(+-)_1$ strand.

\begin{equation}
\begin{tikzcd}
  \C_b      &                    & \C_c      \\ 
       	  &        \C_a  \arrow[lu, "B_1",swap] \arrow[ru,"B_2"]  & \\
	         	  &        {\color[rgb]{0.000000,0.000000,1.000000}\C_o}   \arrow[u, "i"]  &
\end{tikzcd}
\end{equation}
The finite webs for partitions (3) and (1,1,1) of $3$ are shown in Figure \ref{fig:D4plus3D0one}, where the resolutions are either both $(++)$ or $(--)$.

\begin{equation}
\begin{tikzcd}
      \C_c &       &     \\
          &   \C_b \arrow[lu, "B_1",swap] &    \\
          &        & \C_a \arrow[lu, "B_1",swap] \\
	    &        &        {\color[rgb]{0.000000,0.000000,1.000000}\C_o}   \arrow[u, "i"]  
\end{tikzcd}
\end{equation}

Recall that the $B_1$ and $B_2$ matrices obtained from the covering quiver satisfy the $F$-term relation $[B_1,B_2] + ij =0$ by construction. Mnemonically the $F$-term relation is equivalent to stability of the covering quiver if it is lying on a wedge and gravity points down the page. We use this condition to eliminate many finite webs, including the one in Figure \ref{fig:D4plus3D0one} with a mixed labeling of $(++)$ and $(--)$.

The partitions of $k$ of type row or column generalize easily to a family of finite webs with $k$ strands of type $(++)$ or $(--)$. The finite web for those partition of $4$ is shown in Figure \ref{fig:D4plus4D0column}. 

\begin{equation}
\begin{tikzcd}
\C_d  &      &       &     \\
     & \C_c \arrow[lu, "B_1",swap] &       &     \\
     &     &   \C_b \arrow[lu, "B_1",swap] &    \\
     &     &        & \C_a \arrow[lu, "B_1",swap] \\
     &	    &        &         {\color[rgb]{0.000000,0.000000,1.000000}\C_o}  \arrow[u, "i"]  
\end{tikzcd}
\end{equation}

\begin{figure}[htbp]
\hfill\parbox[b]{10cm}{{\it
Why, sometimes I've believed as many as six impossible things before breakfast.} 
(The Queen of Hearts)}
\centering
\includegraphics[width=0.8\textwidth]{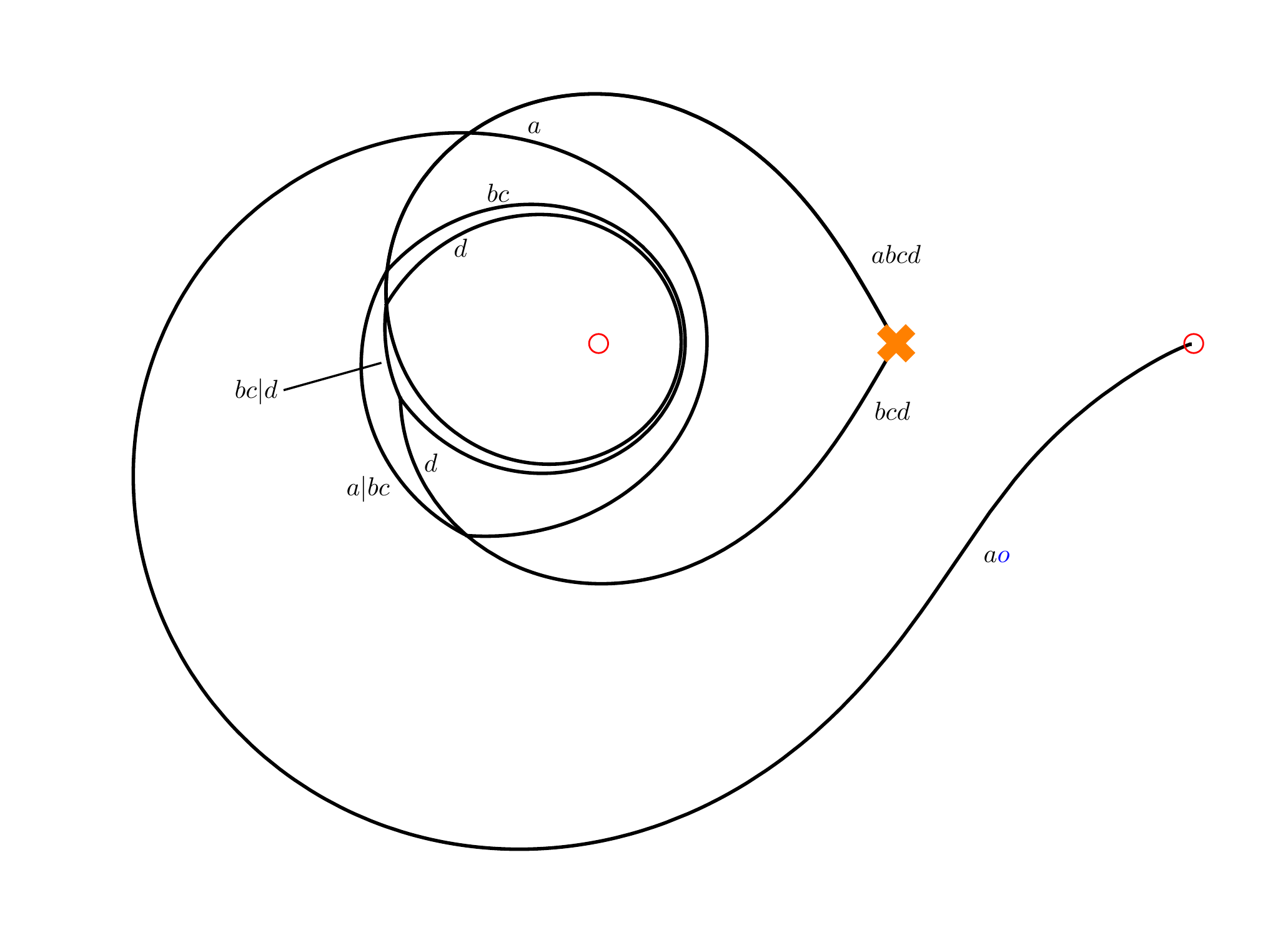}
\caption{D4 + $4$D0, corresponding to the partition (2,2).}
\label{fig:D4plus4D0square}
\end{figure}

The finite web for the square partition, (2,2), of 4 is shown in Figure 
\ref{fig:D4plus4D0square}. There is a $(+-)_1$ strand corresponding to $a|bc$ 
followed by a $(-+)_1$ strand $bc|d$, consistent with the decomposition of the 
covering quiver.  There are no $(++)$ or $(--)$ strands since the (2,2) partition 
is its own transpose.

\begin{equation}
\begin{tikzcd}
          &        \C_d         &   \\
  \C_b    \arrow[ru,"B_2"]   &                    & \C_c    \arrow[lu,"B_1",swap]  \\ 
       	  &        \C_a  \arrow[lu, "B_1",swap] \arrow[ru,"B_2"]  & \\
	         	  &        {\color[rgb]{0.000000,0.000000,1.000000}\C_o}   \arrow[u, "i"]  &
\end{tikzcd}
\end{equation}

\begin{figure}[thbp]
\centering
\includegraphics[width=0.8\textwidth]{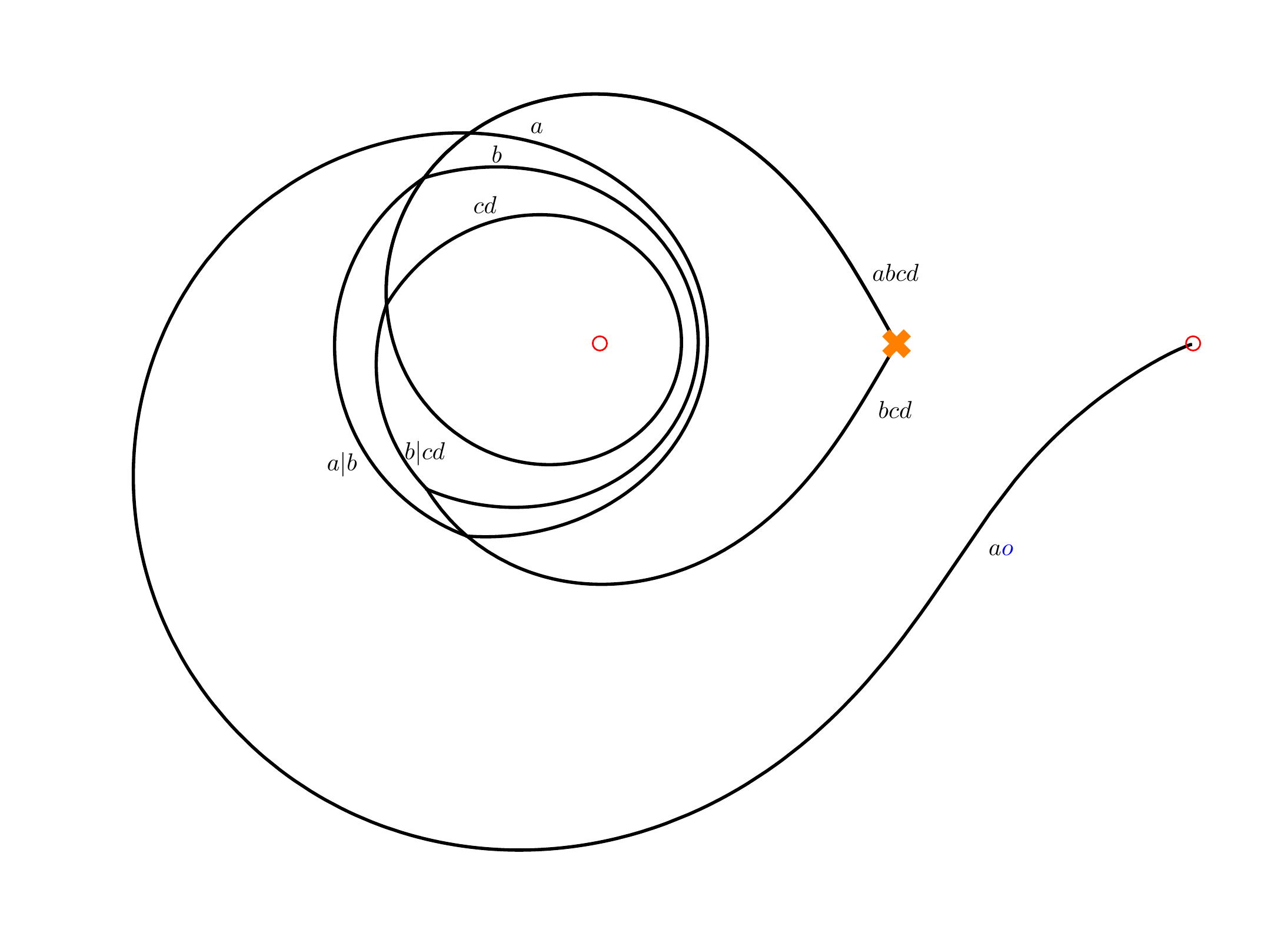}
\caption{D4 + $4$D0, ``L" partition.}
\label{fig:D4plus4D0L}
\end{figure}

The partition (3,1) of 4 is the first one to which we do not know how to associate 
a finite web. There is a reasonable looking candidate shown in Figure \ref{fig:D4plus4D0L}, 
but the associated covering quiver does not satisfy the $F$-term relation.

\bigskip

\newpage

\hfill\parbox[b]{7cm}{{\it
Imagination is the only weapon in the\\ war against reality.} (The Cheshire Cat)}

\ssection{Conclusions}

In this paper, we have studied the geometric B-model description of BPS states in 
$\caln=2$ theories admitting a spectral curve parametrization. One of our main results
is a direct and systematic relationship between geodesic networks and
quiver representations arising from tachyon condensation. Although our main
focus has been on ``exponential networks'' capturing BPS spectra of local 
Calabi-Yau manifolds, our method also produces new results for the ordinary
spectral networks of Gaiotto-Moore-Neitzke. (A related way of associating representations to 
homotopy classes of WKB geodesics in triangulated surfaces appears in 
\cite{Gentle, lf, TinkerToys}.)  We hope that a complete understanding of
the translation between quiver representations and WKB geodesics will
eventually 
resolve the matching between partitions and networks in Section \ref{mirrorADHM}.

Although we have seen that many of the constructions from \cite{gmn0, gmn12} can 
be extended to the mirrors of local Calabi-Yau manifolds, there are many other aspects 
of spectral networks that should have interesting generalizations to exponential 
networks.  In Section \ref{mirrorADHM}, we have identified the mirror of non-compact 
D2- and D4-branes with non-compact cycles on the Riemann surface.  These branes can 
be used to study coupled 3d-5d wall-crossing in a way similar to \cite{gmn11}.  
Another generalization is to extract spin content from an exponential network, 
revealing information about the Hodge polynomial of the moduli space of the corresponding 
quiver quantum mechanics \cite{glw}.  Finally, exponential networks can be used to 
lift 4d gauge theory BPS state counting problems to 5d gauge theory BPS state counting 
problems.  An example of this type of lift is the generalization from non-relativistic 
integrable systems to relativistic integrable systems.
Spectral networks corresponding to the periodic nonrelativistic Toda system were used 
to compute traces of holonomies on the associated cluster variety \cite{Williams}.  
Applying the construction of \cite{GK} to the local Calabi-Yau manifolds $Y^{p,q}$ 
results in a family of generalized periodic relativistic Toda systems \cite{efs}.  
It would be interesting to give a string-theoretic derivation of the Goncharov-Kenyon 
integrable systems \cite{GK} using exponential networks.

One of the most promising avenues is giving concrete B-model descriptions of phenomena 
mirror to well-known A-model constructions.  For instance, it should be possible to 
give a mirror description of the Betti numbers of the moduli space of sheaves on 
$\mathbb{P}^2$ \cite{yo1, yo2} using exponential networks.  A more intriguing challenge 
is to see if there is a B-model explanation for the appearance of mock-modularity in 
these moduli problems.  Perhaps the most pressing challenge is to find a description 
of the D6-brane on the spectral curve side and its effect on the stability of
exponential networks. One of the characteristic properties of the D6-brane
is that it does not arise as the boundary of any A-brane.  Therefore, it seems 
unlikely that the mirror of a D6-brane would be a one-dimensional submanifold of the spectral curve
since any such trajectory could always be terminated at a point on the curve and therefore would in fact bound a B-brane.
We suspect that for a concrete mirror version of Donaldson-Thomas
theory in the local case, one will have to reckon with a description of the 
framing that is non-geometric on the spectral curve.

Finally, there should be a direct proof of the Kontsevich-Soibelman wall-crossing formula 
\cite{ks1} for the mirrors of local Calabi-Yau manifolds directly in terms of exponential 
networks similar to the one in \cite{gmn12}.

\begin{acknowledgments}
We thank Murad Alim for extensive discussions on the material in section \ref{C3} 
and for collaboration on a related project. We also thank Heeyeon Kim,
Wolfgang Lerche, Andy Neitzke, and Nick Warner for valuable discussions. 
The work of S.S.\ was supported in part by a Grant from the 
FQRNT (Fonds Qu\'eb\'ecois de Recherche - Nature et Technologies).
\end{acknowledgments}

\appendix

\section{Coding Advice for Exponential Networks}
\label{sec:coding}
Recall from equation \ref{calibratedeq} the differential
\begin{equation}
\lambda_{(ij)_n}:=\bigl(\log y_j-\log y_i + 2\pi\ii n\bigr) d\log x
\end{equation}
where the $y_i$ is the $i$-th local solution of $H(x,y)=0$.
We now explain some details of solving for the integral geodesics of equation (\ref{geodesiceq})
\begin{equation}
\label{geodapp}
e^{-i\vartheta} \lambda_{(ij)_n}  = dt
\end{equation} 
numerically.
Let $v_{i,j}(t) = \log y_{i,j}(t)$ and absorb the $2 \pi i n$ into the choices of branch 
cuts.  Then in these variables, equation (\ref{geodapp}) becomes
\begin{equation}
\label{appgeo}
\left[v_i(t) - v_j(t) \right] \frac{\dot{x}(t)}{x(t)}  = e^{i \vartheta}.
\end{equation}
Locally, away from a branch point, the system of equations
\begin{equation}
\begin{split}
\dot{x}(t)  & =  \frac{e^{i \vartheta} x(t) }{\left(v_i(t)- v_j(t) \right)}, \\
\dot{v}_i(t) & = - \frac{\partial_x H(x,v)}{\partial_{v} H(x,v)} \dot{x}(t), \\
\dot{v}_j(t) & = - \frac{\partial_x H(x,v)}{\partial_{v} H(x,v)} \dot{x}(t). 
\end{split}
\end{equation}
determine the local evolution of the network.
However, at a branch point, the lifts $v_{i,j}(t)$ coincide, and $v_i(t) = v_j(t).$  
Therefore a more careful analysis is necessary.
Consider the covering of $C$ by $\Sigma$ around a branch point
$p_* = (x_*, y_*)$ of order $n$
\begin{equation}
H(p + p_*) = \pp{H}{x} \Big \vert_{p = p_*}(x - x_*) + \frac{1}{n!} \pp{^nH}{y^n} \Big 
\vert_{p = p_*} (y - y_*)^n.
\end{equation}
We rearrange the equation as
\begin{equation}
(x - x_*) = \kappa (y - y_*)^n
\end{equation}
where
\begin{equation}
\kappa = -\frac{1}{n!} \pp{^nH}{y^n} \frac{1}{\partial H / \partial x}.
\end{equation}
For simplicity, we now restrict our attention to when $n=2$. Solving for $y$,
\begin{equation}
y_{\pm} = y_{*} \pm \frac{\sqrt{z}}{\sqrt{\kappa}},
\end{equation}
where $x \rightarrow z + x_*$.
We then have the approximation
\begin{equation}
\begin{split}
\log y_{+} - \log y_{-} & = \log (1 +  \frac{\sqrt{z(t)}}{\sqrt{\kappa}y_{*}}) - 
\log (1 -  \frac{\sqrt{z(t)}}{\sqrt{\kappa}y_{*}}) \\
& \approx  2 \frac{\sqrt{z(t)}}{\sqrt{\kappa} y_{*}}.
\end{split}
\end{equation}
Substituting into equation (\ref{appgeo}), we have
\begin{equation}
\frac{2 \sqrt{z(t)}}{\sqrt{\kappa} y_{*} x_{*}} \dot{z}(t) = e^{i \vartheta}.
\end{equation}
Integrating from $t = 0$ to $\Delta t$, we have
\begin{equation}
\frac{4}{3} \frac{1}{\sqrt{\kappa} y_* x_*} (x - x_*)^{3/2} = e^{i \vartheta} (\Delta t).
\end{equation}
Solving for $x,$
\begin{equation}
x = x_{*} + \left( \frac{3}{4} \sqrt{\kappa} y_* x_* e^{i \vartheta} (\Delta t) \right)^{2/3}.
\end{equation}
To solve for $y$ with a consistent choice of branches,  we use the following trick:
\begin{equation}
(x - x_*)^{1/2} = \frac{3}{4} \frac{\sqrt{\kappa} y_* x_* e^{i \vartheta} (\Delta t)}{(x - x_*)}.
\end{equation}
We substitute this into
\begin{equation}
\begin{split}
 (y - y_*) & = \pm \frac{(x - x_*)^{1/2}}{\sqrt{\kappa}}, \\
y &= y_* \pm \frac{3}{4} \frac{ y_* x_* e^{i \vartheta} (\Delta t)}{(x - x_*)}.
\end{split}
\end{equation}
The analysis for coverings of higher degree is similar.

\section{Central Charges of Local Calabi-Yau Manifolds}
\label{sec:centralcharge}
The Picard-Fuchs equations for local $\mathbb{P}^2$ are \cite{agm, aspinwallDbranes}
\begin{equation}
\left( q \dd{}{q} \right)^3 \Phi + 27q \left(q \dd{}{q} \right)\left(q \dd{}{q} 
+ \frac{1}{3} \right)\left(q \dd{}{q} + \frac{2}{3}\right) = 0.
\end{equation}
\subsection{Orbifold point}
Near the orbifold point, we switch to the coordinate $\psi,$ where 
$q = (-3 \psi)^{-3}.$  Then a basis of solutions to the Picard-Fuchs equation is
\begin{equation}
\varpi_j = \frac{1}{2 \pi i} \sum_{n=1}^{\infty} \frac{\Gamma(n/3) 
\omega^{n j}}{\Gamma(n+1)\Gamma(1 - n/3)^2} ( 3 \psi)^n
\end{equation}
where $\omega = \exp(2 \pi i/3).$  The central charges of the fractional branes 
near the orbifold point are given by
\begin{equation}
\begin{split}
Z(F_0) & = 1/3 (1 - \varpi_0 + \varpi_1) \\
Z(F_1) & = 1/3 (1 - \varpi_0- 2 \varpi_1) \\
Z(F_2) & = 1/3 (1 + 2 \varpi_0 + \varpi_1),
\end{split}
\end{equation}
and are plotted in Figure \ref{fig:evolution}.

\subsection{Large volume solutions}
Near large volume the periods take the form
\begin{equation}
\begin{split}
\Pi_0(q) & = 1 \\
\Pi_1(q) & = \frac{1}{2 \pi i} \left( a_0(q) \log q + a_1(q) \right) \\
\Pi_2(q) & = \frac{1}{(2 \pi i)^2} \left( a_0(q) \log^2 q + 2 a_1(q) \log q + a_2(q) \right)
\end{split}
\end{equation}
where
\begin{equation}
\begin{split}
a_0(1) & = 1 \\
a_1(q) & = -6 q + 45 q^2 - 560 q^3 + \frac{17325}{2} q^4 + \cdots \\
a_2(q) & = -18 q + \frac{423}{2} q^2 - 2972 q^3 + \frac{389415}{8} q^4 + \cdots
\end{split}
\end{equation}
The central charge of a brane $\fB$ is given by the hemisphere partition function 
\cite{horiromo, ho, sugi}
\begin{equation}
Z_{D^2}(\mathfrak{B}) = d \sigma \int  \Gamma \left(-\frac{3 \epsilon }{2}-3 \sigma \right) 
\Gamma \left(\frac{\epsilon }{2}+\sigma \right)^3 
e^{t \sigma} f_{\mathfrak{B}}(\sigma).
\end{equation}
For the D4-brane $\cO (\pm k)$, 
\begin{equation}
f_{\fB}(\sigma) = e^{-2 \pi  i k \sigma } \left(e^{3 i \pi  \sigma } - e^{- 3 i \pi  \sigma }  \right).
\end{equation}
We find 
\begin{equation}
\begin{split}
Z_{D^2}(\cO)       &= \Pi_2(z) + \Pi_1(z) + 1/2 \\
Z_{D^2}(\cO(-1))  &= \Pi_2(z) - \Pi_1(z) + 1/2 \\
Z_{D^2}(\cO(k))   &=  \Pi_2(z) + (2 k + 1 ) + (k^2 + k + 1/2).
\end{split}
\end{equation}
\section{Representations of Quivers with Superpotential}
\label{appendixC}
In section \ref{conifold} we used the folklore result that the representation theory 
of the conifold quiver effectively reduces to the representation theory of the 
Kronecker-2 quiver.  This means that either only the ``A'' or only the ``B'' 
arrows can be non-zero in an indecomposable representation of the conifold quiver, 
except for when the dimension vector is a multiple of $(1,1)$.  We justify this 
statement in the simple case of dimension vector 
$(2,1)$.  For other representations, the analysis is similar but more involved.
For local $\mathbb{P}^2$ a similar analysis was performed in \cite{dfr}.  In general, it is natural to conjecture that only the representations corresponding to the D0-brane involve all of the fields in the quiver.  For local Calabi-Yau manifolds that arise as cones over surfaces, the representations are expected to restrict to representations of an acyclic quiver corresponding to the surface.  In the example of local $\mathbb{P}^2$ either the $a^i, b^j,$ or $c^k$ fields will be zero in a representation not corresponding to the D0-brane, reducing the representation theory to that of the Beilinson quiver for $\mathbb{P}^2$.

Recall that the conifold quiver has superpotential
$\cW = a_1 b_1 a_2 b_2 - a_1 b_2 a_2 b_1$ and F-term (Jacobian) equations
\begin{equation}
\begin{split}
a_1 b_1 a_2 - a_2 b_1 a_1 & = 0 \\
a_1 b_2 a_2 - a_2 b_2 a_1 & = 0 \\
b_1 a_1 b_2 - b_2 a_1 b_1 & = 0 \\
b_1 a_2 b_2 - b_2 a_2 b_1 & = 0.
\end{split}
\end{equation}
We consider a representation with dimension vector $(2,1)$.  Then a 
representation will take the form
\begin{equation}
\begin{split}
A^1 & = \begin{pmatrix} a^{1}_{1} & a^{1}_{2} \end{pmatrix}, \\
A^2 & = \begin{pmatrix} a^{2}_{1} & a^{2}_{2} \end{pmatrix}, \\
B^1 & = \begin{pmatrix} b^{1}_{1} \\ b^{1}_{2} \end{pmatrix}, \\
B^2 & = \begin{pmatrix} b^{2}_{1} \\ b^{2}_{2} \end{pmatrix}. 
\end{split}
\end{equation}
When expanded out in components, the first two F-term relations yield
\begin{equation}
\begin{split}
\Delta(A)  \begin{pmatrix} -b^{1}_{2} & b^{1}_{1} \end{pmatrix} & = \mathbf{0} \\
\Delta(A)  \begin{pmatrix} -b^{2}_{2} & b^{2}_{1} \end{pmatrix} & = \mathbf{0},
\end{split}
\end{equation}
where  $\Delta(A) = a^{1}_{1} a^{2}_{2} -  a^{1}_{2} a^{2}_{1} $.  Thus 
either $B^1 = B^2 = \mathbf{0}$ or $\Delta(A)  = 0.$
If $\Delta(A)  = 0$, then using the remaining two F-term equations, we find 
$A^1,A^2 = \mathbf{0}$ or $\Delta(B)  = 0.$
So either $\Delta(A)  = \Delta(B)  = 0,$  or $A^1,A^2 = \mathbf{0},$ or $B^1,B^2 = 
\mathbf{0}$. In the first case, we are in the situation we wished to show, that either 
the $A$ or $B$ fields are zero. The other case where $\Delta(A) = \Delta(B)  = 0$ 
corresponds to a bound state of a $(1,1)$ and $(1,0)$ representation.  This follows 
from the fact that $\Delta(B) = 0$ implies that the common image of the maps $B_1, 
B_2$ from $\C$ to $\C^2$ is only one dimensional.

\end{document}